\documentclass[journal]{IEEEtran}
\usepackage[latin9]{inputenc}
\usepackage{color}
\usepackage{units}
\usepackage{tipa}
\usepackage{tipx}
\usepackage{amsthm}
\usepackage{amsmath}
\usepackage{amssymb}
\usepackage{stmaryrd}
\usepackage{graphicx}
\usepackage{esint}
\usepackage[unicode=true,
 bookmarks=false,
 breaklinks=false,pdfborder={0 0 0},backref=false,colorlinks=true]
 {hyperref}
\hypersetup{
 pdfauthor={Nguyen Phan Minh},
 pdfsubject={Information Theory},
 linkcolor=red, citecolor=red, pdfstartview={FitH}}

\makeatletter
\theoremstyle{plain}
\newtheorem{thm}{\protect\theoremname}
\theoremstyle{plain}
\newtheorem{lem}[thm]{\protect\lemmaname}
\theoremstyle{definition}
\newtheorem{defn}[thm]{\protect\definitionname}
\theoremstyle{plain}
\newtheorem{prop}[thm]{\protect\propositionname}

\renewcommand{\footnoterule}{
  \kern -3pt
  \hrule width 1in height 1pt
  \kern 2pt
}

\usepackage{flushend}

\allowdisplaybreaks[3]

\newcounter{MyTempEquationCounter}

\makeatother

\providecommand{\definitionname}{Definition}
\providecommand{\lemmaname}{Lemma}
\providecommand{\propositionname}{Proposition}
\providecommand{\theoremname}{Theorem}

\begin{document}

\title{On Capacity Formulation with Stationary Inputs and Application to
a Bit-Patterned Media Recording Channel Model}

\author{Phan-Minh Nguyen and Marc A. Armand, \textit{Senior Member, IEEE}%
\thanks{The paper forms parts of the first author's 
bachelor thesis submitted to the Department of Electrical and Computer Engineering, 
National University of Singapore, under the supervision of Prof. Marc A. Armand.
This work is supported by the Singapore National Research Foundation
under CRP Award No. NRF-CRP 4-2008-06. 

P.-M. Nguyen was with the Department of Electrical and Computer Engineering,
National University of Singapore, Singapore, and is now with the Department
of Electrical Engineering, Stanford University, Stanford, CA 94305,
USA (email: npminh@stanford.edu).

M.A. Armand is with the Department of Electrical and Computer Engineering,
National University of Singapore, Singapore 117576 (email: eleama@nus.edu.sg).

Copyright (c) 2014 IEEE. Personal use of this material is permitted.  
However, permission to use this material for any other purposes must be 
obtained from the IEEE by sending a request to pubs-permissions@ieee.org.
}}
\maketitle
\begin{abstract}
In this correspondence, we illustrate among other things the use of
the stationarity property of the set of capacity-achieving inputs
in capacity calculations. In particular, as a case study, we consider
a bit-patterned media recording channel model and formulate new lower
and upper bounds on its capacity that yield improvements over existing
results. Inspired by the observation that the new bounds are tight
at low noise levels, we also characterize the capacity of this model
as a series expansion in the low-noise regime.

The key to these results is the realization of stationarity in the
supremizing input set in the capacity formula. While the property
is prevalent in capacity formulations in the ergodic-theoretic literature,
we show that this realization is possible in the Shannon-theoretic
framework where a channel is defined as a sequence of finite-dimensional
conditional probabilities, by defining a new class of consistent stationary
and ergodic channels.\end{abstract}


\section{Background}

The fundamental limit of information transmission through noisy channels,
the channel capacity, has been a holy grail in information theory.
The capacity problem of a general point-to-point channel has been
well resolved with the information-spectrum framework \cite{Verdu-Han-1994}.
Such general formula for the capacity, however, does not lend itself
to computation in general, since it requires one to scrutinize the
distribution of the information density at the limit of infinite block
length. To overcome this problem, a common approach is to find an
alternative expression that, instead of being described by an information-spectrum
quantity, contains a mutual information quantity (or entropy quantities).
While an expression of this kind is not as general, it may cover a
sufficiently large class of channels for many practical purposes.

There are two popular forms of such expression. One is Dobrushin's
information-stable channel capacity \cite{Dobrushin-1963}:
\begin{equation}
C=\lim_{n\to\infty}\frac{1}{n}\sup_{X^{n}}I\left(X^{n};Y^{n}\right)\label{eq:C_of_Info_Stable}
\end{equation}
where the supremum is over all possible sequences of distributions
$\left\{ P^{\left(n\right)}:\; X^{n}\sim P^{\left(n\right)}\right\} $.
This formula holds for the class of information-stable channels. A
similar formula also appears in the context of (decomposable or indecomposable)
finite-state channels \cite{Gallager-1968}. The other form swaps
the supremum and the limit in the above formula, with the supremum
being taken over a smaller set of input distributions with special
structures. This type of formula appears in the ergodic-theoretic
literature of information theory. For example, for $\bar{d}$-continuous
discrete stationary and ergodic (SE) two-sided channels, the capacity
was shown to be \cite{Gray-1979} 
\begin{equation}
C=\sup_{\text{Stationary }\mu}\lim_{n\to\infty}\frac{1}{n}I\left(X^{n};Y^{n}\right)\label{eq:C_of_d_cont_SE_channel}
\end{equation}
where $\mu$ is a probability measure that describes the input process.
(See Section \ref{sub:Definitions} for the distinction between the
two sets being supremized over in the above formulas.)

Such capacity formulations that involve supremization over stationary
inputs are common in the ergodic-theoretic setting, where a channel
only admits infinitely long input sequences, mostly for SE channels
with memory and anticipation \cite{Gray-Entropy,Gray-1979}. In the
Shannon-theoretic framework, where an admissible input is a sequence
of finite-dimensional distributions, capacity formulations similar
to Eq. \eqref{eq:C_of_Info_Stable} are, however, more prevalent%
\footnote{An exception is the standard insertion-deletion channel, where stationary
and ergodic inputs could achieve capacity \cite{Dobrushin-1967,Kanoria-2013}.
This is a specific channel model that is beyond the scope of this
paper.%
}.

Intuitively constraints on the supremizing input set reduce the capacity-achieving
input search space and may become useful. A major aim of this paper
is to illustrate possible ways to exploit the stationarity property
of the supremizing input set in capacity calculations, via a case
study of a bit-patterned media recording (BPMR) channel model. This
model was first introduced in \cite{Iyengar-2011} and subsequently
studied in \cite{Mazumdar-2011}. To achieve ultra-high density magnetic
recording, in BPMR technologies, the data write process takes place
on a new magnetic medium comprising of magnetic islands that are separated
by non-magnetic materials. The difficulty in maintaining synchronization
between the write head's position and the correct island where data
is to be written in is captured by a channel model with paired insertion-deletion
errors, underlied by a first-order Markov process. The channel input
is the actual data to be written and the output is the data as written
on the islands. This channel model is essentially a finite-state channel
with dependent insertion-deletion (DID) errors, and is henceforth
called the DID channel model.

Both works \cite{Iyengar-2011} and \cite{Mazumdar-2011} analyze
the capacity of the DID channel using a formula of the form of Eq.
\eqref{eq:C_of_Info_Stable}. In our case study, we will contrast
this with the use of a formula that involves stationary inputs, which
is able to yield improvements and new results. In particular, the
formula we seek assumes a form similar to Eq. \eqref{eq:C_of_d_cont_SE_channel}.
We consider both Shannon-theoretic and ergodic-theoretic frameworks.
While we analyze specifically the DID channel, we believe techniques
we present to evaluate its capacity could be applicable to similar
BPMR channel models, e.g. the model in \cite{Wu-2013}, and beyond.

We give a high-level view of the calculation techniques. As in Eq.
\eqref{eq:C_of_Info_Stable} and \eqref{eq:C_of_d_cont_SE_channel},
computing the capacity generally involves an infinite-dimensional
optimization problem, which in most cases is intractable. Our idea
towards computational feasibility is to decompose $I\left(X^{n};Y^{n}\right)$
into a sum of finite-dimensional terms of the form
\[
I\left(X^{n};Y^{n}\right)\approx\sum_{i=1}^{n}f\left(\left\llbracket X\right\rrbracket _{i},\left\llbracket Y\right\rrbracket _{i},\left\llbracket Z\right\rrbracket _{i}\right)
\]
where $\left\llbracket X\right\rrbracket _{i}$ indicates a finite-size
group of terms in the neighborhood of $X_{i}$ (i.e. $\left(X_{i-a},X_{i-a+1},\ldots,X_{i+b}\right)$
for some finite non-negative constants $a$ and $b$), and the same
for $\left\llbracket Y\right\rrbracket _{i}$ and $\left\llbracket Z\right\rrbracket _{i}$,
and $f$ is a function mapping the joint distribution of $\left(\left\llbracket X\right\rrbracket _{i},\left\llbracket Y\right\rrbracket _{i},\left\llbracket Z\right\rrbracket _{i}\right)$
to $\mathbb{R}_{+}$ and is independent of the index $i$. Here $\left\llbracket Z\right\rrbracket _{i}$
can be thought of as an external source of randomness introduced by
the channel. The approximation ``$\approx$'' is to be replaced
with an exact inequality (i.e. ``$\leq$'' or ``$\geq$''). Given
the finite-dimensional distribution of each input block $\left\llbracket X\right\rrbracket _{i}$,
$f\left(\left\llbracket X\right\rrbracket _{i},\left\llbracket Y\right\rrbracket _{i},\left\llbracket Z\right\rrbracket _{i}\right)$
can be computed. However the curse of dimensionality is not yet completely
eliminated: for a general input, we have $n$ possibly different distributions
for $\left\llbracket X\right\rrbracket _{1},\left\llbracket X\right\rrbracket _{2},\ldots,\left\llbracket X\right\rrbracket _{n}$.

The problem is resolved when the input is stationary, in which we
reduce the number of said distributions from $n$ to one, which is
the lowest possible. Furthermore, for many classes of channels, input
stationarity implies stationarity of $\left\{ \left\llbracket X\right\rrbracket _{i},\left\llbracket Y\right\rrbracket _{i},\left\llbracket Z\right\rrbracket _{i}\right\} _{i=1}^{n}$.
Effectively,
\[
\frac{1}{n}I\left(X^{n};Y^{n}\right)\approx f\left(\left\llbracket X\right\rrbracket _{i},\left\llbracket Y\right\rrbracket _{i},\left\llbracket Z\right\rrbracket _{i}\right)
\]
which is now computable.

Broadly speaking, input stationarity helps reduce an infinite-dimensional
problem to a finite-dimensional one. This reduction is potentially
useful for constructing upper bounds on the capacity%
\footnote{One can always restrict attention to stationary inputs to lower-bound
the capacity regardless of whether stationary inputs can achieve the
capacity.%
}. With a ``good'' $f$, one may hope $f\left(\left\llbracket X\right\rrbracket _{i},\left\llbracket Y\right\rrbracket _{i},\left\llbracket Z\right\rrbracket _{i}\right)$
provides a tight upper bound. With a matching lower bound, the exact
capacity can be deduced. We make a note that the advantage is not
only computational, but also analytical, since we then only need to
work with a finite number of variables, instead of infinitely many.
We shall illustrate so for the case of the DID channel.

For the rest of this section, after introducing some mathematical
conventions and common definitions, we briefly discussed the two different
channel definitions that correspond to the two frameworks, and their
relation to practical channel models. We then give a brief description
of the DID channel model with known results on its capacity and outline
the main contributions of the paper.

\subsection{Mathematical Conventions}

To denote random variables, uppercase letters (e.g. $V$, $X$, $Y$)
are used, and corresponding lowercase letters (e.g. $v$, $x$, $y$)
are adopted for values they take. Their respective alphabets are in
calligraphic style (e.g. $\mathcal{V}$, $\mathcal{X}$, $\mathcal{Y}$).
$V_{a}^{b}$ denotes the vector $\left(V_{a},V_{a+1},...,V_{b}\right)^{\top}$
for $a\leq b$. At times $V^{b}$ may be used in place of $V_{1}^{b}$
(or $V_{0}^{b}$ where appropriate). Whenever the length is not specified
or is implicitly understood, the vector is written in boldface, e.g.
$\mathbf{V}$ (resp. $\mathbf{v}$), in which case $V_{a}$ (resp.
$v_{a}$) denotes its $a$-th entry. All vectors are understood to
be column vectors.

The notation $\left\{ V_{i}\right\} _{i=1}^{\infty}$ (or $\left\{ V_{i}\right\} $,
depending on the starting index) denotes a one-sided random process.
We only consider processes $\left\{ V_{i}\right\} $ in which $V_{i}\in\mathcal{V}$
$\forall i$. This restriction is technically not an issue, since
we can always expand the alphabet of each $V_{i}$ to the largest
one, with an appropriately modified probability measure that assigns
probability $0$ to any event that involves taking values that are
not in the original alphabets.

An event $E_{V}$ (on a process $\left\{ V_{i}\right\} $) is a set
of values $\mathbf{v}$ that the process takes. At times we use the
notations, e.g. $\left\{ V_{i}=a\right\} $, to refer to an event
$\left\{ \mathbf{v}:\; v_{i}=a\right\} $. A similar meaning applies
to e.g. $\left\{ V_{i}=a\;\text{or}\; V_{i}=b\right\} $. We also
write e.g. $\left\{ V_{a}^{a+b}=v_{1}v_{2}...v_{b+1}\right\} $, where
$v_{1},v_{2},...,v_{b+1}\in{\cal V}$, to mean $\left\{ V_{a}=v_{1},V_{a+1}=v_{2},...,V_{a+b}=v_{b+1}\right\} $.

The letter $X$ is used specifically for the channel input, and the
letter $Y$ is for the channel output. 

The operator $\left|\cdot\right|$ returns the length of a vector,
e.g. $\left|V_{a}^{b}\right|=b-a+1$, the absolute value of a scalar
quantity, or the size of a set. The expectation operator is $\mathbb{E}\left[\cdot\right]$.
The binary entropy function is $h_{2}(x)=-x\log x-(1-x)\log(1-x)$.
We write $\left[\mathbf{x},\mathbf{y}\right]$ to denote the concatenation
of vectors $\mathbf{x}$ and $\mathbf{y}$, i.e. $\left[\mathbf{x},\mathbf{y}\right]=\left(x_{1},...,x_{\left|\mathbf{x}\right|},y_{1},...,y_{\left|\mathbf{y}\right|}\right)^{\top}$.
When the alphabet is binary, we use $\neg\mathbf{x}$ to denote the
vector obtained by flipping all bits in $\mathbf{x}$, and $\neg E$
to denote $\left\{ \neg\mathbf{x}:\;\mathbf{x}\in E\right\} $.

Throughout this paper, $\log$ is understood to be of base $2$.

\subsection{Definitions\label{sub:Definitions}}

Let $T$ denote the left-shift transformation. That is, e.g. for a
vector $\left(v_{1},v_{2},...\right)^{\top}$, we have $T\left(v_{1},v_{2},...\right)^{\top}=\left(v_{2},v_{3},...\right)^{\top}$.
Also, let $T^{-1}E=\left\{ \mathbf{v}:\; T\mathbf{v}\in E\right\} $.

We refer to ``finite-dimensional'' events as cylinders. That is,
a cylinder $E$ on the process $\{V_{n}\}_{n=1}^{\infty}$ takes the
form $c_{t}^{m}\left(G\right)=\left\{ \mathbf{v}:\;\left(v_{t},v_{t+1},\ldots,v_{t+m-1}\right)\in G\right\} $
for some $G\subseteq{\cal V}^{m}$. Also, $\text{cylind}\left(t,m,{\cal V}\right)=\left\{ c_{t}^{m}\left(G\right):\; G\subseteq{\cal V}^{m}\right\} $,
the set of all cylinders with ``starting index'' $t$ and length
$m$ over ${\cal V}$. For finite $m$ and ${\cal V}$, $\text{cylind}\left(t,m,{\cal V}\right)$
is finite-sized.

A random process $\left\{ V_{n}\right\} $, defined over a probability
space $\left(\Omega,\mathcal{F},P\right)$, is said to be stationary
if $\forall E\in\mathcal{F}$, $P\left(E\right)=P\left(T^{-1}E\right)$.
It is $N$-stationary, for some integer $N\geq1$, if $\forall E\in\mathcal{F}$,
$P\left(E\right)=P\left(T^{-N}E\right)$. It is ergodic if $P\left(E\right)=0$
or $P\left(E\right)=1$ for all invariant events%
\footnote{An invariant event $E$ is defined to satisfy $T^{-1}E=E$.%
} $E\in\mathcal{F}$.

The notion of stationarity can be extended to an $n$-dimensional
probability measure $P^{\left(n\right)}$ for finite $n$. That is,
\[
P^{\left(n\right)}\left(E\times{\cal V}\right)=P^{\left(n\right)}\left(T^{-1}E\right)\quad\forall E\subseteq{\cal V}^{n-1}
\]
It can be observed that if $P$ is stationary and $P^{\left(n\right)}$
is an $n$-dimensional marginal of $P$, i.e. $P^{\left(n\right)}\left(E\right)=P\left(\left\{ V_{1+k}^{n+k}\in E\right\} \right)$
$\forall E\subseteq{\cal V}^{n}$ for some $k$, then $P^{\left(n\right)}$
is stationary.

We make a note on Kolmogorov consistency. A sequence $\left\{ V^{n}\right\} $
is consistent, where $V^{n}$ is associated with probability measure
$P^{\left(n\right)}$ for each $n$, if 
\[
\sum_{v_{n+1}\in{\cal V}}P^{\left(n+1\right)}\left(v^{n+1}\right)=P^{\left(n\right)}\left(v^{n}\right)\quad\forall v^{n}\in{\cal V}^{n}
\]
for every $n$. That is, consistent $\left\{ V^{n}\right\} $ can
be described by a single probability measure, whereas inconsistent
$\left\{ V^{n}\right\} $ is described by an infinite number of probability
measures $\left\{ P^{\left(n\right)}\right\} $. Notice that the notation
$\left\{ V_{n}\right\} $ implies a standard random process which
must be consistent by definition, while $\left\{ V^{n}\right\} $
is only a sequence in $n$. The two notations coincide if $\left\{ V^{n}\right\} $
is consistent. We also note that if $\left\{ V^{n}\right\} $ is stationary,
$N$-stationary or ergodic, it must be specified by a single probability
measure and therefore consistent.

\subsubsection{Channel Definitions}

We discuss the channel definition in two settings: the ergodic-theoretic
framework (see e.g. \cite{Gray-Entropy}) and the Shannon-theoretic
framework (see e.g. \cite{Verdu-Han-1994}). 
\begin{itemize}
\item In the ergodic-theoretic setting, a channel is defined as a list of
probability measures $\left\{ \nu_{\mathbf{x}}:\;\mathbf{x}\in{\cal X}^{\infty}\right\} $
which act on output events $E_{Y}\subseteq{\cal Y}^{\infty}$, where
each $\mathbf{x}$ is either a bi-infinite sequence (i.e. $\mathbf{x}=\left(\ldots,x_{-1},x_{0},x_{1},\ldots\right)$)
or a uni-infinite sequence (i.e. $\mathbf{x}=\left(x_{k},x_{k+1},\ldots\right)$
for some finite $k\in\mathbb{Z}$). This definition only admits inputs
that are consistent random processes. The joint input-output distribution
$\omega$ for an input process $\mu$ is determined by
\begin{equation}
\omega\left(E_{X},E_{Y}\right)=\int_{E_{X}}\nu_{\mathbf{x}}\left(E_{Y}\right)d\mu\left(\mathbf{x}\right)\label{eq:Eq_input_output_dist_channel_def_ergodic}
\end{equation}
for any events $E_{X}$ and $E_{Y}$ on the channel input and output
respectively. Here we only consider uni-infinite input sequences only,
which correspond to one-sided processes and one-sided channels. The
literature on bi-infinite inputs (and correspondingly two-sided channels)
is more extensive, but lacks the descriptive power for the channel
model of application in this paper, i.e. the DID channel.
\item The Shannon-theoretic definition of a channel is a sequence of finite-dimensional
conditional probability measures $\left\{ P_{\left.\mathbf{Y}\middle|X^{n}\right.}\left(\cdot\middle|\mathbf{x}\right),\;\mathbf{x}\in\mathcal{X}^{n}\right\} _{n=1}^{\infty}$,
in which $\mathbf{Y}$ can take any abstract alphabet. This definition
avoids placing any restrictions on the channel and allows inputs that
are not necessarily consistent and take the form $\left\{ X^{n}\right\} $.
The joint input-output distribution is determined for each block length
$n$, i.e.
\begin{equation}
P_{X^{n}\mathbf{Y}}\left(x^{n},\mathbf{y}\right)=P_{\left.\mathbf{Y}\middle|X^{n}\right.}\left(\mathbf{y}\middle|x^{n}\right)Q^{\left(n\right)}\left(x^{n}\right)\label{eq:Eq_input_output_dist_channel_def_Shannon}
\end{equation}
where $X^{n}\sim Q^{\left(n\right)}$, and it is not necessarily consistent
throughout all $n$'s.
\end{itemize}
Compatibility between the two definitions may not be entirely immediate.
On one hand, the sequence in the Shannon-theoretic definition may
not be viewed as converging to a limit equal to the ergodic-theoretic
definition without careful justification of such a limit. On the other
hand, the lack of channel laws $P_{\left.\mathbf{Y}\middle|X^{n}\right.}\left(\cdot\middle|\mathbf{x}\right)$
for finite block lengths $n$ in the ergodic-theoretic framework has
been noted to pose technical difficulties in proving coding theorems
\cite{Gray-1979}. The difference in the channel definition, as noted
in \cite{Verdu-Han-1994}, also leads to a difference in how the error
probability is defined. Despite this discrepancy, the capacities under
the two frameworks are of little difference in the overall implication
to reliable communications in the asymptotic regime of infinite block
length. Depending on the actual channel model, one may find either
framework suitable. In general, we shall treat the two frameworks
separately.

\subsubsection{From Channel Definitions to Channel Models}

The channel definitions abstract away from specific channel models
and form the basis under which reliable communications is defined.
Practical channel models, however, are usually not described in terms
of probability measures as in the channel definitions. When the ergodic-theoretic
channel definition is applicable to a channel model, we mean that
there exists a list of probability measures $\left\{ \nu_{\mathbf{x}}:\;\mathbf{x}\in{\cal X}^{\infty}\right\} $
such that for every consistent input, the joint input-output distribution
can be described by Eq. \eqref{eq:Eq_input_output_dist_channel_def_ergodic}.
Likewise, when the Shannon-theoretic channel definition is applicable
to a channel model, there exists a sequence of finite-dimensional
conditional probability measures $\left\{ P_{\left.\mathbf{Y}\middle|X^{n}\right.}\left(\cdot\middle|\mathbf{x}\right),\;\mathbf{x}\in\mathcal{X}^{n}\right\} _{n=1}^{\infty}$
such that for every input $\left\{ X^{n}\sim Q^{\left(n\right)}\right\} $,
the joint input-output distributions can be described by Eq. \eqref{eq:Eq_input_output_dist_channel_def_Shannon}.

While the two frameworks associated with the two channel definitions
are handled separately, when both channel definitions are applicable,
the observation that the measures $\nu_{\mathbf{x}}$ and $P_{\left.\mathbf{Y}\middle|X^{n}\right.}\left(\cdot\middle|\mathbf{x}\right)$
coexist for a channel model can be exploited. In particular, in such
cases, the joint input-output distribution can be described either
ways, thereby allowing certain results from one framework to be used
in another.

\subsection{DID Channel Model\label{sub:DID-Channel-Model}}

The DID channel model is described by
\[
Y_{i}=X_{i-Z_{i}}
\]
where $\left\{ Z_{i}\right\} _{i=1}^{\infty}$ is a first-order binary
Markov process, independent of the binary input. Note that the starting
indices of the state process and the output are $1$ and that of the
input is $0$. Here $P\left(Z_{i}=1\middle|Z_{i-1}=0\right)=p_{i}$,
the insertion probability, and $P\left(Z_{i}=0\middle|Z_{i-1}=1\right)=p_{d}$,
the deletion probability.

\begin{figure}
\begin{centering}
\includegraphics[width=0.6\columnwidth]{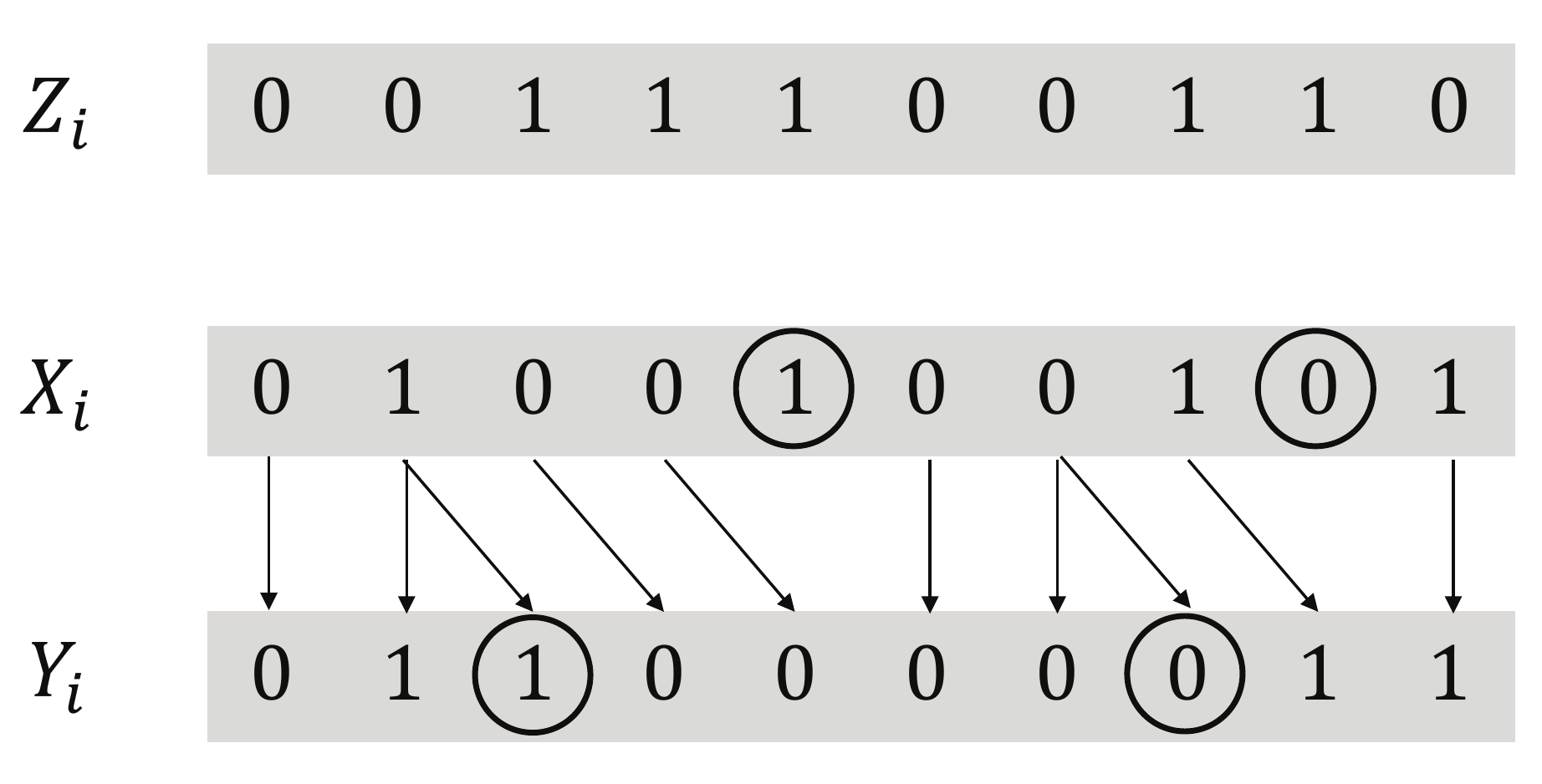}
\par\end{centering}

\protect\caption{An illustration of the DID channel. The circled input bits are deleted,
and the circled output bits are inserted. Each $Z_{i}=0\rightarrow Z_{i+1}=1$
transition induces an insertion, in which the inserted bit is the
same as the last input bit, whereas each $Z_{i}=1\rightarrow Z_{i+1}=0$
transition induces a deletion.}

\label{Fig_DID}
\end{figure}

An illustration is given in Fig. \ref{Fig_DID}. It could be observed
that every insertion must be followed by a deletion and vice versa,
i.e. insertions and deletions are paired; hence the name dependent
insertion-deletion channel.

This channel describes a simplified model for the BPMR write process,
capturing certain key features of the errors introduced in this process.
As mentioned, the aim of this technology is to achieve ultra-high
storage densities, envisioned at $10\text{ Tb/in}^{2}$ and beyond
\cite{Shiroishi-TMAG-2009}. Conventional magnetic recording media
comprise of successive magnetic units, each of which is written with
one data bit, represented by the magnetic state of the unit. As the
density increases, interference from adjacent units produces graver
effects on the magnetic state, hence degrading the reliability of
the process and placing a limit on these media. BPMR, as one of the
responses to this problem, separates these units with non-magnetic
materials in between. However, as another challenge with increasing
densities, the write head does not shrink proportionally with the
size of each magnetic unit; in fact, it may span over multiple units.
Difficulties thereby arise in controlling the position of the write
head relative to the units, leading to timing mismatches. There are
two typical erroneous scenarios: one is when the write head lags behind
the next unit and fails to write the data bit on this unit, and the
other is when it advances beyond the intended unit which is hence
written with the past data bit. The first error is modeled by the
transition $Z_{i-1}=1\to Z_{i}=0$, and the second error corresponds
to $Z_{i-1}=0\to Z_{i}=1$. 

For more details about the BPMR, see \cite{Richter-TMAG-2006,Shiroishi-TMAG-2009,Zhang-TMAG-2011,Keele-PhDThesis-2012}.
A more informative justification of the DID channel model can be found
in \cite{Iyengar-2011,Mazumdar-2011}. It has been pointed out that
this model does not capture all error types in the BPMR write process,
which motivates another channel model in \cite{Wu-2013}. They are
however similar from an information-theoretic perspective. Since this
paper concerns with the capacity aspect of these channels, the simplicity
of the DID channel is appealing, making the model suitable for illustrating
our information-theoretic results and capacity-evaluating techniques.

A number of bounds on the DID channel capacity have been established
in \cite{Iyengar-2011} and \cite{Mazumdar-2011}. In particular,
in \cite{Iyengar-2011}, an upper bound, given by $1-p_{i}p_{d}/\left(p_{i}+p_{d}\right)$
and termed \textit{genie-erasure upper bound}, was derived, and a
numerical simulation-based lower bound, which is in fact the achievable
rate $C_{iud}$ with independent and uniformly distributed (i.u.d.)
input, was computed for all channel parameters $p_{i}$ and $p_{d}$.
The specific case of $p_{d}=1$ was analyzed in \cite{Mazumdar-2011},
which provided a finite-lettered expression of $C_{iud}$ and also
the genie-erasure upper bound. However these upper and lower bounds,
as seen in \cite[Fig. 10, 11]{Iyengar-2011} and \cite[Fig. 2]{Mazumdar-2011},
are relatively distant. (We reproduce their bounds, plotted as dashed
curves, in Fig. \ref{Fig_Capacity_Markov_Lower_Bound} and \ref{Fig_Capacity_Bound_Pd}
below, for greater convenience of the readers.) We emphasize that
both works relied on the capacity formula for indecomposable channels%
\footnote{Although indecomposable channels are under the Shannon-theoretic framework,
with respect to our ergodic-theoretic capacity formulation in Section
\ref{sec:Ergodic-theoretic_Capacity_Formulation}, the achievable
rate $C_{iud}$ is in fact the same, and the genie-erasure upper bound
can be easily established using the same argument as in \cite{Iyengar-2011}.%
}.

In later sections, we restrict our attention to $p_{i}\in(0,1)$ and
$p_{d}\in(0,1)$. A simple continuity argument yields results for
special cases where $p_{i}\in\{0,1\}$ or $p_{d}\in\{0,1\}$. Henceforth
we say e.g. $p_{d}=1$ to mean $p_{d}$ is very close to $1$.

\subsection{Summary of Contributions and Structure}

As said, a main contribution of this work is the realization of the
stationarity condition in the supremizing input set. In Sections \ref{sec:Ergodic-theoretic_Capacity_Formulation}
and \ref{sec:Shannon-Theoretic-Capacity-Formulation}, we explore
this in capacity formulations for the DID channel model in both ergodic-theoretic
and Shannon-theoretic frameworks.
\begin{itemize}
\item The ergodic-theoretic literature on such formulation is vast. However
to the best of our knowledge, the theory developed for one-sided channels
contains solely forward coding theorems, which establish achievability
results on the rate \cite{Gray-Entropy}. Converse theorems, which
concern with inachievability, are unfortunately missing%
\footnote{On the contrary, the theory for two-sided channels is quite complete.
We point to some representative references \cite{Gray-1979,Gray-Entropy}
for excellent summaries of the results. In particular, \cite{Gray-Entropy}
contains results that are applicable to both one-sided and two-sided
channels. References \cite{Feinstein-1959,Adler-1961,Parthasarathy-1961,Sujan-1981,Fontana-1981}
contain results that are proven in the two-sided setting and correspond
to tools used in this paper.%
}. In Section \ref{sec:Ergodic-theoretic_Capacity_Formulation}, we
verify that a ``capacity'' formula is applicable to the DID channel.
This formula encompasses other known achievable rate formulas for
one-sided channels. For simplicity, we shall refer to this formula
as a capacity formula.
\item To realize input stationarity in the Shannon-theoretic framework,
we introduce new definitions on consistent, stationary and ergodic
channels, and prove that SE inputs can achieve the capacity of the
finite-alphabet class of these channels in Theorem \ref{thm:Theorem_consistent_SE_Channel_Capacity (Shannon-theoretic)}.
We also verify its applicability to the DID channel. Note that the
formula obtained here is the true capacity. These are done in Section
\ref{sec:Shannon-Theoretic-Capacity-Formulation}.
\end{itemize}
Further towards the aim of computational suitability, a portion of
these two sections introduces the notion of bit-symmetry and proves
that we can further restrict the input search space to bit-symmetric
inputs in Proposition \ref{prop:Proposition_Bit-symmetry (ergodic)}
(and to a lesser extent, Proposition \ref{prop:Proposition_Bit-symmetry (Shannon-theoretic)})
when the channel is, in addition, bit-symmetric.

While it is not conclusive whether the ergodic-theoretic formula is
the true capacity, it is the same as the one in the Shannon-theoretic
framework. Subsequent sections are devoted to evaluations of this
single formula. In Section \ref{sec:DID-Capacity-Lower-Bound}, a
new lower bound is derived analytically in a finite-lettered computable
form. A series of new computation-based upper bounds is formulated
in Section \ref{sec:DID-Capacity-Upper-Bound}. These bounds are shown
to yield improvements over those in \cite{Iyengar-2011} and \cite{Mazumdar-2011}
and are tight at low noise levels. Inspired by this observation, we
then characterize the DID channel capacity in the low-noise regime
in Section \ref{sec:DID-Capacity-at-Low-Noise} for the case $p_{i}=p_{d}=p_{id}$,
given by
\[
C=1-\left[\sum_{k=1}^{\infty}\frac{1}{2^{k+1}}h_{2}\left(\frac{1}{2}+\frac{1}{2}\left(1-2p_{id}\right)^{k}\right)\right]+{\cal O}\left(p_{id}^{2}\right)
\]
This characterization is achieved (up to the order of $p_{id}^{2}$)
by the i.u.d. input. The paper concludes with Section \ref{sec:Conclusions}.

As shown in \cite{Keele-PhDThesis-2012}, in models that closely mimic
the actual BPMR write channel, the occurrence frequency should be
almost the same for both insertion and deletion errors. For this reason,
while our bounds are formulated for any $p_{i}$ and $p_{d}$, we
illustrate the numerical evaluation of the DID channel capacity mostly
for the case $p_{i}=p_{d}=p_{id}$. It is also shown in \cite{Keele-PhDThesis-2012}
that the occurrence frequency could be as low as $10^{-4}$, which
justifies our interest in the low-noise regime.

\section{Ergodic-Theoretic Capacity Formulation\label{sec:Ergodic-theoretic_Capacity_Formulation}}

In this section, we find a ``capacity'' formula that involves stationary
inputs for the DID channel under the ergodic-theoretic framework.
As a reminder, this formula yields an achievable rate, not the true
capacity, and we consider one-sided channels only. Our main reference
is the work \cite{Gray-Entropy}. We also develop the notion of bit-symmetry
in this setting. Roughly speaking, bit-symmetry for binary inputs
is the property in which the probability of drawing a binary string
$\mathbf{x}$ is equal to the probability of drawing $\neg\mathbf{x}$.
Likewise, a bit-symmetric (binary) channel is one in which the posterior
probability of the output $\mathbf{y}$ given the input $\mathbf{x}$
is equal to that of $\neg\mathbf{y}$ given $\neg\mathbf{x}$. Intuitively
a bit-symmetric channel should attain its capacity for some bit-symmetric
inputs. We show that this is true for certain channels.

\subsection{Ergodic-Theoretic Capacity\label{sub:Ergodic-theoretic Capacity}}

We review some definitions under the ergodic-theoretic framework.
\begin{itemize}
\item A random process $\{V_{n}\}_{n=1}^{\infty}$, defined over a probability
space $(\Omega,\mathcal{F},P)$, is said to be asymptotically mean
stationary (AMS) if $\forall E\in\mathcal{F}$, the limit 
\[
P_{AMS}\left(E\right)=\lim_{n\to\infty}\frac{1}{n}\sum_{i=0}^{n-1}P\left(T^{-i}E\right)
\]
exists. $P_{AMS}$ is called the stationary mean. A stationary processes
is also AMS.
\item A channel $\left\{ \nu_{\mathbf{x}}:\;\mathbf{x}\in{\cal X}^{\infty}\right\} $
is stationary if $\nu_{\mathbf{x}}\left(T^{-1}E_{Y}\right)=\nu_{T\mathbf{x}}\left(E_{Y}\right)$
for any output event $E_{Y}\subseteq{\cal Y}^{\infty}$. It is known
that the joint input-output distribution $\omega$ (and hence the
output distribution $\eta\left(E_{Y}\right)=\omega\left({\cal X}^{\infty},E_{Y}\right)$)
is stationary given a stationary input $\mu$ and a stationary channel
\cite[Lemma 9.3.1]{Gray-Entropy}.
\item A channel is said to be AMS if given an AMS input, $\omega$ is also
AMS. A stationary channel is also AMS \cite[Lemma 9.3.2]{Gray-Entropy}.
\item A stationary (resp. AMS) channel is ergodic if $\omega$ is ergodic,
given any stationary (resp. AMS) and ergodic input distribution $\mu$.
\end{itemize}
Let us first consider stationary channels, in which the developments
are more straightforward. In the ergodic-theoretic literature, coding
theorems are established for finite-alphabet stationary and ergodic
(SE) channels with various assumptions on the channel memory and anticipation.
A simple class is the class of finite-input-memory and causal channels.
A channel is said to have finite input memory if there exists a natural
number $m$ such that for any $n$,
\[
\nu_{\mathbf{x}}\left(T^{-n}E_{Y}\right)=\nu_{\hat{\mathbf{x}}}\left(T^{-n}E_{Y}\right)
\]
for any $\mathbf{x}$ and $\hat{\mathbf{x}}$ such that $x_{n-m}^{\infty}=\hat{x}_{n-m}^{\infty}$
and any event $E_{Y}$. A channel is causal (or without anticipation)
if for any $n$,
\[
\nu_{\mathbf{x}}\left(\left\{ Y^{n}=y^{n}\right\} \right)=\nu_{\hat{\mathbf{x}}}\left(\left\{ Y^{n}=y^{n}\right\} \right)
\]
for any $\mathbf{x}$ and $\hat{\mathbf{x}}$ such that $x^{n}=\hat{x}^{n}$
and any $y^{n}$.

For the DID channel, we use $P_{X_{0}^{\infty}}$, $P_{X_{0}^{\infty}\mathbf{Y}}$
and $P_{\left.\mathbf{Y}\middle|X_{0}^{\infty}\right.}\left(\cdot\middle|\mathbf{x}\right)$
in place of $\mu$, $\omega$ and $\nu_{\mathbf{x}}$ respectively.
The model can be described under the ergodic-theoretic channel definition
as follows. For an output sequence $\mathbf{Y}=\mathbf{y}$, given
an input sequence $X_{0}^{\infty}=\mathbf{x}$, each triplet $\left(y_{i},x_{i},x_{i-1}\right)$
corresponding to the pair $(\mathbf{x},\mathbf{y})$ uniquely determines
the occurrence of either one of the four events $\left\{ Z_{i}=0\right\} $,
$\left\{ Z_{i}=1\right\} $, $\left\{ Z_{i}=0\;\text{or}\; Z_{i}=1\right\} $
and $\left\{ Z_{i}\in\emptyset\right\} $. Let us denote the determined
event ${\cal E}\left(Z_{i};y_{i},x_{i},x_{i-1}\right)$. Also, for
some output event $E_{Y}$, let us define 
\[
{\cal E}\left(\mathbf{x},E_{Y}\right)=\bigcup_{\mathbf{y}\in E_{Y}}\bigcap_{i=1}^{\infty}{\cal E}\left(Z_{i};y_{i},x_{i},x_{i-1}\right)
\]
Then the channel law $P_{\left.\mathbf{Y}\middle|X_{0}^{\infty}\right.}\left(\cdot\middle|\mathbf{x}\right)$
is determined by
\begin{align}
P_{\left.\mathbf{Y}\middle|X_{0}^{\infty}\right.}\left(E_{Y}\middle|\mathbf{x}\right) & =\int_{E_{Y}}dP_{\left.\mathbf{Y}\middle|X_{0}^{\infty}\right.}\left(\mathbf{y}\middle|\mathbf{x}\right)\nonumber \\
 & =\int_{E_{Y}}dP_{\mathbf{Z}}\left(\bigcap_{i=1}^{\infty}{\cal E}\left(Z_{i};y_{i},x_{i},x_{i-1}\right)\right)\nonumber \\
 & \stackrel{\left(a\right)}{=}P_{\mathbf{Z}}\left({\cal E}\left(\mathbf{x},E_{Y}\right)\right)\label{eq:Eq_channelLaw_ergodic}
\end{align}
where $P_{\mathbf{Z}}$ is the probability measure of the $\left\{ Z_{i}\right\} _{i=1}^{\infty}$
process. The step $\left(a\right)$ is by the following reason: ${\cal E}\left(Z_{i};y_{i},x_{i},x_{i-1}\right)$
and ${\cal E}\left(Z_{i};y_{i}',x_{i},x_{i-1}\right)$ are disjoint
for $y_{i}\neq y_{i}'$, so $\bigcap_{i=1}^{\infty}{\cal E}\left(Z_{i};y_{i},x_{i},x_{i-1}\right)$
and $\bigcap_{i=1}^{\infty}{\cal E}\left(Z_{i};y_{i}',x_{i},x_{i-1}\right)$
are disjoint for $\mathbf{y}\neq\mathbf{y}'$.

It is easy to see that the DID channel model is finite-input-memory
and causal. Combining \cite[Theorem 12.6.1]{Gray-Entropy} and \cite[Lemma 12.4.2]{Gray-Entropy},
the capacity of the DID channel is given by%
\footnote{The statement of \cite[Theorem 12.6.1]{Gray-Entropy} applies to AMS
$\bar{d}$-continuous channels. Finite-input-memory and causal channels
are a special case of $\bar{d}$-continuous channels. Also, as mentioned,
a stationary channel is AMS. %
}
\begin{align}
C & =\sup_{\text{\emph{\emph{Stationary} }}P_{X_{0}^{\infty}}}\lim_{n\to\infty}\frac{1}{n}I\left(X_{0}^{n};Y_{1}^{n}\right)\label{eq:Equation_C_0}\\
 & =\sup_{\text{\emph{\emph{Stationary} }}P_{X_{0}^{\infty}}}\lim_{n\to\infty}\frac{1}{n}\left[H\left(Y^{n}\right)-H\left(Y^{n}\middle|X_{0}^{n}\right)\right]\label{eq:Equation_C_1}\\
 & =\sup_{\text{\emph{\emph{Stationary} }}P_{X_{0}^{\infty}}}\Big[\lim_{n\rightarrow\infty}H\left(Y_{n}\middle|Y^{n-1}\right)\nonumber \\
 & \qquad\qquad\qquad\qquad-\lim_{n\rightarrow\infty}H\left(Y_{n}\middle|Y^{n-1},X_{0}^{n}\right)\Big]\nonumber 
\end{align}
if it could be shown to be SE. The third equation follows from causality
of the channel and the fact that the joint input-output distribution
is stationary. We note that the starting indices in the formula are
chosen to suit the DID channel description, although this is not critical
for the following reason. For any finite non-negative $a$ and $b$,
\begin{align*}
I\left(X_{-b}^{n+a};Y_{1}^{n}\right) & =I\left(X_{0}^{n};Y_{1}^{n}\right)+I\left(X_{-b}^{-1},X_{n+1}^{n+a};Y_{1}^{n}\middle|X_{0}^{n}\right)\\
 & \leq I\left(X_{0}^{n};Y_{1}^{n}\right)+\left(a+b\right)\log\left|{\cal X}\right|
\end{align*}
and that $I\left(X_{-b}^{n+a};Y_{1}^{n}\right)\geq I\left(X_{0}^{n};Y_{1}^{n}\right)$.
Then:
\[
\lim_{n\rightarrow\infty}\frac{1}{n}I\left(X_{0}^{n};Y_{1}^{n}\right)=\lim_{n\rightarrow\infty}\frac{1}{n}I\left(X_{-b}^{n+a};Y_{1}^{n}\right)
\]
since $\left(a+b\right)\log\left|{\cal X}\right|$ is finite.

We now argue that the DID channel is stationary under the condition
that $\left\{ Z_{i}\right\} _{i=1}^{\infty}$ is stationary, which
can be achieved with a suitable initialization: $P\left(Z_{1}=0\right)=p_{d}/\left(p_{i}+p_{d}\right)$
and $P\left(Z_{1}=1\right)=p_{i}/\left(p_{i}+p_{d}\right)$. For any
output event $E_{Y}$, 
\begin{align*}
P_{\left.\mathbf{Y}\middle|X_{0}^{\infty}\right.}\left(E_{Y}\middle|T\mathbf{x}\right) & =P_{\mathbf{Z}}\left({\cal E}\left(T\mathbf{x},E_{Y}\right)\right)\\
 & \stackrel{\left(a\right)}{=}P_{\mathbf{Z}}\left(T^{-1}{\cal E}\left(T\mathbf{x},E_{Y}\right)\right)\\
 & =P_{\mathbf{Z}}\left(\bigcup_{\mathbf{y}\in E_{Y}}\bigcap_{i=1}^{\infty}{\cal E}\left(Z_{i+1};y_{i},x_{i+1},x_{i}\right)\right)\\
 & =P_{\mathbf{Z}}\left(\bigcup_{\mathbf{y}\in E_{Y}}\bigcap_{i=2}^{\infty}{\cal E}\left(Z_{i};y_{i-1},x_{i},x_{i-1}\right)\right)\\
 & =P_{\mathbf{Z}}\left(\bigcup_{\mathbf{y}\in T^{-1}E_{Y}}\bigcap_{i=1}^{\infty}{\cal E}\left(Z_{i};y_{i},x_{i},x_{i-1}\right)\right)\\
 & =P_{\left.\mathbf{Y}\middle|X_{0}^{\infty}\right.}\left(T^{-1}E_{Y}\middle|\mathbf{x}\right)
\end{align*}
 where $\left(a\right)$ is because $P_{\mathbf{Z}}$ is stationary.
Stationarity of the DID channel is thus proven.

To prove ergodicity, we refer to the following definition. A channel
is said to be output strongly mixing if for every $\mathbf{x}\in{\cal X}^{\infty}$
and cylinders $F_{1}$ and $F_{2}$ on the output,
\[
\lim_{n\to\infty}\left|\nu_{\mathbf{x}}\left(T^{-n}F_{1}\cap F_{2}\right)-\nu_{\mathbf{x}}\left(T^{-n}F_{1}\right)\nu_{\mathbf{x}}\left(F_{2}\right)\right|=0
\]

\begin{lem}
\label{lem:Lemma_Ergodicity_by_Output_Strongly_Mixing}A stationary
channel is ergodic if it is output strongly mixing.
\end{lem}
Lemma \ref{lem:Lemma_Ergodicity_by_Output_Strongly_Mixing} is an
easy consequence of \cite[Lemma 9.4.3]{Gray-Entropy}. Now for any
two cylinders $F_{1}$ and $F_{2}$ on the output, for sufficiently
large $n$, we have:
\begin{align*}
 & \Big|P_{\left.\mathbf{Y}\middle|X_{0}^{\infty}\right.}\left(T^{-n}F_{1}\cap F_{2}\middle|\mathbf{x}\right)-\\
 & \qquad\qquad P_{\left.\mathbf{Y}\middle|X_{0}^{\infty}\right.}\left(T^{-n}F_{1}\middle|\mathbf{x}\right)P_{\left.\mathbf{Y}\middle|X_{0}^{\infty}\right.}\left(F_{2}\middle|\mathbf{x}\right)\Big|\\
 & =\Big|P_{\mathbf{Z}}\left({\cal E}\left(\mathbf{x},T^{-n}F_{1}\cap F_{2}\right)\right)-\\
 & \qquad\qquad P_{\mathbf{Z}}\left({\cal E}\left(\mathbf{x},T^{-n}F_{1}\right)\right)P_{\mathbf{Z}}\left({\cal E}\left(\mathbf{x},F_{2}\right)\right)\Big|\\
 & \stackrel{\left(a\right)}{=}\Big|P_{\mathbf{Z}}\left(T^{-n}{\cal E}\left(T^{n}\mathbf{x},F_{1}\right)\cap{\cal E}\left(\mathbf{x},F_{2}\right)\right)-\\
 & \qquad\qquad P_{\mathbf{Z}}\left(T^{-n}{\cal E}\left(T^{n}\mathbf{x},F_{1}\right)\right)P_{\mathbf{Z}}\left({\cal E}\left(\mathbf{x},F_{2}\right)\right)\Big|\\
 & \xrightarrow{n\to\infty}0
\end{align*}
To see $\left(a\right)$, suppose $F_{1}$ and $F_{2}$ take the forms
$c_{t_{1}}^{m_{1}}\left(G_{1}\right)$ and $c_{t_{2}}^{m_{2}}\left(G_{2}\right)$
for, respectively, some $G_{1}\subseteq{\cal Y}^{m_{1}}$ and $G_{2}\subseteq{\cal Y}^{m_{2}}$.
Then for $n\geq t_{2}+m_{2}-t_{1}$:
\begin{align*}
 & {\cal E}\left(\mathbf{x},T^{-n}F_{1}\cap F_{2}\right)\\
 & =\bigcup_{\mathbf{y}\in T^{-n}F_{1}\cap F_{2}}\bigcap_{i=1}^{\infty}{\cal E}\left(Z_{i};y_{i},x_{i},x_{i-1}\right)\\
 & =\left[\bigcup_{\tilde{\mathbf{y}}\in G_{1}}\bigcap_{i=n+t_{1}}^{n+t_{1}+m_{1}-1}{\cal E}\left(Z_{i};\tilde{y}_{i-n-t_{1}+1},x_{i},x_{i-1}\right)\right]\\
 & \qquad\cap\left[\bigcup_{\tilde{\mathbf{y}}\in G_{2}}\bigcap_{i=t_{2}}^{t_{2}+m_{2}-1}{\cal E}\left(Z_{i};\tilde{y}_{i-t_{2}+1},x_{i},x_{i-1}\right)\right]\\
 & =T^{-n}\left[\bigcup_{\tilde{\mathbf{y}}\in G_{1}}\bigcap_{i=t_{1}}^{t_{1}+m_{1}-1}{\cal E}\left(Z_{i};\tilde{y}_{i-t_{1}+1},x_{n+i},x_{n+i-1}\right)\right]\\
 & \qquad\cap\left[\bigcup_{\tilde{\mathbf{y}}\in G_{2}}\bigcap_{i=t_{2}}^{t_{2}+m_{2}-1}{\cal E}\left(Z_{i};\tilde{y}_{i-t_{2}+1},x_{i},x_{i-1}\right)\right]\\
 & =T^{-n}\left[\bigcup_{\mathbf{y}\in F_{1}}\bigcap_{i=1}^{\infty}{\cal E}\left(Z_{i};y_{i},x_{n+i},x_{n+i-1}\right)\right]\\
 & \qquad\cap\left[\bigcup_{\mathbf{y}\in F_{2}}\bigcap_{i=1}^{\infty}{\cal E}\left(Z_{i};y_{i},x_{i},x_{i-1}\right)\right]\\
 & =T^{-n}{\cal E}\left(T^{n}\mathbf{x},F_{1}\right)\cap{\cal E}\left(\mathbf{x},F_{2}\right)
\end{align*}
Replacing $F_{2}$ by $c_{t_{2}}^{m_{2}}\left({\cal Y}^{m_{2}}\right)$
in the above, we also obtain ${\cal E}\left(\mathbf{x},T^{-n}F_{1}\right)=T^{-n}{\cal E}\left(T^{n}\mathbf{x},F_{1}\right)$;
hence step $\left(a\right)$ is shown. The convergence with $n\to\infty$
in the last step is justified as follows. Under the aforementioned
distribution of $Z_{1}$, the process $\left\{ Z_{i}\right\} _{i=1}^{\infty}$
is mixing \cite{Gray-Probability}, i.e. for any events $E_{1}$ and
$E_{2}$ and any $\epsilon>0$, there exists a finite $n_{\epsilon}\left(E_{1},E_{2}\right)$
such that $\forall n\geq n_{\epsilon}\left(E_{1},E_{2}\right)$, 
\[
\left|P_{\mathbf{Z}}\left(T^{-n}E_{1}\cap E_{2}\right)-P_{\mathbf{Z}}\left(T^{-n}E_{1}\right)P_{\mathbf{Z}}\left(E_{2}\right)\right|<\epsilon
\]
With $F_{1}=c_{t_{1}}^{m_{1}}\left(G_{1}\right)$ and $F_{2}=c_{t_{2}}^{m_{2}}\left(G_{2}\right)$
as above, let
\begin{align*}
N_{\epsilon} & =\Big\{ n_{\epsilon}\left(E_{1},E_{2}\right)\Big|E_{1}\in\text{cylind}\left(t_{1},m_{1},{\cal Z}\right),\\
 & \qquad\qquad\qquad\qquad E_{2}\in\text{cylind}\left(t_{2},m_{2},{\cal Z}\right)\Big\}
\end{align*}
which is a finite set, since $\text{cylind}\left(t_{1},m_{1},{\cal Z}\right)$
and $\text{cylind}\left(t_{2},m_{2},{\cal Z}\right)$ are finite.
Hence $\max N_{\epsilon}$ exists and is finite. Then:
\begin{align*}
 & \Big|P_{\mathbf{Z}}\left(T^{-n}{\cal E}\left(T^{n}\mathbf{x},F_{1}\right)\cap{\cal E}\left(\mathbf{x},F_{2}\right)\right)-\\
 & \quad P_{\mathbf{Z}}\left(T^{-n}{\cal E}\left(T^{n}\mathbf{x},F_{1}\right)\right)P_{\mathbf{Z}}\left({\cal E}\left(\mathbf{x},F_{2}\right)\right)\Big|<\epsilon\quad\forall n\geq\max N_{\epsilon}
\end{align*}
since ${\cal E}\left(T^{n}\mathbf{x},F_{1}\right)\in\text{cylind}\left(t_{1},m_{1},{\cal Z}\right)$
and ${\cal E}\left(\mathbf{x},F_{2}\right)\in\text{cylind}\left(t_{2},m_{2},{\cal Z}\right)$.
This completes the establishment of the DID channel's ergodicity.

With other initializations, can we still achieve a rate equal to that
in Eq. \eqref{eq:Equation_C_1}? The answer is positive, even though
the DID channel turns out to be AMS in this case. This is shown in
\nameref{sec:Appendix A}, which is an interesting application of
the connection between the ergodic-theoretic and Shannon-theoretic
frameworks.

\subsection{Bit-Symmetric Channels: Ergodic-Theoretic Setting}
\begin{defn}
A binary probability measure $\mu$ (i.e. one that acts on $\mathrm{GF}\left(2\right)$)
is bit-symmetric if $\mu\left(E\right)=\mu\left(\neg E\right)$ $\forall E\subseteq\mathrm{GF}\left(2\right)^{\infty}$.
\end{defn}

\begin{defn}
A binary channel, in which $X_{i},Y_{i}\in\mathrm{GF}\left(2\right)$,
is bit-symmetric if $\nu_{\mathbf{x}}\left(E_{Y}\right)=\nu_{\neg\mathbf{x}}\left(\neg E_{Y}\right)$
$\forall\mathbf{x}\in\mathrm{GF}\left(2\right)^{\infty},E_{Y}\subseteq\mathrm{GF}\left(2\right)^{\infty}$.\end{defn}
\begin{prop}
\label{prop:Proposition_Bit-symmetry (ergodic)}For any bit-symmetric
binary channel,
\begin{align*}
\sup_{\text{\emph{Stationary} }\mu}\lim_{n\to\infty}\frac{1}{n}I\left(X_{0}^{n};Y_{1}^{n}\right) & =\sup_{\substack{\text{\emph{Bit-symmetric,}}\\
\emph{stationary }\mu
}
}\lim_{n\to\infty}\frac{1}{n}I\left(X_{0}^{n};Y_{1}^{n}\right)
\end{align*}

\end{prop}
As before, as long as the starting indices are finite and the ending
indices are within finite differences from $n$, they do not affect
the result.
\begin{IEEEproof}
The complete proof is given in \nameref{sec:Appendix-Proof_Bit-symmetry}.
The main idea is to define an input distribution $\mu_{0}$ based
on a stationary input $\mu$ such that $\mu_{0}\left(E\right)=\left(\mu\left(E\right)+\mu\left(\neg E\right)\right)/2$,
then prove that the mutual information corresponding to $\mu_{0}$
is at least that of $\mu$. Also, $\mu_{0}$ can be shown to be stationary
and bit-symmetric.
\end{IEEEproof}
It is easy to see that the DID channel is bit-symmetric. Then from
the above proposition,
\begin{align}
C & =\sup_{\substack{\text{Bit-symmetric,}\\
\text{stationary }P_{X_{0}^{\infty}}
}
}\Big[\lim_{n\rightarrow\infty}H\left(Y_{n}\middle|Y^{n-1}\right)\nonumber \\
 & \qquad\qquad\qquad\qquad-\lim_{n\rightarrow\infty}H\left(Y_{n}\middle|Y^{n-1},X_{0}^{n}\right)\Big]\label{eq:Equation_C_2}
\end{align}
This equation is pivotal to subsequent calculations of the DID channel
capacity.

The implication of the proposition is broader: capacity of a bit-symmetric
channel whose form is similar to that of the DID channel can be attained
with some bit-symmetric inputs. Although the bit-symmetry condition
may not be as helpful to tightening capacity bounds as the stationarity
condition as shall be seen in later sections, it offers an analytical
advantage by reducing the number of variables representing an input
by a half.

\subsection{Two-sided Channels}

Although the DID channel is naturally modeled as a one-sided channel,
we make a note here on the applicability of the theory for two-sided
channels. In this setting, all processes are two-sided. Roughly speaking,
this means that the starting index of the processes is $-\infty$.
The difficulty of casting the DID channel as a two-sided one is that
the state process $\left\{ Z_{i}\right\} $ has an initialization.

Consider a different model of the state process: $\left\{ Z_{i}\right\} $
(or more explicitly, $\left\{ Z_{i}\right\} _{i=-\infty}^{+\infty}$)
is a (two-sided) stationary and ergodic binary first-order Markov
process, in which $P\left(Z_{i}=1\middle|Z_{i-1}=0\right)=p_{i}$
and $P\left(Z_{i}=0\middle|Z_{i-1}=1\right)=p_{d}$. This implies
that for any $i\in\mathbb{Z}$, $P\left(Z_{i}=0\right)=p_{d}/\left(p_{i}+p_{d}\right)$
and $P\left(Z_{i}=1\right)=p_{i}/\left(p_{i}+p_{d}\right)$. Then
the DID channel falls under the two-sided setting:
\[
P_{\left.\mathbf{Y}\middle|X_{-\infty}^{+\infty}\right.}\left(E_{Y}\middle|\mathbf{x}\right)=P_{\mathbf{Z}}\left(\bigcup_{\mathbf{y}\in E_{Y}}\bigcap_{i=-\infty}^{+\infty}{\cal E}\left(Z_{i};y_{i},x_{i},x_{i-1}\right)\right)
\]
where $\mathbf{x}$ is a bi-infinite input sequence and $E_{Y}$ is
a subset of the space of all bi-infinite sequences $\left(\ldots,y_{-1},y_{0},y_{1},\ldots\right)$,
$y_{i}\in{\cal Y}$. All properties we have showed in the one-sided
case (stationarity, ergodicity, bit-symmetry) can be proven in a similar
fashion. Its true capacity is given by \cite{Gray-1979}
\begin{align*}
{\cal C} & =\sup_{\text{Stationary }P_{X_{-\infty}^{+\infty}}}\lim_{n\to\infty}\frac{1}{n}I\left(X_{0}^{n};Y_{1}^{n}\right)\\
 & =\sup_{\substack{\text{Bit-symmetric,}\\
\text{stationary }P_{X_{-\infty}^{+\infty}}
}
}\lim_{n\to\infty}\frac{1}{n}I\left(X_{0}^{n};Y_{1}^{n}\right)
\end{align*}
which is essentially the same as Eq. \eqref{eq:Equation_C_0} and
\eqref{eq:Equation_C_2}.

\section{Shannon-Theoretic Capacity Formulation\label{sec:Shannon-Theoretic-Capacity-Formulation}}

In this section, we depart from the ergodic-theoretic setting and
formulate a capacity formula that involves input stationarity in the
Shannon-theoretic framework. To this end, we define a new class of
consistent, stationary and ergodic channels. Like the previous section,
we also introduce the notion of bit-symmetry under the Shannon-theoretic
framework with a similar result. We verify that the results are applicable
to the DID channel model.

\subsection{Consistent, Stationary and Ergodic Channels}

Consider the following channel definition. A channel is specified
by a sequence of conditional probabilities $\left\{ P_{\left.Y^{n}\middle|X_{1-\lambda_{-}}^{n+\lambda_{+}}\right.}\left(\cdot\middle|\mathbf{x}\right),\;\mathbf{x}\in\mathcal{X}^{n+\lambda_{-}+\lambda_{+}}\right\} _{n=1}^{\infty}$,
in which $\lambda_{-}$ and $\lambda_{+}$ are two finite non-negative
integer constants specified by the specific channel model. Here $P_{\left.Y^{n}\middle|X_{1-\lambda_{-}}^{n+\lambda_{+}}\right.}\left(\cdot\middle|\mathbf{x}\right)$
is a (finite-dimensional) probability measure on $Y^{n}$ and hence
can admit any events on $Y^{n}$ (e.g. $\left\{ Y_{k}^{n}=\mathbf{y}\right\} $
for $1\leq k\leq n$). As usual, 
\[
P_{\left.Y^{n}\middle|X_{1-\lambda_{-}}^{n+\lambda_{+}}\right.}\left(\mathbf{y}\middle|\mathbf{x}\right)=P_{\left.Y^{n}\middle|X_{1-\lambda_{-}}^{n+\lambda_{+}}\right.}\left(\left\{ Y^{n}=\mathbf{y}\right\} \middle|\mathbf{x}\right)
\]
Also, the channel admits input sequences $\left\{ X_{1-\lambda_{-}}^{n+\lambda_{+}}:\; X_{1-\lambda_{-}}^{n+\lambda_{+}}\sim P^{\left(n\right)}\right\} _{n=1}^{\infty}$
that are not necessarily consistent. In this context, $P^{\left(n\right)}$
is understood to be $\left(n+\lambda_{+}+\lambda_{-}\right)$-dimensional.
This channel definition can be easily seen to be a subclass of the
Shannon-theoretic definition in Section \ref{sub:Definitions}. It
allows us to focus on channel models whose operation is described
for each length-$n$ output sequence $Y^{n}$ (of starting index $1$)
given an input sequence $X_{1-\lambda_{-}}^{n+\lambda_{+}}$. In the
case of the DID channel, $\lambda_{-}=1$ and $\lambda_{+}=0$.

We expect that capacity-achieving inputs would be stationary for channels
that behave similarly to those SE channels of the ergodic-theoretic
setting. This motivates the following definitions.
\begin{defn}
A channel is consistent if for every consistent input sequence $\left\{ X_{1-\lambda_{-}}^{n+\lambda_{+}}\right\} _{n=1}^{\infty}$,
the channel induces consistent $\left\{ X_{1-\lambda_{-}}^{n+\lambda_{+}},Y^{n}\right\} _{n=1}^{\infty}$.
\end{defn}

\begin{defn}
A consistent channel is weakly stationary if for any $N\geq1$, any
$N$-stationary input distribution $P_{X_{1-\lambda_{-}}^{\infty}}$
(or $N$-stationary input sequence $\left\{ X_{1-\lambda_{-}}^{n+\lambda_{+}}\right\} _{n=1}^{\infty}$)
induces a joint input-output sequence $\left\{ X_{1-\lambda_{-}}^{n+\lambda_{+}},Y^{n}\right\} _{n=1}^{\infty}$
that is $N$-stationary. A consistent weakly stationary channel is
ergodic if any SE input distribution induces a joint input-output
sequence that is ergodic.
\end{defn}
By defining a consistent channel, we legitimize our definition of
weakly stationary and ergodic channels. It can be seen that for a
consistent weakly stationary channel, if the finite-dimensional input
$X_{1-\lambda_{-}}^{n+\lambda_{+}}$ satisfies the stationarity condition,
i.e. 
\[
P_{X_{1-\lambda_{-}}^{n+\lambda_{+}}}\left(E\times{\cal X}\right)=P_{X_{1-\lambda_{-}}^{n+\lambda_{+}}}\left(T^{-1}E\right)\quad\forall E\subseteq{\cal X}^{n+\lambda_{+}+\lambda_{-}-1}
\]
the corresponding finite-dimensional input-output pair $\left(X_{1-\lambda_{-}}^{n+\lambda_{+}},Y^{n}\right)$
is stationary as well, since one can easily construct a stationary
input $P_{X_{1-\lambda_{-}}^{\infty}}$ that retains the same marginal
distribution on $X_{1-\lambda_{-}}^{n+\lambda_{+}}$. Therefore our
definition of weakly stationary channels captures the stationary behavior
for every block length. This is however not the case for ergodic channels,
since ergodicity cannot be described by finite-dimensional measures.
Our definition of ergodic channels consequently remains almost the
same as the ergodic-theoretic definition.
\begin{defn}
A channel is stationary if for any $n,k\geq1$,
\[
P_{\left.Y^{n+k}\middle|X_{1-\lambda_{-}}^{n+k+\lambda_{+}}\right.}\left(\left\{ Y_{k+1}^{n+k}=\mathbf{y}\right\} \middle|\left[\mathbf{\tilde{x}},\mathbf{x}\right]\right)=P_{\left.Y^{n}\middle|X_{1-\lambda_{-}}^{n+\lambda_{+}}\right.}\left(\mathbf{y}\middle|\mathbf{x}\right)
\]
for any $\mathbf{x}\in{\cal X}^{n+\lambda_{+}+\lambda_{-}}$, $\mathbf{\tilde{x}}\in{\cal X}^{k}$,
$\mathbf{y}\in{\cal Y}^{n}$.
\end{defn}
Our definitions of stationary and weakly stationary channels can be
viewed as Shannon-theoretic counterparts of, respectively, the classical
and general definitions of the ergodic-theoretic stationary channel
in \cite{Gray-Entropy,Fontana-1981}. The following two lemmas are
useful facts concerning these channels.
\begin{lem}
\label{lem:Lemma_Kolmogorov_Consistency_Check}A sequence $\left\{ P_{\left.Y^{n}\middle|X_{1-\lambda_{-}}^{n+\lambda_{+}}\right.}\left(\cdot\middle|\mathbf{x}\right)\right\} _{n=1}^{\infty}$
forms a consistent channel if $\forall y^{n},x_{1-\lambda_{-}}^{n+1+\lambda_{+}}$
for every $n\geq1$, there exists a probability measure $\Phi_{n+1}\left(\cdot\middle|y^{n},x_{1-\lambda_{-}}^{n+1+\lambda_{+}}\right)$
on $Y_{n+1}$ such that
\begin{align*}
 & P_{\left.Y^{n+1}\middle|X_{1-\lambda_{-}}^{n+1+\lambda_{+}}\right.}\left(\left[y^{n},y_{n+1}\right]\middle|x_{1-\lambda_{-}}^{n+1+\lambda_{+}}\right)\\
 & =\Phi_{n+1}\left(y_{n+1}\middle|y^{n},x_{1-\lambda_{-}}^{n+1+\lambda_{+}}\right)P_{\left.Y^{n}\middle|X_{1-\lambda_{-}}^{n+\lambda_{+}}\right.}\left(y^{n}\middle|x_{1-\lambda_{-}}^{n+\lambda_{+}}\right)
\end{align*}
for any $y_{n+1}$.\end{lem}
\begin{IEEEproof}
For any consistent $\left\{ X_{1-\lambda_{-}}^{n+\lambda_{+}}\sim P^{\left(n\right)}\right\} _{n=1}^{\infty}$,
\begin{align*}
 & \sum_{\substack{x_{n+1+\lambda_{+}}\\
y_{n+1}
}
}P_{X_{1-\lambda_{-}}^{n+1+\lambda_{+}}Y^{n+1}}\left(x_{1-\lambda_{-}}^{n+1+\lambda_{+}},y^{n+1}\right)\\
 & =\sum_{\substack{x_{n+1+\lambda_{+}}\\
y_{n+1}
}
}P^{\left(n+1\right)}\left(x_{1-\lambda_{-}}^{n+1+\lambda_{+}}\right)\times\\
 & \qquad\qquad P_{\left.Y^{n+1}\middle|X_{1-\lambda_{-}}^{n+1+\lambda_{+}}\right.}\left(y^{n+1}\middle|x_{1-\lambda_{-}}^{n+1+\lambda_{+}}\right)\\
 & =\sum_{\substack{x_{n+1+\lambda_{+}}\\
y_{n+1}
}
}P^{\left(n+1\right)}\left(x_{1-\lambda_{-}}^{n+1+\lambda_{+}}\right)\times\\
 & \qquad\quad\Phi_{n+1}\left(y_{n+1}\middle|y^{n},x_{1-\lambda_{-}}^{n+1+\lambda_{+}}\right)P_{\left.Y^{n}\middle|X_{1-\lambda_{-}}^{n+\lambda_{+}}\right.}\left(y^{n}\middle|x_{1-\lambda_{-}}^{n+\lambda_{+}}\right)\\
 & \stackrel{\left(a\right)}{=}\sum_{x_{n+1+\lambda_{+}}}P^{\left(n+1\right)}\left(x_{1-\lambda_{-}}^{n+1+\lambda_{+}}\right)P_{\left.Y^{n}\middle|X_{1-\lambda_{-}}^{n+\lambda_{+}}\right.}\left(y^{n}\middle|x_{1-\lambda_{-}}^{n+\lambda_{+}}\right)\\
 & \stackrel{\left(b\right)}{=}P_{X_{1-\lambda_{-}}^{n+\lambda_{+}}Y^{n}}\left(x_{1-\lambda_{-}}^{n+\lambda_{+}},y^{n}\right)
\end{align*}
$\forall x_{1-\lambda_{-}}^{n+\lambda_{+}},y^{n}$, for every $n\geq1$,
where $\left(a\right)$ is because $\sum_{y}\Phi_{n+1}\left(y\middle|y^{n},x_{1-\lambda_{-}}^{n+1+\lambda_{+}}\right)=1$,
and $\left(b\right)$ is because 
\[
\sum_{x_{n+1+\lambda_{+}}}P^{\left(n+1\right)}\left(x_{1-\lambda_{-}}^{n+1+\lambda_{+}}\right)=P^{\left(n\right)}\left(x_{1-\lambda_{-}}^{n+\lambda_{+}}\right)
\]
thanks to consistency of the input.\end{IEEEproof}
\begin{lem}
For a consistent channel, if it is stationary, it is also weakly stationary.\end{lem}
\begin{IEEEproof}
For any $k$, consider a $k$-stationary input $P_{X_{1-\lambda_{-}}^{\infty}}$,
inducing a joint input-output distribution $P_{X_{1-\lambda_{-}}^{\infty}\mathbf{Y}}$.
For any $n$ and any $\mathbf{x}\in{\cal X}^{n+\lambda_{+}+\lambda_{-}}$,
$\mathbf{y}\in{\cal Y}^{n}$,
\begin{align*}
 & P_{X_{1-\lambda_{-}}^{\infty}\mathbf{Y}}\left(T^{-k}\left\{ X_{1-\lambda_{-}}^{n+\lambda_{+}}=\mathbf{x},Y^{n}=\mathbf{y}\right\} \right)\\
 & =\sum_{\mathbf{\tilde{x}}\in{\cal X}^{k}}P_{\left.Y^{n+k}\middle|X_{1-\lambda_{-}}^{n+k+\lambda_{+}}\right.}\left(\left\{ Y_{k+1}^{n+k}=\mathbf{y}\right\} \middle|\left[\mathbf{\tilde{x}},\mathbf{x}\right]\right)\\
 & \qquad\qquad\times P_{X_{1-\lambda_{-}}^{\infty}}\left(\left\{ X_{1-\lambda_{-}}^{n+k+\lambda_{+}}=\left[\mathbf{\tilde{x}},\mathbf{x}\right]\right\} \right)\\
 & =\sum_{\mathbf{\tilde{x}}\in{\cal X}^{k}}P_{\left.Y^{n}\middle|X_{1-\lambda_{-}}^{n+\lambda_{+}}\right.}\left(\mathbf{y}\middle|\mathbf{x}\right)P_{X_{1-\lambda_{-}}^{\infty}}\left(\left\{ X_{1-\lambda_{-}}^{n+k+\lambda_{+}}=\left[\mathbf{\tilde{x}},\mathbf{x}\right]\right\} \right)\\
 & =P_{\left.Y^{n}\middle|X_{1-\lambda_{-}}^{n+\lambda_{+}}\right.}\left(\mathbf{y}\middle|\mathbf{x}\right)P_{X_{1-\lambda_{-}}^{\infty}}\left(\left\{ X_{1+k-\lambda_{-}}^{n+k+\lambda_{+}}=\mathbf{x}\right\} \right)\\
 & \stackrel{\left(a\right)}{=}P_{\left.Y^{n}\middle|X_{1-\lambda_{-}}^{n+\lambda_{+}}\right.}\left(\mathbf{y}\middle|\mathbf{x}\right)P_{X_{1-\lambda_{-}}^{\infty}}\left(\left\{ X_{1-\lambda_{-}}^{n+\lambda_{+}}=\mathbf{x}\right\} \right)\\
 & =P_{X_{1-\lambda_{-}}^{\infty}\mathbf{Y}}\left(\left\{ X_{1-\lambda_{-}}^{n+\lambda_{+}}=\mathbf{x},Y^{n}=\mathbf{y}\right\} \right)
\end{align*}
where $\left(a\right)$ is because the input is $k$-stationary. Hence
$P_{X_{1-\lambda_{-}}^{\infty}\mathbf{Y}}$ is also $k$-stationary.
\end{IEEEproof}

\subsection{Capacity Theorem}

Let 
\begin{align*}
 & i^{\left(n\right)}\left(x_{1-\lambda_{-}}^{n+\lambda_{+}};y^{n}\right)=\\
 & \log\frac{P_{\left.Y^{n}\middle|X_{1-\lambda_{-}}^{n+\lambda_{+}}\right.}\left(y^{n}\middle|x_{1-\lambda_{-}}^{n+\lambda_{+}}\right)}{{\displaystyle \sum_{\hat{x}_{1-\lambda_{-}}^{n+\lambda_{+}}}P_{\left.Y^{n}\middle|X_{1-\lambda_{-}}^{n+\lambda_{+}}\right.}\left(y^{n}\middle|\hat{x}_{1-\lambda_{-}}^{n+\lambda_{+}}\right)P^{\left(n\right)}\left(\hat{x}_{1-\lambda_{-}}^{n+\lambda_{+}}\right)}}
\end{align*}
be the information density of $x_{1-\lambda_{-}}^{n+\lambda_{+}}$
and $y^{n}$ under the input $\left\{ X_{1-\lambda_{-}}^{n+\lambda_{+}}\sim P^{\left(n\right)}\right\} _{n=1}^{\infty}$.
The mutual information is denoted by
\[
I^{\left(n\right)}\left(X_{1-\lambda_{-}}^{n+\lambda_{+}};Y^{n}\right)=\mathbb{E}^{\left(n\right)}\left[i^{\left(n\right)}\left(X_{1-\lambda_{-}}^{n+\lambda_{+}};Y^{n}\right)\right]
\]
The superscript $\left(n\right)$ implies the calculation is w.r.t.
$P_{\left.Y^{n}\middle|X_{1-\lambda_{-}}^{n+\lambda_{+}}\right.}$
and $P^{\left(n\right)}$, and it is dropped without ambiguity when
the input and the channel are both consistent.

With respect to the channel definition stated previously, define the
following capacity formula:
\[
C=\sup_{\left\{ X_{1-\lambda_{-}}^{n+\lambda_{+}}\right\} _{n=1}^{\infty}}\underline{\mathbf{I}}\left(\mathbf{X};\mathbf{Y}\right)
\]
where 
\begin{align*}
 & \underline{\mathbf{I}}\left(\mathbf{X};\mathbf{Y}\right)=\\
 & \sup\left\{ \alpha\in\mathbb{R}:\;\lim_{n\rightarrow\infty}\Pr\left(\frac{1}{n}i^{\left(n\right)}\left(X_{1-\lambda_{-}}^{n+\lambda_{+}};Y^{n}\right)\leq\alpha\right)=0\right\} 
\end{align*}
It is still an open question whether there is only one input that
attains the supremum in $C$, so we generally do not assume such.
As shown by Verd\textipa{\'{u}} and Han in \cite{Verdu-Han-1994},
the above capacity is equal to the operational capacity, which holds
with full generality for any point-to-point channel under the considered
definition. Suppose that we restrict our attention to the class of
SE inputs. We then have the following definition.
\begin{defn}
The stationary and ergodic capacity of a channel is
\[
C_{SE}=\sup_{\text{SE }\left\{ X_{1-\lambda_{-}}^{n+\lambda_{+}}\right\} _{n=1}^{\infty}}\underline{\mathbf{I}}\left(\mathbf{X};\mathbf{Y}\right)
\]
where the supremum is taken over all SE inputs.\end{defn}
\begin{thm}
\label{thm:Theorem_consistent_SE_Channel_Capacity (Shannon-theoretic)}The
capacity of a consistent SE channel with finite alphabets is 
\begin{align*}
C & =C_{SE}\\
 & =\sup_{\text{\emph{SE} }\left\{ X_{1-\lambda_{-}}^{n+\lambda_{+}}\right\} _{n=1}^{\infty}}\lim_{n\to\infty}\frac{1}{n}I\left(X_{1-\lambda_{-}}^{n+\lambda_{+}};Y^{n}\right)
\end{align*}
i.e. SE inputs can achieve its capacity.\end{thm}
\begin{IEEEproof}
The proof is a combination of ergodic-theoretic techniques, information-spectrum
results and manipulations on information-theoretic quantities. We
provide a sketch here; the complete proof is given in \nameref{sec:Appendix-Proof_of_SE_Capacity_Shannon-theoretic_Framework}.
We first establish the second equality via the Shannon-McMillan-Breiman
theorem, in which the information density normalized by $n$ converges
to the mutual information rate. To show that $C=C_{SE}$, note that
$C_{SE}\leq C$ trivially. We then have to show that the mutual information
rate quantity is an upper bound on $C$. To do so, we use a technique
in \cite{Feinstein-1959} to construct an SE input $\hat{\mu}$ from
an arbitrary finite-dimensional distribution on $X_{1-\lambda_{-}}^{r+\lambda_{+}}$,
for some $r\in\mathbb{N}^{+}$. We show that
\[
\lim_{n\rightarrow\infty}\frac{1}{n}I_{\hat{\mu}}\left(X_{1-\lambda_{-}}^{n+\lambda_{+}};Y^{n}\right)\geq\frac{1}{r+\lambda_{+}+\lambda_{-}}I^{\left(r\right)}\left(X_{1-\lambda_{-}}^{r+\lambda_{+}};Y_{1}^{r}\right)
\]
where $I_{\hat{\mu}}$ is the mutual information under $\hat{\mu}$.
The right-hand side term is an upper bound on $\underline{\mathbf{I}}\left(\mathbf{X};\mathbf{Y}\right)$
in the limit $r\to\infty$. Since $\hat{\mu}$ is SE, for any such
$\hat{\mu}$,
\[
C_{SE}\geq\lim_{n\rightarrow\infty}\frac{1}{n}I_{\hat{\mu}}\left(X_{1-\lambda_{-}}^{n+\lambda_{+}};Y^{n}\right)
\]
which completes the proof.
\end{IEEEproof}

\subsection{Bit-Symmetric Channels: Shannon-Theoretic Setting\label{sub:Bit_Symmetry (Shannon-theoretic)}}
\begin{defn}
A binary input $\left\{ X_{1-\lambda_{-}}^{n+\lambda_{+}}\right\} _{n=1}^{\infty}$
, where $X_{1-\lambda_{-}}^{n+\lambda_{+}}\sim P^{\left(n\right)}$,
is bit-symmetric if for any $n\geq1$, $P^{\left(n\right)}(\mathbf{x})=P^{\left(n\right)}(\neg\mathbf{x})$
$\forall\mathbf{x}\in\mathrm{GF}(2)^{n+\lambda_{+}+\lambda_{-}}$.
\end{defn}

\begin{defn}
A binary channel $\left\{ P_{\left.Y^{n}\middle|X_{1-\lambda_{-}}^{n+\lambda_{+}}\right.}\left(\cdot\middle|\mathbf{x}\right)\right\} _{n=1}^{\infty}$
is bit-symmetric if for any $n\geq1$, $P_{\left.Y^{n}\middle|X_{1-\lambda_{-}}^{n+\lambda_{+}}\right.}\left(\mathbf{y}\middle|\mathbf{x}\right)=P_{\left.Y^{n}\middle|X_{1-\lambda_{-}}^{n+\lambda_{+}}\right.}\left(\neg\mathbf{y}\middle|\neg\mathbf{x}\right)$
$\forall\mathbf{x},\mathbf{y}$.\end{defn}
\begin{prop}
\label{prop:Proposition_Bit-symmetry (Shannon-theoretic)}For any
bit-symmetric binary channel,
\begin{align}
 & \sup_{\text{\emph{Stationary} }\left\{ X_{1-\lambda_{-}}^{n+\lambda_{+}}\right\} _{n=1}^{\infty}}\lim_{n\to\infty}\frac{1}{n}I\left(X_{1-\lambda_{-}}^{n+\lambda_{+}};Y^{n}\right)\nonumber \\
 & =\sup_{\text{\emph{Stationary, bit-symmetric} }\left\{ X_{1-\lambda_{-}}^{n+\lambda_{+}}\right\} _{n=1}^{\infty}}\lim_{n\to\infty}\frac{1}{n}I\left(X_{1-\lambda_{-}}^{n+\lambda_{+}};Y^{n}\right)\label{eq:Eq_Proposition_Bit-symmetry (Shannon-theoretic)}
\end{align}

\end{prop}
The proof is given in \nameref{sec:Appendix-Proof_Bit-symmetry}.
The idea is similar to the proof of Proposition \ref{prop:Proposition_Bit-symmetry (ergodic)}.

It is an open question whether the supremizing input set in Proposition
\ref{prop:Proposition_Bit-symmetry (Shannon-theoretic)} could further
be reduced to the SE input set to match with Theorem \ref{thm:Theorem_consistent_SE_Channel_Capacity (Shannon-theoretic)},
thereby allowing us to obtain a Shannon-theoretic result in parallel
with Eq. \eqref{eq:Equation_C_2}. In the specific case of the DID
channel, we shall answer this in the positive in the next section.
Nevertheless, since the left-hand side of Eq. \eqref{eq:Eq_Proposition_Bit-symmetry (Shannon-theoretic)}
is an upper bound on the capacity $C$ given in Theorem \ref{thm:Theorem_consistent_SE_Channel_Capacity (Shannon-theoretic)},
the proposition helps in calculating this upper bound in general.

\subsection{Applicability to the DID Channel\label{sub:Shannon-Applicability-DID}}

Reusing the notation ${\cal E}\left(Z_{i};y_{i},x_{i},x_{i-1}\right)$
in Section \ref{sub:Ergodic-theoretic Capacity}, define
\[
{\cal E}\left(x_{0}^{n},\left\{ Y^{n}=y^{n}\right\} \right)=\bigcap_{i=1}^{n}{\cal E}\left(Z_{i};y_{i},x_{i},x_{i-1}\right)
\]
Then the channel law, under the Shannon-theoretic setting, is specified
by:
\[
P_{\left.Y^{n}\middle|X_{0}^{n}\right.}\left(y^{n}\middle|x_{0}^{n}\right)=P_{\mathbf{Z}}\left({\cal E}\left(x_{0}^{n},\left\{ Y^{n}=y^{n}\right\} \right)\right)
\]

We first verify that the DID channel is consistent. Noticing $\left\{ Z_{i}\right\} _{i=1}^{\infty}$
is consistent, one easily deduces that:
\begin{align*}
 & P_{\left.Y^{n+1}\middle|X_{0}^{n+1}\right.}\left(y^{n+1}\middle|x_{0}^{n+1}\right)\\
 & =P_{\left.Y^{n}\middle|X_{0}^{n}\right.}\left(y^{n}\middle|x_{0}^{n}\right)P_{\left.Z_{n+1}\middle|Z_{1}^{n}\right.}\Big({\cal E}\left(Z_{n+1};y_{n+1},x_{n+1},x_{n}\right)\Big|\\
 & \qquad\qquad\qquad\qquad\qquad\qquad\qquad{\cal E}\left(x_{0}^{n},\left\{ Y^{n}=y^{n}\right\} \right)\Big)
\end{align*}
But for any $y^{n}$ and $x_{0}^{n+1}$,
\begin{align*}
 & \sum_{y_{n+1}\in\mathrm{GF}\left(2\right)}P_{\left.Z_{n+1}\middle|Z_{1}^{n}\right.}\Big({\cal E}\left(Z_{n+1};y_{n+1},x_{n+1},x_{n}\right)\Big|\\
 & \qquad\qquad\qquad\qquad\qquad{\cal E}\left(x_{0}^{n},\left\{ Y^{n}=y^{n}\right\} \right)\Big)=1
\end{align*}
and therefore, 
\[
P_{\left.Z_{n+1}\middle|Z_{1}^{n}\right.}\left({\cal E}\left(Z_{n+1};\cdot,x_{n+1},x_{n}\right)\middle|{\cal E}\left(x_{0}^{n},\left\{ Y^{n}=y^{n}\right\} \right)\right)
\]
 is equivalent to a probability measure $\Phi_{n+1}\left(\cdot\middle|y^{n},x_{0}^{n+1}\right)$
on $Y_{n+1}$. By Lemma \ref{lem:Lemma_Kolmogorov_Consistency_Check},
the DID channel is consistent.

Like Section \ref{sub:Ergodic-theoretic Capacity}, under the initialization
$P\left(Z_{1}=0\right)=p_{d}/\left(p_{i}+p_{d}\right)$ and $P\left(Z_{1}=1\right)=p_{i}/\left(p_{i}+p_{d}\right)$,
we have:
\begin{align*}
 & P_{\left.Y^{n+k}\middle|X_{0}^{n+k}\right.}\left(\left\{ Y_{k+1}^{n+k}=\mathbf{y}\right\} \middle|x_{0}^{n+k}\right)\\
 & =P_{\mathbf{Z}}\left(\bigcap_{i=1}^{n}{\cal E}\left(Z_{k+i};y_{i},x_{k+i},x_{k+i-1}\right)\right)\\
 & =P_{\mathbf{Z}}\left(\bigcap_{i=1}^{n}{\cal E}\left(Z_{i};y_{i},x_{k+i},x_{k+i-1}\right)\right)\\
 & =P_{\left.Y^{n}\middle|X_{0}^{n}\right.}\left(\mathbf{y}\middle|x_{k}^{n+k}\right)
\end{align*}
which establishes the DID channel's stationarity. It is also easy
to see that the DID channel is bit-symmetric in the Shannon-theoretic
sense. Finally, we recall that the channel law $P_{\left.\mathbf{Y}\middle|X_{0}^{\infty}\right.}\left(\cdot\middle|\mathbf{x}\right)$
exists for the DID channel under the ergodic-theoretic framework.
As such, for consistent inputs, the joint input-output distribution
of this model can always be described by Eq. \eqref{eq:Eq_input_output_dist_channel_def_ergodic}.
In Section \ref{sub:Ergodic-theoretic Capacity}, we have shown that
for the DID channel model, under the aforementioned initialization,
the channel law $P_{\left.\mathbf{Y}\middle|X_{0}^{\infty}\right.}\left(\cdot\middle|\mathbf{x}\right)$
is ergodic (in the ergodic-theoretic sense), i.e. an SE input induces
an ergodic joint input-output distribution. By our Shannon-theoretic
definition of ergodic channels, the DID channel's ergodicity is thus
validated.

Now we argue that initializations are irrelevant to the DID channel
capacity under the Shannon-theoretic framework. The following capacity
formula for indecomposable finite-state channels (FSC), which are
those with the initial FSC state's effect vanishing with time, is
well-known \cite{Gallager-1968}:
\begin{equation}
C={\displaystyle \lim_{n\rightarrow\infty}\frac{1}{n}\sup_{X_{0}^{n}}I\left(X_{0}^{n};Y^{n}\right)}\label{eq:Equation_C_indecomposableFSC}
\end{equation}
It can be proven that the DID channel is an indecomposable FSC for
any $p_{i},p_{d}\in(0,1)$ by an easy extension of the argument of
the case $p_{d}=1$ presented in \cite[Proposition 11]{Mazumdar-2011}.
The initial FSC state is $\left(X_{0},Z_{0}\right)$ as in that argument,
where we extend the DID channel state sequence to $Z_{0}$ without
affecting $Z_{1},Z_{2},\ldots$, which is possible for Markov processes.
Since the DID channel is indecomposable, its capacity is the same
for all distributions on $\left(X_{0},Z_{0}\right)$. Hence for given
$p_{i}$ and $p_{d}$, it can be calculated w.r.t. $P\left(Z_{0}=0\right)=p_{d}/\left(p_{i}+p_{d}\right)$
and $P\left(Z_{0}=1\right)=p_{i}/\left(p_{i}+p_{d}\right)$. But this
distribution on $Z_{0}$ induces the aforementioned distribution on
$Z_{1}$, which concludes our argument.

Finally the observation that the ergodic-theoretic channel definition
also applies to the DID channel model leads to the following result:
\[
C=\sup_{\substack{\text{Stationary, bit-symmetric}\\
\left\{ X_{0}^{n}\right\} _{n=1}^{\infty}
}
}\lim_{n\to\infty}\frac{1}{n}I\left(X_{0}^{n};Y^{n}\right)
\]
To see this, we first note that since the channel law $P_{\left.\mathbf{Y}\middle|X_{0}^{\infty}\right.}\left(\cdot\middle|\mathbf{x}\right)$
exists and is shown to define an ergodic-theoretic stationary (one-sided)
channel, and the joint input-output distribution obeys Eq. \eqref{eq:Eq_input_output_dist_channel_def_ergodic},
applying \cite[Lemma 12.4.2]{Gray-Entropy}, we have:
\[
\sup_{\substack{\text{Stationary}\\
\left\{ X_{0}^{n}\right\} _{n=1}^{\infty}
}
}\lim_{n\to\infty}\frac{1}{n}I\left(X_{0}^{n};Y^{n}\right)=\sup_{\text{SE }\left\{ X_{0}^{n}\right\} _{n=1}^{\infty}}\lim_{n\to\infty}\frac{1}{n}I\left(X_{0}^{n};Y^{n}\right)
\]
The formula for $C$ then follows immediately from Proposition \ref{prop:Proposition_Bit-symmetry (Shannon-theoretic)}
and Theorem \ref{thm:Theorem_consistent_SE_Channel_Capacity (Shannon-theoretic)}.

The formula we obtain is hence the same under both frameworks. In
subsequent sections, we shall explore various ways to evaluate the
right-hand side of Eq. \eqref{eq:Equation_C_2}, thereby yielding
the same results for both frameworks.

\section{DID Channel Capacity Lower Bound\label{sec:DID-Capacity-Lower-Bound}}

\begin{figure}
\begin{centering}
\includegraphics[width=1\columnwidth]{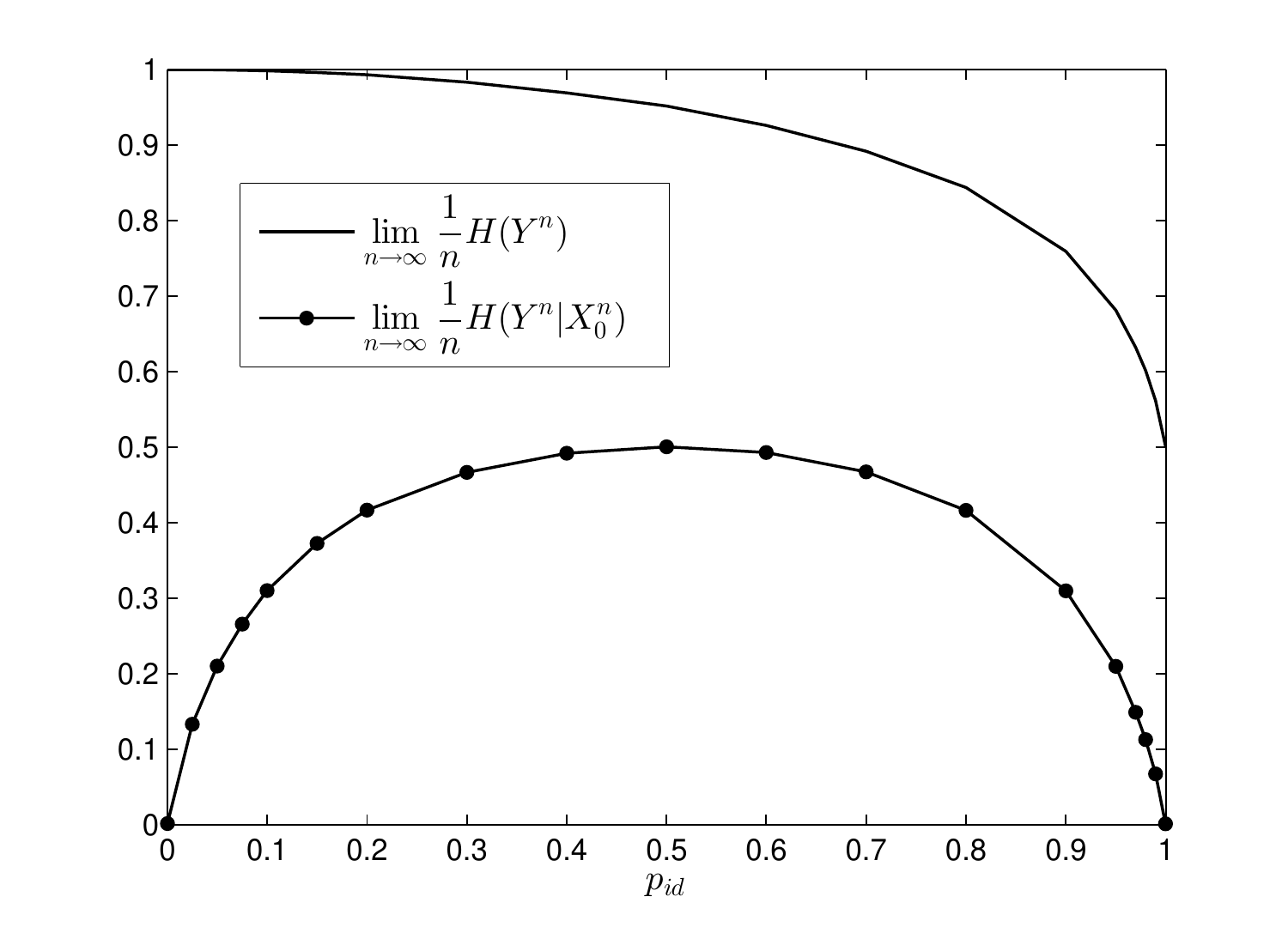}
\par\end{centering}

\protect\caption{The two terms in Eq. \eqref{eq:Equation_C_1} for the i.u.d. input.}

\label{Fig_Capacity_Entropy_Term}
\end{figure}

\begin{figure*}
\normalsize
\setcounter{MyTempEquationCounter}{\value{equation}}
\setcounter{equation}{11}

\begin{equation}
H\left(Y_{n}\middle|Y_{n-1},X_{n-2},Z_{n-2}\right)={\displaystyle \frac{\alpha p_{d}}{p_{i}+p_{d}}h_{2}(\alpha-\alpha p_{i})}+\frac{p_{i}(1-\alpha p_{d})}{p_{i}+p_{d}}h_{2}\left(A_{1}\right){\displaystyle +\frac{p_{d}(1+\alpha p_{i}-\alpha)}{p_{i}+p_{d}}h_{2}\left(A_{2}\right)}\label{eq:Lower_bound_term_1}
\end{equation}
\begin{align*}
A_{1} & ={\displaystyle \frac{1-\alpha-2\alpha p_{d}+\alpha p_{d}^{2}+3\alpha^{2}p_{d}-2\alpha^{2}p_{d}^{2}+\alpha p_{i}p_{d}-\alpha^{2}p_{i}p_{d}}{1-\alpha p_{d}}}\\
A_{2} & ={\displaystyle \frac{1+2\alpha p_{i}-2\alpha-\alpha p_{i}^{2}-2\alpha^{2}p_{i}+\alpha^{2}+\alpha^{2}p_{i}^{2}-\alpha p_{d}p_{i}+2\alpha^{2}p_{i}p_{d}}{1+\alpha p_{i}-\alpha}}
\end{align*}
\vspace*{4pt}

\rule[0.5ex]{1\textwidth}{1pt}

\setcounter{equation}{12}
\begin{align}
 & {\displaystyle \lim_{n\to\infty}H\left(Y_{n}\middle|Y^{n-1},X_{0}^{n}\right)}=\nonumber \\
 & {\displaystyle \sum_{k=1}^{\infty}\alpha^{2}\left(1-\alpha\right)^{k-1}\left[{\displaystyle \frac{p_{d}}{p_{i}+p_{d}}h_{2}\left(\frac{p_{d}+p_{i}\left(1-p_{i}-p_{d}\right)^{k}}{p_{i}+p_{d}}\right)}{\displaystyle +\frac{p_{i}}{p_{i}+p_{d}}h_{2}\left(\frac{p_{i}+p_{d}\left(1-p_{i}-p_{d}\right)^{k}}{p_{i}+p_{d}}\right)}\right]}\label{eq:Lower_bound_term_2}
\end{align}
\setcounter{equation}{\value{MyTempEquationCounter}}\vspace*{4pt}

\rule[0.5ex]{1\textwidth}{1pt}
\end{figure*}

To derive a good lower bound, behaviors of the two terms in Eq. \eqref{eq:Equation_C_1}
have to be analyzed. Fig. \ref{Fig_Capacity_Entropy_Term} plots the
terms for the case of i.u.d. input, using the method in \cite{Arnold-2006},
when $p_{i}=p_{d}=p_{id}$. It can be observed that $\lim_{n\to\infty}\frac{1}{n}H\left(Y^{n}\right)$
decreases slowly for low $p_{id}$ (e.g. $p_{id}<0.5$), whereas $\lim_{n\to\infty}\frac{1}{n}H\left(Y^{n}\middle|X_{0}^{n}\right)$
varies drastically. As such, if we are to approximate the capacity
for practical values of $p_{id}$, the term $\lim_{n\to\infty}\frac{1}{n}H\left(Y^{n}\middle|X_{0}^{n}\right)$
should be carefully handled, and an estimation of $\lim_{n\to\infty}\frac{1}{n}H\left(Y^{n}\right)$
might be sufficient.

To gain more insights into how the terms could be evaluated, from
Eq. \eqref{eq:Equation_C_2}, notice the following correspondence:
\begin{align*}
\lim_{n\to\infty}\frac{1}{n}H\left(Y^{n}\right) & ={\displaystyle \lim_{n\to\infty}H\left(Y_{n}\middle|Y^{n-1}\right)}\\
\lim_{n\to\infty}\frac{1}{n}H\left(Y^{n}\middle|X_{0}^{n}\right) & =\lim_{n\to\infty}H\left(Y_{n}\middle|Y^{n-1},X_{0}^{n}\right)
\end{align*}
Since the input is i.u.d., it can be shown that $H\left(Y_{n}\right)=1$
for any $p_{id}$. With the fact that the first term is close to $1$
at low $p_{id}$ whereas the second term differs, one may say, only
a few most recent past outputs in $Y^{n-1}$ carry a majority of knowledge
about $Y_{n}$ for low $p_{id}$; however, when given $X_{0}^{n}$,
farther past outputs are able to resolve more uncertainty about $Y_{n}$
and thus may not be ignored. While this discussion pertains to the
i.u.d. input only, the (only) capacity-achieving input is i.u.d. when
$p_{id}=0$ (i.e. the channel is noiseless) and so by a continuity
argument, at low $p_{id}$, the inputs that achieve the capacity should
behave nearly i.u.d.-like and the insights drawn are thus expected
to be useful.

Recall that the lower bound established in \cite{Iyengar-2011,Mazumdar-2011}
is the achievable rate with i.u.d. input $C_{iud}$. Now to derive
an analytical lower bound that improves on $C_{iud}$, not only do
we need to consider a more complex input distribution, but it also
cannot be too complicated to analyze. An immediate candidate is its
fortiori, a stationary bit-symmetric first-order Markovian input:
$P_{\left.X_{n+1}\middle|X_{n}\right.}(1|0)=P_{\left.X_{n+1}\middle|X_{n}\right.}(0|1)=\alpha$
and $P_{X_{0}}(0)=P_{X_{0}}(1)=0.5$. We evaluate each term in Eq.
\eqref{eq:Equation_C_2} in the following.

\subsection{The first term}

As discussed, an estimation of this term that retains the first few
recent past outputs is sufficient. We lower-bound the term with $Y^{n-2}$
to be discarded: 
\begin{align}
\lim_{n\to\infty}H\left(Y_{n}\middle|Y^{n-1}\right) & \geq\lim_{n\to\infty}H\left(Y_{n}\middle|Y^{n-1},X_{0}^{n-2},Z^{n-2}\right)\nonumber \\
 & \stackrel{\left(a\right)}{=}\lim_{n\to\infty}H\left(Y_{n}\middle|Y_{n-1},X_{0}^{n-2},Z^{n-2}\right)\nonumber \\
 & \stackrel{\left(b\right)}{=}\lim_{n\to\infty}H\left(Y_{n}\middle|Y_{n-1},X_{n-2},Z_{n-2}\right)\nonumber \\
 & \stackrel{\left(c\right)}{=}H\left(Y_{n}\middle|Y_{n-1},X_{n-2},Z_{n-2}\right)\label{eq:Eq_Supplemental_Lower_Bound_First_Term_Equality_3}
\end{align}
where $\left(a\right)$ is because $Y_{i}$ is a function of $Z_{i}$
and $X_{i-1}^{i}$, $\left(b\right)$ is because 
\[
\left(X_{0}^{n-3},Z^{n-3}\right)\to\left(X_{n-2},Z_{n-2}\right)\to\left(X_{n-2}^{n},Z_{n-1}^{n}\right)\to Y_{n-1}^{n}
\]
 i.e. they form a Markov chain in that order, and $\left(c\right)$
is due to the following:
\begin{align*}
 & P_{Y_{n}Y_{n-1}X_{n-2}Z_{n-2}}\left(y_{1},y_{2},x,z\right)\\
 & =\sum_{\substack{x_{1},x_{2}\\
z_{1},z_{2}
}
}\left\{ \begin{array}{c}
P_{\left.Y_{n}\middle|X_{n}X_{n-1}Z_{n}\right.}\left(y_{1}\middle|x_{1},x_{2},z_{1}\right)\times\\
P_{\left.Y_{n-1}\middle|X_{n-1}X_{n-2}Z_{n-1}\right.}\left(y_{2}\middle|x_{2},x,z_{2}\right)\times\\
P_{Z_{n-2}^{n}}\left(z_{1},z_{2},z\right)P_{X_{n-2}^{n}}\left(x_{1},x_{2},x\right)
\end{array}\right\} \\
 & =\sum_{\substack{x_{1},x_{2}\\
z_{1},z_{2}
}
}\left\{ \begin{array}{c}
P_{\left.Y_{n+1}\middle|X_{n+1}X_{n}Z_{n+1}\right.}\left(y_{1}\middle|x_{1},x_{2},z_{1}\right)\times\\
P_{\left.Y_{n}\middle|X_{n}X_{n-1}Z_{n}\right.}\left(y_{2}\middle|x_{2},x,z_{2}\right)\times\\
P_{Z_{n-1}^{n+1}}\left(z_{1},z_{2},z\right)P_{X_{n-1}^{n+1}}\left(x_{1},x_{2},x\right)
\end{array}\right\} \\
 & =P_{Y_{n+1}Y_{n}X_{n-1}Z_{n-1}}\left(y_{1},y_{2},x,z\right)
\end{align*}
for any $\left(y_{1},y_{2},x,z\right)$ and $n\geq3$, where we make
use of the fact that $Y_{n}$ is a time-invariant function of $\left(X_{n},X_{n-1},Z_{n}\right)$
for any $n$, the considered input is stationary, and the capacity
is computed w.r.t. stationary $\left\{ Z_{n}\right\} _{n=1}^{\infty}$,
as mentioned in Section \ref{sub:Ergodic-theoretic Capacity}. $H\left(Y_{n}\middle|Y_{n-1},X_{n-2},Z_{n-2}\right)$
is then given by Eq. \eqref{eq:Lower_bound_term_1}\addtocounter{equation}{1},
given that $\left\{ Z_{n}\right\} _{n=1}^{\infty}$ attains its stationary
distribution as stated.

For a better approximation, repeating the same argument above, we
have:
\begin{align*}
\lim_{n\to\infty}H\left(Y_{n}\middle|Y^{n-1}\right) & \geq H\left(Y_{n}\middle|Y_{n-\kappa}^{n-1},X_{n-\kappa-1},Z_{n-\kappa-1}\right)
\end{align*}
for some finite $\kappa\geq1$. However at relatively low noise, $\kappa=1$
suffices.

\subsection{The second term}

\begin{figure*}
\normalsize
\setcounter{MyTempEquationCounter}{\value{equation}}
\setcounter{equation}{13}
\begin{equation}
\left[\begin{array}{cc}
P_{Z_{n}|Z_{n-k}}(0|0) & P_{Z_{n}|Z_{n-k}}(0|1)\\
P_{Z_{n}|Z_{n-k}}(1|0) & P_{Z_{n}|Z_{n-k}}(1|1)
\end{array}\right]\;=\;\mathfrak{M}^{k}\;=\;{\displaystyle \frac{1}{p_{i}+p_{d}}\left[\begin{array}{cc}
p_{d}+p_{i}\left(1-p_{i}-p_{d}\right)^{k} & p_{d}-p_{d}\left(1-p_{i}-p_{d}\right)^{k}\\
p_{i}-p_{i}\left(1-p_{i}-p_{d}\right)^{k} & p_{i}+p_{d}\left(1-p_{i}-p_{d}\right)^{k}
\end{array}\right]}\label{eq:Eq_P(Zn, Zn-k) by eigen decomposition}
\end{equation}

\vspace*{4pt}

\rule[0.5ex]{1\textwidth}{1pt}

\setcounter{equation}{14}
\begin{equation}
H\left(Z_{n}\middle|Z_{n-k}\right)={\displaystyle \frac{p_{d}}{p_{i}+p_{d}}h_{2}\left(\frac{p_{d}+p_{i}\left(1-p_{i}-p_{d}\right)^{k}}{p_{i}+p_{d}}\right)+\frac{p_{i}}{p_{i}+p_{d}}h_{2}\left(\frac{p_{i}+p_{d}\left(1-p_{i}-p_{d}\right)^{k}}{p_{i}+p_{d}}\right)}\label{eq:Eq_H(Zn | Zn-k)}
\end{equation}

\vspace*{4pt}

\rule[0.5ex]{1\textwidth}{1pt}\setcounter{equation}{15}
\begin{equation}
C_{Mkv}^{lb}=\max_{\alpha\in\left[0,1\right]}\left[H\left(Y_{n}\middle|Y_{n-1},X_{n-2},Z_{n-2}\right)-\lim_{n\to\infty}H\left(Y_{n}\middle|Y^{n-1},X_{0}^{n}\right)\right]_{{\displaystyle X_{0}^{\infty}\sim\mbox{Markov}\left(\alpha\right)}}\label{eq:Lower_bound}
\end{equation}

\setcounter{equation}{\value{MyTempEquationCounter}}\vspace*{4pt}

\rule[0.5ex]{1\textwidth}{1pt}
\end{figure*}

\begin{lem}
\label{lem:Lemma_Second_Term_Formula_for_Stationary_Input}$\lim_{n\to\infty}H\left(Y_{n}\middle|Y^{n-1},X_{0}^{n}\right)$
is given by Eq. \eqref{eq:Lower_bound_term_2}\addtocounter{equation}{1},
in which $\alpha=P\left(X_{i}\neq X_{i-1}\right)$, for any stationary
input.
\end{lem}

\begin{IEEEproof}
$\alpha$ exists thanks to stationarity of the input. We make a few
observations:

(Ob.1) Given $X_{i}\neq X_{i-1}$, knowledge (or respectively, uncertainty)
about $Y_{i}$ is equivalent to knowledge (or respectively, uncertainty)
about $Z_{i}$.

(Ob.2) Given $X_{i}=X_{i-1}$, uncertainty about $Y_{i}$ is completely
resolved, and knowledge of $Y_{i}$ provides no information about
$Z_{i}$.

(Ob.3) Without knowing $Y_{i}$, knowledge of $X_{i}$ also provides
no information about $Z_{i}$, since the input $X_{0}^{n}$ and the
channel state $Z^{n}$ are independent.

(Ob.4)\label{(Ob.4) Observation 4 - Lower Bound} As for resolving
uncertainty about $Z_{i}$, knowledge of $Y^{i-1}$ and $X_{0}^{i-1}$
is only as good as to provide (partial) knowledge of $Z^{i-1}$. Furthermore,
since $Z^{n}$ is a first-order Markov process, in the resolution
of uncertainty about $Z_{i}$, for $q\leq k\leq i-1$, knowing $Z_{q}^{k}$
(without knowledge of $Z_{k+1}^{i-1}$, if $k<i-1$) is the same as
knowing $Z_{k}$.

We then have:
\begin{align*}
 & H\left(Y_{n}\middle|Y^{n-1},X_{0}^{n}\right)\\
 & =H\left(Y_{n}\middle|Y^{n-1},X_{0}^{n},X_{n}=X_{n-1}\right)P\left(X_{n}=X_{n-1}\right)\\
 & \quad+H\left(Y_{n}\middle|Y^{n-1},X_{0}^{n},X_{n}\neq X_{n-1}\right)P\left(X_{n}\neq X_{n-1}\right)\\
 & =\alpha H\left(Z_{n}\middle|Y^{n-1},X_{0}^{n},X_{n}\neq X_{n-1}\right)\\
 & =\alpha H\left(Z_{n}\middle|Y^{n-1},X_{0}^{n-1}\right)
\end{align*}
\begin{align*}
 & H\left(Z_{n}\middle|Y^{n-1},X_{0}^{n-1}\right)\\
 & =\alpha H\left(Z_{n}\middle|Y^{n-1},X_{0}^{n-1},X_{n-1}\neq X_{n-2}\right)\\
 & \quad+\left(1-\alpha\right)H\left(Z_{n}\middle|Y^{n-1},X_{0}^{n-1},X_{n-1}=X_{n-2}\right)\\
 & =\alpha H\left(Z_{n}\middle|Z_{n-1}\right)+\left(1-\alpha\right)H\left(Z_{n}\middle|Y^{n-2},X_{0}^{n-2}\right)
\end{align*}
Similarly:
\begin{align*}
 & H\left(Z_{n}\middle|Y^{n-2},X_{0}^{n-2}\right)\\
 & =\alpha H\left(Z_{n}\middle|Z_{n-2}\right)+\left(1-\alpha\right)H\left(Z_{n}\middle|Y^{n-3},X_{0}^{n-3}\right)\\
 & \ldots\\
 & H\left(Z_{n}\middle|Y_{1},X_{0}^{1}\right)=\alpha H\left(Z_{n}\middle|Z_{1}\right)+\left(1-\alpha\right)H\left(Z_{n}\right)
\end{align*}
We are then left with evaluating $H\left(Z_{n}\middle|Z_{n-k}\right)$
for $k=1,\ldots,n-1$. Consider 
\begin{align*}
\mathfrak{M} & =\left[\begin{array}{cc}
P_{Z_{n}|Z_{n-1}}(0|0) & P_{Z_{n}|Z_{n-1}}(0|1)\\
P_{Z_{n}|Z_{n-1}}(1|0) & P_{Z_{n}|Z_{n-1}}(1|1)
\end{array}\right]\\
 & =\left[\begin{array}{cc}
1-p_{i} & p_{d}\\
p_{i} & 1-p_{d}
\end{array}\right]
\end{align*}
We then obtain Eq. \eqref{eq:Eq_P(Zn, Zn-k) by eigen decomposition}\addtocounter{equation}{1}.
Subsequently $H\left(Z_{n}\middle|Z_{n-k}\right)$ is given by Eq.
\eqref{eq:Eq_H(Zn | Zn-k)}\addtocounter{equation}{1}. Putting everything
together, we obtain the lemma.
\end{IEEEproof}

When the input is the aforementioned first-order Markovian process,
$\alpha$ given in Eq. \eqref{eq:Lower_bound_term_2}\addtocounter{equation}{1}
is also $P_{\left.X_{n+1}\middle|X_{n}\right.}(1|0)=P_{\left.X_{n+1}\middle|X_{n}\right.}(0|1)$.

\begin{figure}
\begin{centering}
\includegraphics[width=1\columnwidth]{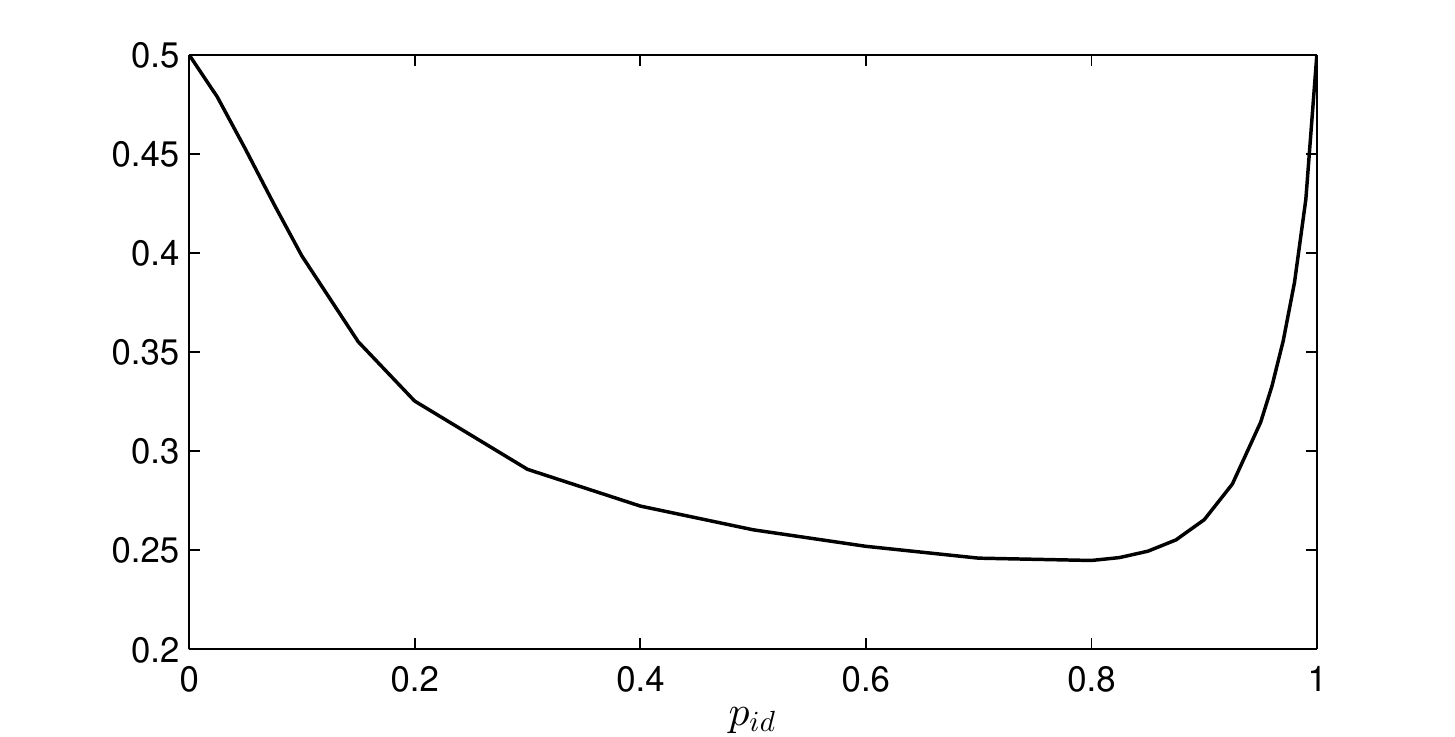}
\par\end{centering}

\protect\caption{Optimized $\alpha$ for $C_{Mkv}^{lb}$.}

\label{Fig_Capacity_Markov_Optimized}
\end{figure}

\begin{figure}
\begin{centering}
\includegraphics[width=1\columnwidth]{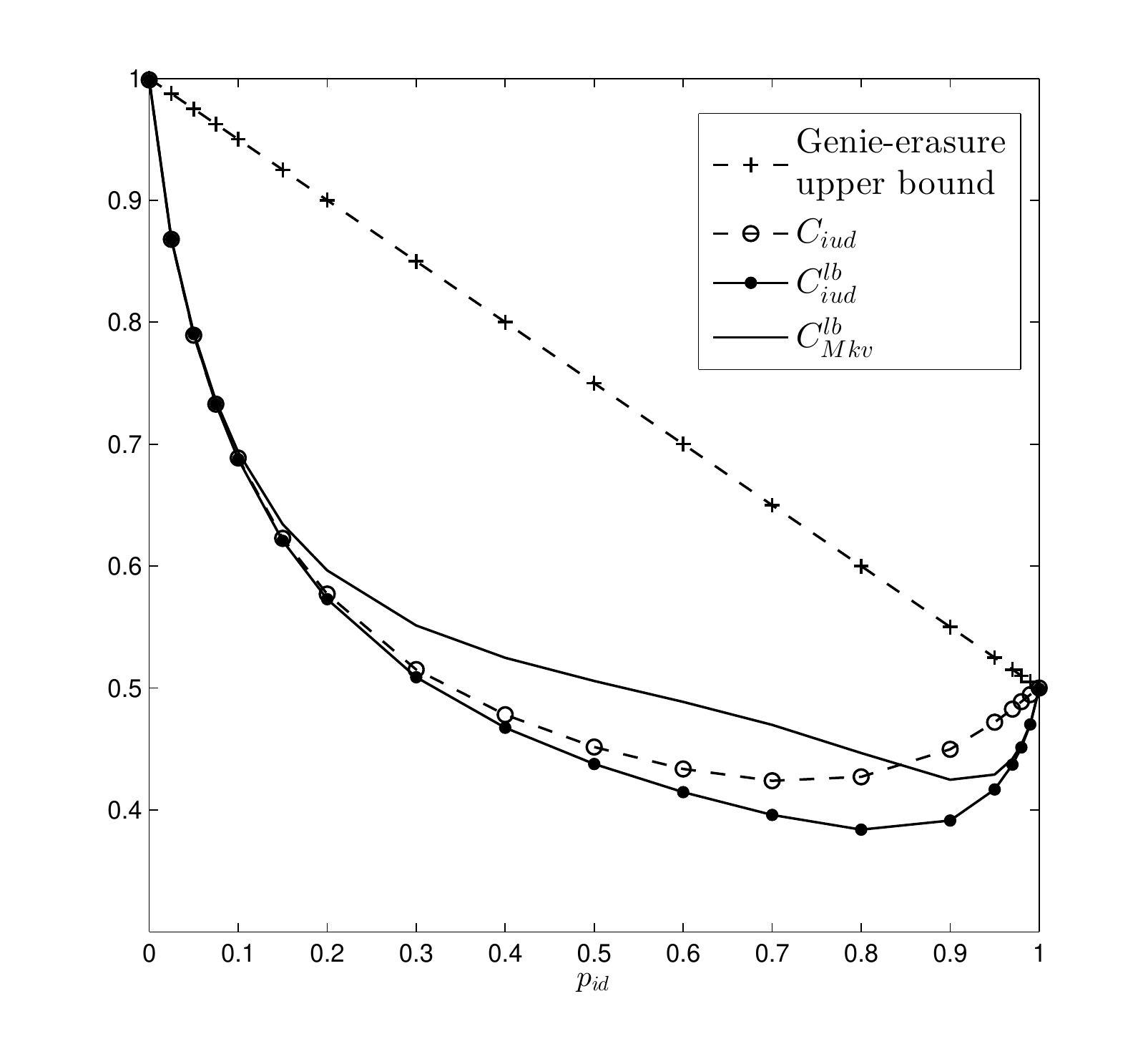}
\par\end{centering}

\protect\caption{New lower bound $C_{Mkv}^{lb}$. Here $C_{iud}^{lb}$ is $C_{Mkv}^{lb}$
with unoptimized $\alpha=0.5$ (i.e. i.u.d. input) for all $p_{id}$.}

\label{Fig_Capacity_Markov_Lower_Bound}
\end{figure}

Now combining Eq. \eqref{eq:Lower_bound_term_1} and \eqref{eq:Lower_bound_term_2},
the new lower bound $C_{Mkv}^{lb}$ is then given by Eq. \eqref{eq:Lower_bound}.
The maximization is reduced to a univariate one (i.e. maximization
in $\alpha$), which can be computed efficiently. The results are
given in Fig. \ref{Fig_Capacity_Markov_Optimized} and \ref{Fig_Capacity_Markov_Lower_Bound}.
It can be observed that $C_{Mkv}^{lb}$ has improved over $C_{iud}$,
given in \cite{Iyengar-2011}, for most values of $p_{id}$. We also
note, the fact that $C_{iud}^{lb}$ (which is $C_{Mkv}^{lb}$ with
unoptimized $\alpha=0.5$) is close to $C_{iud}$ shows that the estimation
in Eq. \eqref{eq:Eq_Supplemental_Lower_Bound_First_Term_Equality_3}
is not a bad one as previously expected.

The bounding technique here relies on the exact establishment of the
second entropy term. While Lemma \ref{lem:Lemma_Second_Term_Formula_for_Stationary_Input}
is heavily channel-dependent, we make a short note on how the strategy
could be extended to calculate the entropy term when $Z_{i}$ has
a larger alphabet. For example, consider ${\cal Z}=\left\{ 0,1,2\right\} $,
and the term is then $\lim_{n\to\infty}H\left(Y_{n}\middle|Y^{n-1},X_{-1}^{n}\right)$.
Consider the following $4$ variables:
\begin{align*}
\alpha_{1} & =P\left(X_{n}=X_{n-1}=X_{n-2}\right)\\
\alpha_{2} & =P\left(X_{n}=X_{n-1}\neq X_{n-2}\right)\\
\alpha_{3} & =P\left(X_{n}\neq X_{n-1}=X_{n-2}\right)\\
\alpha_{4} & =P\left(X_{n-1}\neq X_{n}=X_{n-2}\right)
\end{align*}
$\left\{ \alpha_{1},\alpha_{2},\alpha_{3},\alpha_{4}\right\} $ plays
the role as $\alpha$ in Lemma \ref{lem:Lemma_Second_Term_Formula_for_Stationary_Input}
and can be reduced in size via the stationarity condition and also
the fact $\alpha_{1}+\alpha_{2}+\alpha_{3}+\alpha_{4}=1$. The key
to the extension is to realize that similar to (Ob.1) and (Ob.2),
given an event that corresponds to one among $\alpha_{1},\alpha_{2},\alpha_{3},\alpha_{4}$,
the uncertainty of $Y_{i}$ could be partially reduced to that of
$Z_{i}$, and knowledge of $Y_{i}$ is partially equivalent to that
of $Z_{i}$. For example, 
\begin{align*}
 & H\left(Y_{n}\middle|Y^{n-1},X_{-1}^{n},X_{n}\neq X_{n-1}=X_{n-2}\right)\\
 & =H\left(Z_{n}^{\left(12\right)}\middle|Y^{n-1},X_{-1}^{n},X_{n}\neq X_{n-1}=X_{n-2}\right)\\
 & =H\left(Z_{n}^{\left(12\right)}\middle|Y^{n-1},X_{-1}^{n-1}\right)\\
 & H\left(Z_{n}^{\left(12\right)}\middle|Y^{n-1},X_{-1}^{n-1},X_{n-2}\neq X_{n-1}=X_{n-3}\right)\\
 & =H\left(Z_{n}^{\left(12\right)}\middle|Z_{n-1}^{\left(02\right)},Y^{n-1},X_{-1}^{n-1},X_{n-2}\neq X_{n-1}=X_{n-3}\right)\\
 & =H\left(Z_{n}^{\left(12\right)}\middle|Z_{n-1}^{\left(02\right)},Y^{n-2},X_{-1}^{n-2}\right)
\end{align*}
where $Z_{n}^{\left(jk\right)}=\mathbb{I}\left(Z_{n}\in\left\{ j,k\right\} \right)$.
Here $\mathbb{I}\left(A\right)$ is $1$ if $A$ is true and $0$
otherwise.

\section{DID Channel Capacity Upper Bound\label{sec:DID-Capacity-Upper-Bound}}

In this section, we formulate a new series of computable upper bounds
with a parameter ${\cal L}$. The upper bounds improve as ${\cal L}$
increases, and match up with the developed lower bound at low noise
levels. We also discuss the crucial role of the input stationarity
condition in the new upper bounds, without which the bounds could
be trivialized.

\subsection{Formulation and Computation}

An upper bound is usually difficult to establish analytically, since
it involves maximization over all input distributions, unlike lower
bounds. We shall rely on computational methods. To do so, we assume
stationary inputs and bound each term in Eq. \eqref{eq:Equation_C_2}
as follows. Let $\mathcal{L}\geq1$ be a given finite parameter to
control the accuracy of the to-be-formulated upper bound. For any
$n\geq1$, we have:
\[
H\left(Y_{n+{\cal L}}\middle|Y^{n+{\cal L}-1}\right)\leq H\left(Y_{n+{\cal L}}\middle|Y_{n+1}^{n+{\cal L}-1}\right)
\]
Note that the right-hand side is independent of $n$ by stationarity.
As such, 
\begin{equation}
\lim_{n\to\infty}H\left(Y_{n}\middle|Y^{n-1}\right)\leq H\left(Y_{n+{\cal L}}\middle|Y_{n+1}^{n+{\cal L}-1}\right)\label{eq:Eq_Upper Bound, Bound on Firsrt Term}
\end{equation}
Next, notice that in the resolution of uncertainty in $Y_{i}$, given
$X_{i-1}^{i}$, other random variables can help at best by providing
information about $Z_{i}$. Together with observation (Ob.4) made
earlier, we have the Markov chain 
\[
\left(Y^{n},X_{0}^{n-1}\right)\to\left(Y_{n+1}^{n+{\cal L}-1},X_{n}^{n+{\cal L}},Z_{n}\right)\to Y_{n+{\cal L}}
\]
Therefore:
\begin{align*}
H\left(Y_{n+{\cal L}}\middle|Y^{n+{\cal L}-1},X_{0}^{n+{\cal L}}\right) & \geq H\left(Y_{n+{\cal L}}\middle|Y^{n+{\cal L}-1},X_{0}^{n+{\cal L}},Z_{n}\right)\\
 & =H\left(Y_{n+{\cal L}}\middle|Y_{n+1}^{n+{\cal L}-1},X_{n}^{n+{\cal L}},Z_{n}\right)
\end{align*}
whose right-hand side is independent of $n$ for the same reason that
leads to Eq. \eqref{eq:Eq_Supplemental_Lower_Bound_First_Term_Equality_3}.
Then: 
\begin{equation}
\lim_{n\to\infty}H\left(Y_{n}\middle|Y^{n-1},X_{0}^{n}\right)\geq H\left(Y_{n+{\cal L}}\middle|Y_{n+1}^{n+{\cal L}-1},X_{n}^{n+{\cal L}},Z_{n}\right)\label{eq:Eq_Upper Bound, Bound on Second Term}
\end{equation}
By Eq. \eqref{eq:Eq_Upper Bound, Bound on Firsrt Term} and \eqref{eq:Eq_Upper Bound, Bound on Second Term},
letting 
\begin{align*}
 & C_{\mathcal{L}}^{ub}\left(P_{X_{n}^{n+{\cal L}}}\right)\\
 & =H\left(Y_{n+{\cal L}}\middle|Y_{n+1}^{n+{\cal L}-1}\right)-H\left(Y_{n+{\cal L}}\middle|Y_{n+1}^{n+{\cal L}-1},X_{n}^{n+{\cal L}},Z_{n}\right)
\end{align*}
we have 
\[
C\leq C_{\mathcal{L}}^{ub}\equiv\sup_{\substack{\text{Stationary, bit-symmetric}\\
P_{X_{n}^{n+{\cal L}}}
}
}C_{\mathcal{L}}^{ub}\left(P_{X_{n}^{n+{\cal L}}}\right)
\]
Then $C_{\mathcal{L}}^{ub}$ gives a capacity upper bound, controlled
by $\mathcal{L}$.

To compute this bound, we turn to the following lemma, which is a
straightforward exercise.
\begin{lem}
Given vectors $w_{i},q_{i}\in\mathbb{R}^{n}$ and variable $u\in\mathbb{R}^{n}$,
define a function $f(u)=-\sum_{i}w_{i}^{\top}u\log\left(w_{i}^{\top}u\left/q_{i}^{\top}u\right)\right.$
where $\mbox{dom}(f)=\{u:\; w_{i}^{\top}u\geq0\;\forall i\}$. Then
$f$ is a concave function.
\end{lem}
It is easy to see that $H\left(Y_{n+{\cal L}}\middle|Y_{n+1}^{n+{\cal L}-1},X_{n}^{n+{\cal L}},Z_{n}\right)$
is affine in $P_{X_{n}^{n+{\cal L}}}$, and $H\left(Y_{n+{\cal L}}\middle|Y_{n+1}^{n+{\cal L}-1}\right)$
takes the form of function $f$ in the above lemma with $P_{X_{n}^{n+{\cal L}}}$
being the variable $u$. Then $C_{\mathcal{L}}^{ub}\left(P_{X_{n}^{n+{\cal L}}}\right)$
is a concave function. We note the following:
\begin{itemize}
\item $P_{X_{n}^{n+{\cal L}}}$ being a valid probability measure means
$\sum_{\mathbf{x}}P_{X_{n}^{n+{\cal L}}}\left(\mathbf{x}\right)=1$
(unity-sum condition) and $P_{X_{n}^{n+{\cal L}}}\left(\mathbf{x}\right)\geq0$
(non-negativity condition) $\forall\mathbf{x}\in\mbox{GF}(2)^{\mathcal{L}+1}$.
\item The stationarity condition is linear and hence can be converted into
the form $Sp_{X}=0$, where $p_{X}=\left[\begin{array}{ccc}
... & P_{X_{n}^{n+{\cal L}}}\left(\mathbf{x}\right) & ...\end{array}\right]^{\top}$ and $S$ is a matrix. For example, when $\mathcal{L}=1$, we have:
\[
p_{X}=\left[\begin{array}{c}
P_{X_{n}^{n+1}}\left(0,0\right)\\
P_{X_{n}^{n+1}}\left(0,1\right)\\
P_{X_{n}^{n+1}}\left(1,0\right)\\
P_{X_{n}^{n+1}}\left(1,1\right)
\end{array}\right]\quad S=\left[\begin{array}{clcc}
0 & 1 & -1 & 0\end{array}\right]
\]
which represents $P_{X_{n}}\left(0\right)=P_{X_{n+1}}\left(0\right)$
and $P_{X_{n}}\left(1\right)=P_{X_{n+1}}\left(1\right)$. In general,
$S$ can be constructed efficiently by computations. The following
lemma indicates that the number of rows in $S$ is at most $2^{\mathcal{L}}-1$,
i.e. it grows linearly with the size of $p_{X}$ and so using the
stationarity condition in computations is not too costly.\end{itemize}
\begin{lem}
\label{lem:Lemma_Stationarity Condition}For any finite $n$ and random
vector $V^{n}\in\text{\normalfont GF}(2)^{n}$ with probability distribution
$P$, it is stationary if and only if, for any $\mathbf{v}\in\text{\normalfont GF}(2)^{n-1}$,
\begin{equation}
P\left(V_{1}^{n-1}=\mathbf{v}\right)=P\left(V_{2}^{n}=\mathbf{v}\right)\label{eq:Eq_Lemma_Stationarity Condition}
\end{equation}
In fact, if Eq. \eqref{eq:Eq_Lemma_Stationarity Condition} is satisfied
for any $2^{n-1}-1$ vectors $\mathbf{v}$ from $\text{\normalfont GF}(2)^{n-1}$,
it is also satisfied for the other $\mathbf{v}\in\text{\normalfont GF}(2)^{n-1}$.\end{lem}
\begin{IEEEproof}
Let us consider the first claim. The forward is immediate. We prove
the converse. Since $V^{n}$ is defined over $\mbox{GF}(2)$, any
event on it implies that a subset of entries in $\{V_{i}:\; i=1,...,n\}$
takes up a specific value. Consider an arbitrary event $E=\left\{ V_{i}=a_{i},\; i\in N_{E}\right\} $
for any $N_{E}\subseteq\left\{ 1,...,n-1\right\} $. Then if for any
such event, $P(E)=P(T^{-s}E)$ for $s=1,...,n-\max N_{E}$, $P$ is
stationary. Assume that for any $\mathbf{v}\in\mbox{GF}(2)^{n-1}$,
$P\left(V_{1}^{n-1}=\mathbf{v}\right)=P\left(V_{2}^{n}=\mathbf{v}\right)$.
We have:
\begin{align*}
P(T^{-s}E) & =\sum_{\substack{v_{i+s-1}=a_{i},\; i\in N_{E}\\
v_{j+s-1}\in\text{GF}(2),\; j\notin N_{E}
}
}P\left(V_{2}^{n}=\mathbf{v}\right)\\
 & =\sum_{\substack{v_{i+s-1}=a_{i},\; i\in N_{E}\\
v_{j+s-1}\in\text{GF}(2),\; j\notin N_{E}
}
}P\left(V_{1}^{n-1}=\mathbf{v}\right)\\
 & =\sum_{\substack{v_{i+s-2}=a_{i},\; i\in N_{E}\\
v_{j+s-2}\in\text{GF}(2),\; j\notin N_{E}
}
}P\left(V_{2}^{n}=\mathbf{v}\right)\\
 & =\sum_{\substack{v_{i+s-2}=a_{i},\; i\in N_{E}\\
v_{j+s-2}\in\text{GF}(2),\; j\notin N_{E}
}
}P\left(V_{1}^{n-1}=\mathbf{v}\right)\\
 & \ldots\\
 & =\sum_{\substack{v_{i}=a_{i},\; i\in N_{E}\\
v_{j}\in\text{GF}(2),\; j\notin N_{E}
}
}P\left(V_{2}^{n}=\mathbf{v}\right)\\
 & =\sum_{\substack{v_{i}=a_{i},\; i\in N_{E}\\
v_{j}\in\text{GF}(2),\; j\notin N_{E}
}
}P\left(V_{1}^{n-1}=\mathbf{v}\right)\\
 & =P(E)
\end{align*}
 The first claim is hence proven.

To see the second claim, notice Eq. \eqref{eq:Eq_Lemma_Stationarity Condition}
can be written as
\[
P_{V^{n}}\left(\left[\mathbf{v},0\right]\right)+P_{V^{n}}\left(\left[\mathbf{v},1\right]\right)=P_{V^{n}}\left(\left[0,\mathbf{v}\right]\right)+P_{V^{n}}\left(\left[1,\mathbf{v}\right]\right)
\]
Summing both sides over all $\mathbf{v}\in\mbox{GF}(2)^{n-1}$, we
then obtain a trivial equation, which implies one redundant equation
in the system.
\end{IEEEproof}

\begin{figure}
\begin{centering}
\includegraphics[width=1\columnwidth]{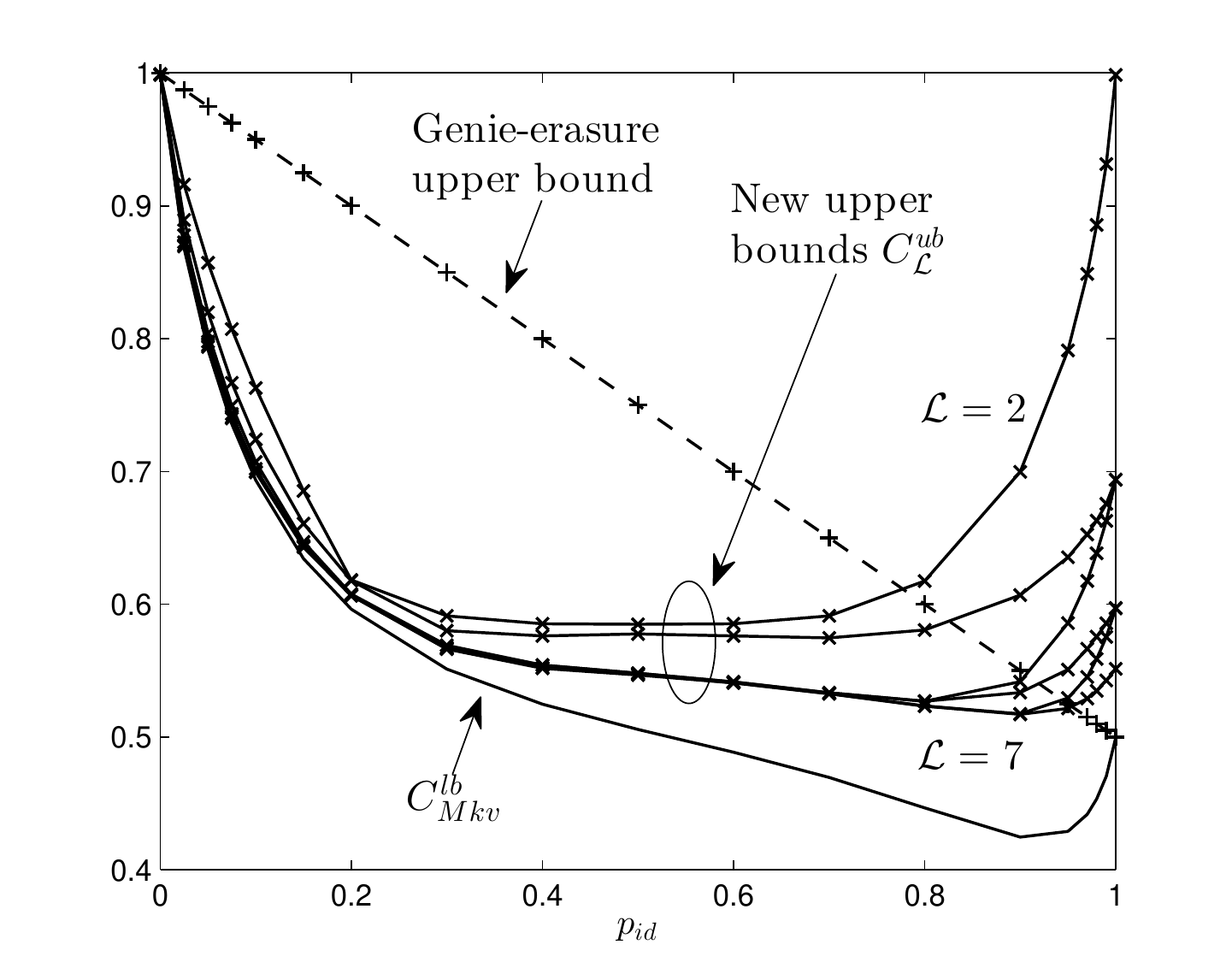}
\par\end{centering}

\protect\caption{New upper bounds $C_{\mathcal{L}}^{ub}$ for $\mathcal{L}=2,3,...,7$.
The higher curve corresponds to the lower $\mathcal{L}$.}

\label{Fig_Capacity_Upper_Bound}
\end{figure}

\begin{itemize}
\item The condition of bit-symmetry is also equivalent to a set of linear
equations $P_{X_{n}^{n+{\cal L}}}\left(\mathbf{x}\right)=P_{X_{n}^{n+{\cal L}}}\left(\neg\mathbf{x}\right)$
$\forall\mathbf{x}\in\mbox{GF}(2)^{\mathcal{L}+1}$. In fact, this
condition is not even needed: one can easily establish a result for
$C_{\mathcal{L}}^{ub}$ similar to Proposition \ref{prop:Proposition_Bit-symmetry (ergodic)}
using the same technique, i.e.
\[
\sup_{\substack{\text{Stationary}\\
P_{X_{n}^{n+{\cal L}}}
}
}C_{\mathcal{L}}^{ub}\left(P_{X_{n}^{n+{\cal L}}}\right)=\sup_{\substack{\text{Bit-symmetric,}\\
\text{stationary }P_{X_{n}^{n+{\cal L}}}
}
}C_{\mathcal{L}}^{ub}\left(P_{X_{n}^{n+{\cal L}}}\right)
\]

\end{itemize}
With those points made above, we conclude that finding $C_{\mathcal{L}}^{ub}$
is a convex optimization problem, which can be efficiently solved
using various computational methods \cite{Boyd-2004}. The results
are shown in Fig. \ref{Fig_Capacity_Upper_Bound} for $\mathcal{L}=2,3,...,7$.
It can be observed that the series of upper bounds improves over the
genie-erasure upper bound, given in \cite{Iyengar-2011}, for most
values of $p_{id}$ and approaches closely to the new lower bound
$C_{Mkv}^{lb}$ at low $p_{id}$.

The complexity of the program to find $C_{\mathcal{L}}^{ub}$ scales
rapidly with ${\cal L}$: it increases polynomially with the number
of variables $2^{{\cal L}+1}$ and the number of equality constraints
$2^{{\cal L}}$ (including the stationarity condition and the unity-sum
condition, excluding the bit-symmetry condition). Nevertheless Fig.
\ref{Fig_Capacity_Upper_Bound} suggests that the upper bounds converge
quickly with ${\cal L}$, so small ${\cal L}$ is sufficient to produce
decent results.

\begin{figure}
\begin{centering}
\includegraphics[width=1\columnwidth]{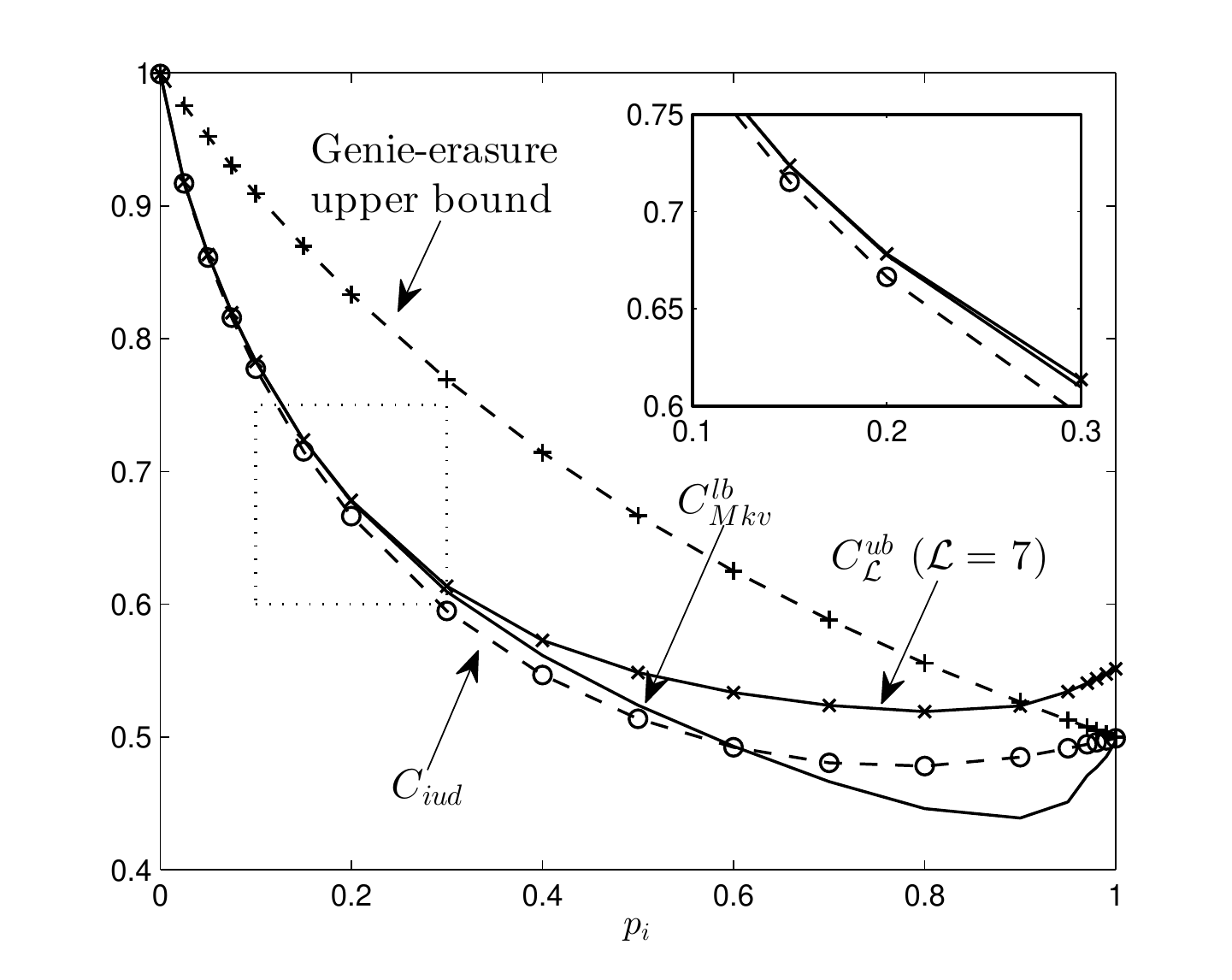}
\par\end{centering}

\protect\caption{New bounds for the case $p_{d}=1$ in comparison with bounds derived
in \cite{Mazumdar-2011} (dashed curves).}

\label{Fig_Capacity_Bound_Pd}
\end{figure}

In Fig. \ref{Fig_Capacity_Bound_Pd}, we also compare our new bounds
to those in \cite{Mazumdar-2011}, which were derived for the case
$p_{d}=1$. Improvements are observed.

As a note, Lemma \ref{lem:Lemma_Second_Term_Formula_for_Stationary_Input}
should not be used in place of Eq. \eqref{eq:Eq_Upper Bound, Bound on Second Term};
otherwise the maximization problem would be a non-convex one.

\subsection{Tightness of the New Upper Bounds\label{sub:Tightness-of-Upper Bound}}

It is easy to see that the bound is improved monotonically with increasing
$\mathcal{L}$. Indeed,
\begin{align*}
H\left(Y_{n+{\cal L}+1}\middle|Y_{n+1}^{n+{\cal L}}\right) & =H\left(Y_{n+{\cal L}}\middle|Y_{n}^{n+{\cal L}-1}\right)\\
 & \leq H\left(Y_{n+{\cal L}}\middle|Y_{n+1}^{n+{\cal L}-1}\right)
\end{align*}
where the first equation is because of the channel output's stationarity,
given that the channel and the input are stationary. We also have:
\begin{align*}
 & H\left(Y_{n+{\cal L}+1}\middle|Y_{n+1}^{n+{\cal L}},X_{n}^{n+{\cal L}+1},Z_{n}\right)\\
 & =H\left(Y_{n+{\cal L}}\middle|Y_{n}^{n+{\cal L}-1},X_{n-1}^{n+{\cal L}},Z_{n-1}\right)\\
 & \geq H\left(Y_{n+{\cal L}}\middle|Y_{n}^{n+{\cal L}-1},X_{n-1}^{n+{\cal L}},Z_{n-1},Z_{n}\right)\\
 & =H\left(Y_{n+{\cal L}}\middle|Y_{n+1}^{n+{\cal L}-1},X_{n}^{n+{\cal L}},Z_{n}\right)
\end{align*}
where the first and last equalities are by arguing in a similar manner
to the derivation of Eq. \eqref{eq:Eq_Supplemental_Lower_Bound_First_Term_Equality_3}
and \eqref{eq:Eq_Upper Bound, Bound on Second Term} respectively.
These show that
\[
C_{\mathcal{L}+1}^{ub}\left(P_{X_{n}^{n+{\cal L}+1}}\right)\leq C_{\mathcal{L}}^{ub}\left(P_{X_{n}^{n+{\cal L}}}\right)
\]
where $P_{X_{n}^{n+{\cal L}}}$ is a marginal distribution of $P_{X_{n}^{n+{\cal L}+1}}$.
Taking the supremum, we obtain $C\leq C_{\mathcal{L}+1}^{ub}\leq C_{\mathcal{L}}^{ub}$.

We next discuss the gap between $C$ and $C_{\infty}^{ub}$. Here
$C_{\infty}^{ub}$ exists since $C_{{\cal L}}^{ub}$ decreases with
increasing ${\cal L}$ and is bounded from below. We first connect
$C$ and $C_{\mathcal{L}}^{ub}\left(P_{X_{n}^{n+{\cal L}}}\right)$
by establishing the following:
\begin{align*}
 & C_{\mathcal{L}}^{ub}\left(P_{X_{n}^{n+{\cal L}}}\right)\xrightarrow{\mathcal{L}\to\infty}\\
 & \left[\lim_{n\to\infty}H\left(Y_{n}\middle|Y^{n-1}\right)-\lim_{n\to\infty}H\left(Y_{n}\middle|Y^{n-1},X_{0}^{n}\right)\right]_{X_{0}^{\infty}\sim P_{X_{0}^{\infty}}}
\end{align*}
where the right-hand side is evaluated w.r.t. a stationary input distribution
$P_{X_{0}^{\infty}}$ such that for every ${\cal L}$, $P_{X_{n}^{n+{\cal L}}}$
is a marginal distribution of $P_{X_{0}^{\infty}}$. Let $g\left(P_{X_{0}^{\infty}}\right)$
denote the right-hand side. Since $P_{X_{0}^{\infty}}$ is stationary,
all $\left({\cal L}+1\right)$-dimensional marginals of $P_{X_{0}^{\infty}}$
are the same. We can then regard $C_{\mathcal{L}}^{ub}\left(\cdot\right)$
as a function of $P_{X_{0}^{\infty}}$. With an abuse of notation,
we write $C_{\mathcal{L}}^{ub}\left(P_{X_{0}^{\infty}}\right)$ to
mean $C_{\mathcal{L}}^{ub}\left(P_{X_{n}^{n+{\cal L}}}\right)$. It
is easy to see that
\begin{align*}
 & C_{\mathcal{L}}^{ub}\left(P_{X_{0}^{\infty}}\right)\\
 & =H\left(Y_{n+{\cal L}}\middle|Y_{n+1}^{n+{\cal L}-1}\right)+H\left(Z_{n}\middle|Y_{n+1}^{n+{\cal L}-1},X_{n}^{n+{\cal L}}\right)\\
 & \quad-H\left(Y_{n+{\cal L}}\middle|Y_{n+1}^{n+{\cal L}-1},X_{n}^{n+{\cal L}}\right)-H\left(Z_{n}\middle|Y_{n+1}^{n+{\cal L}},X_{n}^{n+{\cal L}}\right)
\end{align*}
Notice that 
\begin{align*}
H\left(Z_{n}\middle|Y_{n+1}^{n+{\cal L}-1},X_{n}^{n+{\cal L}}\right) & \leq H\left(Z_{n}\middle|Y_{n+1}^{n+{\cal L}-1},X_{n}^{n+{\cal L}-1}\right)\\
H\left(Z_{n}\middle|Y_{n+1}^{n+{\cal L}-1},X_{n}^{n+{\cal L}}\right) & \geq H\left(Z_{n}\middle|Y_{n+1}^{n+{\cal L}},X_{n}^{n+{\cal L}}\right)
\end{align*}
whereas
\begin{align*}
 & \lim_{\mathcal{L\to\infty}}H\left(Z_{n}\middle|Y_{n+1}^{n+{\cal L}-1},X_{n}^{n+{\cal L}-1}\right)\\
 & \quad=\lim_{\mathcal{L\to\infty}}H\left(Z_{n}\middle|Y_{n+1}^{n+{\cal L}},X_{n}^{n+{\cal L}}\right)
\end{align*}
Furthermore, thanks to stationarity of the channel output and the
joint input-output distribution,
\begin{align*}
\lim_{\mathcal{L\to\infty}}H\left(Y_{n+{\cal L}}\middle|Y_{n+1}^{n+{\cal L}-1}\right) & ={\displaystyle \lim_{n\to\infty}H\left(Y_{n}\left|Y^{n-1}\right)\right.}\\
\lim_{\mathcal{L\to\infty}}H\left(Y_{n+{\cal L}}\middle|Y_{n+1}^{n+{\cal L}-1},X_{n}^{n+{\cal L}}\right) & =\lim_{n\to\infty}H\left(Y_{n}\middle|Y^{n-1},X_{0}^{n}\right)
\end{align*}
Therefore, $C_{\mathcal{L}}^{ub}\left(P_{X_{0}^{\infty}}\right)\to g\left(P_{X_{0}^{\infty}}\right)$
as ${\cal L}\to\infty$, for any stationary distribution $P_{X_{0}^{\infty}}$.
Now notice that
\[
C=\sup_{\text{Stationary }P_{X_{0}^{\infty}}}g\left(P_{X_{0}^{\infty}}\right)=\sup_{\text{Stationary }P_{X_{0}^{\infty}}}\lim_{{\cal L}\to\infty}C_{\mathcal{L}}^{ub}\left(P_{X_{0}^{\infty}}\right)
\]
whereas
\[
C_{\infty}^{ub}=\lim_{{\cal L}\to\infty}\sup_{\text{Stationary }P_{X_{0}^{\infty}}}C_{\mathcal{L}}^{ub}\left(P_{X_{0}^{\infty}}\right)
\]
The gap hence lies in the order of the limit and the supremum. This
is a consequence of computational feasibility: restriction to finite-dimensional
distributions $P_{X_{n}^{n+{\cal L}}}$ is required for computations,
whereas $C$ is inherently a quantity with infinite block lengths.
We conjecture that no other upper-bounding methods that directly modify
the entropy terms in Eq. \eqref{eq:Equation_C_2} (or the mutual information
term in Eq. \eqref{eq:Equation_C_0}) are better than the presented
one. If the discussed gap is non-trivial and one seeks a better upper
bound, a different formulation of the capacity might be called for.

In fact, in the Shannon-theoretic framework, it shall be proven that
the gap is trivial, i.e. $C_{\infty}^{ub}=C$! We have:
\begin{align*}
 & {\cal L}C_{{\cal L}}^{ub}\left(P_{X_{0}^{\infty}}\right)\stackrel{\left(a\right)}{\leq}\sum_{j=1}^{{\cal L}}C_{j}^{ub}\left(P_{X_{0}^{\infty}}\right)\\
 & =\sum_{j=1}^{{\cal L}}H\left(Y_{n+j}\middle|Y_{n+1}^{n+j-1}\right)-H\left(Y_{n+j}\middle|Y_{n+1}^{n+j-1},X_{n}^{n+j},Z_{n}\right)\\
 & \leq\sum_{j=1}^{{\cal L}}H\left(Y_{n+j}\middle|Y_{n+1}^{n+j-1}\right)-H\left(Y_{n+j}\middle|Y_{n+1}^{n+j-1},X_{n}^{n+{\cal L}},Z_{n}\right)\\
 & =H\left(Y_{n+1}^{n+{\cal L}}\right)-H\left(Y_{n+1}^{n+{\cal L}}\middle|X_{n}^{n+{\cal L}},Z_{n}\right)\\
 & =I\left(X_{n}^{n+{\cal L}};Y_{n+1}^{n+{\cal L}}\right)+I\left(Z_{n};Y_{n+1}^{n+{\cal L}}\middle|X_{n}^{n+{\cal L}}\right)
\end{align*}
and therefore
\begin{align*}
C_{\infty}^{ub} & \leq\lim_{{\cal L}\to\infty}\frac{1}{{\cal L}}\sup_{\text{Stationary }P_{X_{0}^{\infty}}}I\left(X_{n}^{n+{\cal L}};Y_{n+1}^{n+{\cal L}}\right)\\
 & \quad+\lim_{{\cal L}\to\infty}\frac{1}{{\cal L}}\sup_{\text{Stationary }P_{X_{0}^{\infty}}}I\left(Z_{n};Y_{n+1}^{n+{\cal L}}\middle|X_{n}^{n+{\cal L}}\right)\\
 & \stackrel{\left(b\right)}{=}\lim_{{\cal L}\to\infty}\frac{1}{{\cal L}}\sup_{\text{Stationary }P_{X_{0}^{\infty}}}I\left(X_{n}^{n+{\cal L}};Y_{n+1}^{n+{\cal L}}\right)\\
 & \stackrel{\left(c\right)}{=}\lim_{n\to\infty}\frac{1}{n}\sup_{\text{Stationary }P_{X_{0}^{\infty}}}I\left(X_{0}^{n};Y_{1}^{n}\right)\\
 & \leq\lim_{n\to\infty}\frac{1}{n}\sup_{X_{0}^{n}}I\left(X_{0}^{n};Y_{1}^{n}\right)\\
 & \stackrel{\left(d\right)}{=}C
\end{align*}
Here $\left(a\right)$ is because $C_{j+1}^{ub}\left(P_{X_{0}^{\infty}}\right)\leq C_{j}^{ub}\left(P_{X_{0}^{\infty}}\right)$
as proven above, $\left(b\right)$ is because
\[
0\leq\frac{1}{{\cal L}}I\left(Z_{n};Y_{n+1}^{n+{\cal L}}\middle|X_{n}^{n+{\cal L}}\right)\leq\frac{1}{{\cal L}}\log\left|{\cal Z}\right|=\frac{1}{{\cal L}}\to0
\]
$\left(c\right)$ is because of stationarity, and $\left(d\right)$
is Eq. \eqref{eq:Equation_C_indecomposableFSC}, stemming from the
fact the DID channel is an indecomposable FSC in the Shannon-theoretic
framework. Finally, since $C\leq C_{\infty}^{ub}$, we have $C_{\infty}^{ub}=C$.

Aside from increasing ${\cal L}$, one may expect to obtain a better
upper bound by adding more constraints on the input search space.
Note, however, not all constraints would help. One example is bit-symmetry
as pointed out earlier. An opposite example is the stationarity condition,
which is discussed next.

\subsection{Importance of the Stationary Condition\label{sub:Importance-of-Stationary (Upper bounds)}}

We shed light on why the stationarity condition is crucial to developing
meaningful upper bounds. In particular, without this condition, they
could be trivialized, i.e. equal to $1$. To this end, let us consider
the following quantity:
\[
U_{\mathcal{L}}\equiv\sup_{P_{X_{n}^{n+{\cal L}}}}C_{\mathcal{L}}^{ub}\left(P_{X_{n}^{n+{\cal L}}}\right)
\]
which is essentially $C_{\mathcal{L}}^{ub}$ without input stationarity.
Notice that when the input is not stationary, $C_{\mathcal{L}}^{ub}\left(P_{X_{n}^{n+{\cal L}}}\right)$
is implicitly a function of $n$, denoted by $\gamma\left(n\right)$
to emphasize this dependency. In a similar manner to the derivation
of Eq. \eqref{eq:Eq_Supplemental_Lower_Bound_First_Term_Equality_3},
it is easy to see that if $P_{X_{n}^{n+{\cal L}}}=P^{*}$ yields the
supremum of $\gamma\left(n\right)$ then $P_{X_{n+1}^{n+{\cal L}+1}}=P^{*}$
yields the supremum of $\gamma\left(n+1\right)$. As such, the supremum
in $U_{\mathcal{L}}$ is independent of $n$, i.e. $U_{\mathcal{L}}$
is not a function of $n$.

First, we need to justify that $U_{\mathcal{L}}$ is a valid upper
bound on $C$. This is indeed the case since $U_{\mathcal{L}}\geq C_{\mathcal{L}}^{ub}\geq C$.
Alternatively this can be shown without resorting to $C_{\mathcal{L}}^{ub}$
as follows. We modify Eq. \eqref{eq:Eq_Upper Bound, Bound on Firsrt Term}
and \eqref{eq:Eq_Upper Bound, Bound on Second Term}: 
\begin{align*}
 & \lim_{n\to\infty}H\left(Y_{n}\middle|Y^{n-1}\right)\leq\liminf_{n\to\infty}H\left(Y_{n+{\cal L}}\middle|Y_{n+1}^{n+{\cal L}-1}\right)\\
 & \lim_{n\to\infty}H\left(Y_{n}\middle|Y^{n-1},X_{0}^{n}\right)\\
 & \quad\geq\limsup_{n\to\infty}H\left(Y_{n+{\cal L}}\middle|Y_{n+1}^{n+{\cal L}-1},X_{n}^{n+{\cal L}},Z_{n}\right)
\end{align*}
which are valid without the restriction to stationary inputs. Then:
\begin{align*}
C & \leq\sup\left\{ \begin{array}{c}
{\displaystyle \liminf_{n\to\infty}H\left(Y_{n+{\cal L}}\middle|Y_{n+1}^{n+{\cal L}-1}\right)-}\\
{\displaystyle \limsup_{n\to\infty}H\left(Y_{n+{\cal L}}\middle|Y_{n+1}^{n+{\cal L}-1},X_{n}^{n+{\cal L}},Z_{n}\right)}
\end{array}\right\} \\
 & \leq\sup\liminf_{n\to\infty}C_{\mathcal{L}}^{ub}\left(P_{X_{n}^{n+{\cal L}}}\right)\\
 & \leq\liminf_{n\to\infty}\sup C_{\mathcal{L}}^{ub}\left(P_{X_{n}^{n+{\cal L}}}\right)\stackrel{\left(a\right)}{=}U_{\mathcal{L}}
\end{align*}
where the suprema are over all valid inputs, and $\left(a\right)$
is because $U_{{\cal L}}$ is independent of $n$.

We now show that $U_{{\cal L}}=1$ in the case $p_{i}=p_{d}$ for
any $\mathcal{L}\geq2$. Consider a specific input distribution $P_{X_{n}^{n+{\cal L}}}^{*}$
such that
\[
P\left(X_{n}^{n+{\cal L}}=...1010100\right)=P\left(X_{n}^{n+{\cal L}}=...0101011\right)=0.5
\]
i.e. only two input strings with identical first two bits and alternating
bits thereafter are assigned probability $0.5$. It is easy to see
that $P_{X_{n}^{n+{\cal L}}}^{*}$ is not stationary. With probability
$1$, we have $X_{n+{\cal L}}=X_{n+{\cal L}-1}$, which completely
resolves uncertainty in $Y_{n+{\cal L}}$, and so $H\left(Y_{n+{\cal L}}\middle|Y_{n+1}^{n+{\cal L}-1},X_{n}^{n+{\cal L}},Z_{n}\right)=0$.
For any $\mathbf{y}\in\mbox{GF}(2)^{\mathcal{L}-1}$, 
\begin{align*}
 & P\left(Y_{n+{\cal L}}=0\middle|Y_{n+1}^{n+{\cal L}-1}=\mathbf{y}\right)P\left(Y_{n+1}^{n+{\cal L}-1}=\mathbf{y}\right)\\
 & \quad=0.5P\left(Y_{n+1}^{n+{\cal L}-1}=\mathbf{y}\middle|X_{n}^{n+{\cal L}-1}=...1010\right)\\
 & P\left(Y_{n+{\cal L}}=1\middle|Y_{n+1}^{n+{\cal L}-1}=\mathbf{y}\right)P\left(Y_{n+1}^{n+{\cal L}-1}=\mathbf{y}\right)\\
 & \quad=0.5P\left(Y_{n+1}^{n+{\cal L}-1}=\mathbf{y}\middle|X_{n}^{n+{\cal L}-1}=...0101\right)
\end{align*}
As a property of the DID channel, for any output sequence $\mathbf{y}\in\text{GF}\left(2\right)^{{\cal L}-1}$,
there exists only one $\mathbf{z}\in\text{GF}\left(2\right)^{{\cal L}-1}$
such that the event $\left\{ Y_{n+1}^{n+{\cal L}-1}=\mathbf{y}\middle|X_{n}^{n+{\cal L}-1}=...1010\right\} $
is equivalent to $\left\{ Z_{n+1}^{n+{\cal L}-1}=\mathbf{z}\right\} $.
We also have that the event $\left\{ Y_{n+1}^{n+{\cal L}-1}=\mathbf{y}\middle|X_{n}^{n+{\cal L}-1}=...0101\right\} $
is equivalent to $\left\{ Z_{n+1}^{n+{\cal L}-1}=\neg\mathbf{z}\right\} $.
But we have 
\[
P\left(Z_{n+1}^{n+{\cal L}-1}=\mathbf{z}\right)=P\left(Z_{n+1}^{n+{\cal L}-1}=\neg\mathbf{z}\right)
\]
since $p_{i}=p_{d}$. As a result, 
\[
P\left(Y_{n+{\cal L}}=0\middle|Y_{n+1}^{n+{\cal L}-1}=\mathbf{y}\right)=P\left(Y_{n+{\cal L}}=1\middle|Y_{n+1}^{n+{\cal L}-1}=\mathbf{y}\right)
\]
which implies $Y_{n+{\cal L}}$ is independent of $Y_{n+1}^{n+{\cal L}-1}$
and $P\left(Y_{n+{\cal L}}=0\right)=P\left(Y_{n+{\cal L}}=1\right)$.
Then $H\left(Y_{n+{\cal L}}\middle|Y_{n+1}^{n+{\cal L}-1}\right)=1$,
and consequently $C_{\mathcal{L}}^{ub}\left(P_{X_{n}^{n+{\cal L}}}^{*}\right)=1$.
We thus have $C_{\mathcal{L}}^{ub}\left(P_{X_{n}^{n+{\cal L}}}^{*}\right)\leq U_{{\cal L}}\leq1$,
in which the latter inequality is the trivial upper bound $1$. Therefore,
$U_{{\cal L}}=1$ for any $\mathcal{L}\geq2$.

When ${\cal L}=1$, it is easy to construct the same $P_{X_{n}^{n+{\cal L}}}^{*}$
and show that $C_{\mathcal{L}}^{ub}\left(P_{X_{n}^{n+{\cal L}}}^{*}\right)=1$.
Notice that $P_{X_{n}^{n+{\cal L}}}^{*}$ satisfies the stationarity
condition in this case. Then both $C_{\mathcal{L}}^{ub}$ and $U_{{\cal L}}$
are equal to $1$. We see why ${\cal L}\geq2$ has been used to obtain
good upper bounds so far.

\begin{figure*}
\begin{centering}
\includegraphics[width=2\columnwidth]{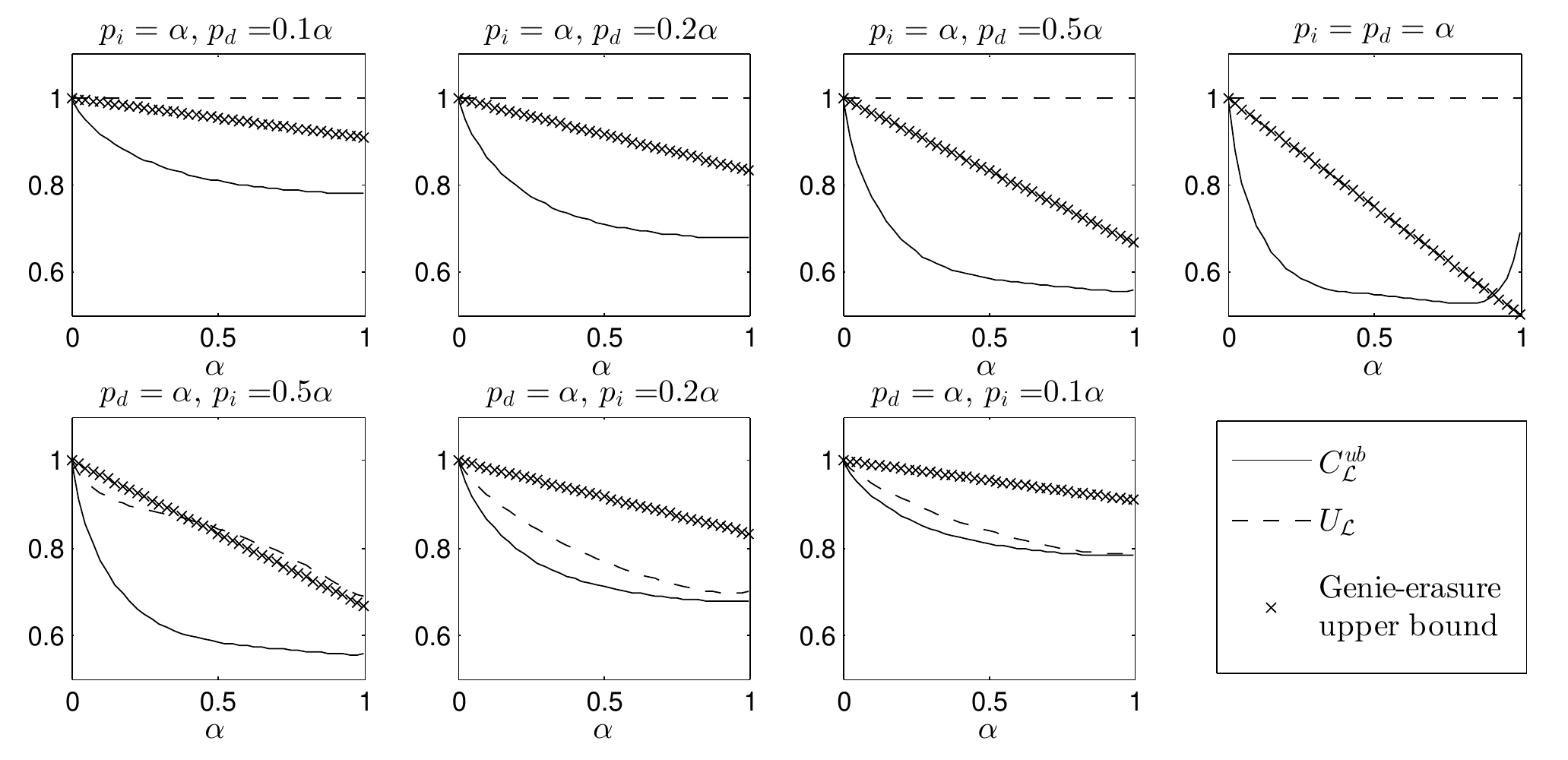}
\par\end{centering}

\protect\caption{Comparison of the upper bounds for different $p_{i}$ and $p_{d}$.
Here $\mathcal{L}=4$.}

\label{Fig_Capacity_3_Upper_Bounds}
\end{figure*}

In the general case where $p_{i}\neq p_{d}$, we wish to find some
similar distribution $P_{X_{n}^{n+{\cal L}}}^{*}$. We may choose
one such that $P\left(X_{n+{\cal L}}\neq X_{n+{\cal L}-1}\right)=0$,
i.e. input strings whose $X_{n+{\cal L}-1}^{n+{\cal L}}$ is either
$10$ or $01$ are not allowed, and hence the term $H\left(Y_{n+{\cal L}}\middle|Y_{n+1}^{n+{\cal L}-1},X_{n}^{n+{\cal L}},Z_{n}\right)$
is eliminated. In order that $H\left(Y_{n+{\cal L}}\middle|Y_{n+1}^{n+{\cal L}-1}\right)=1$,
we seek to have $P\left(Y_{n+{\cal L}}=0\middle|Y_{n+1}^{n+{\cal L}-1}=\mathbf{y}\right)=P\left(Y_{n+{\cal L}}=1\middle|Y_{n+1}^{n+{\cal L}-1}=\mathbf{y}\right)$
for any $\mathbf{y}\in\mbox{GF}(2)^{\mathcal{L}-1}$. This is equivalent
to finding $P_{X_{n}^{n+{\cal L}}}^{*}$ that satisfies 
\begin{align*}
 & \sum_{\mathbf{x}}\left\{ \begin{array}{c}
P\left(X_{n+{\cal L}-1}^{n+{\cal L}}=00,X_{n}^{n+{\cal L}-2}=\mathbf{x}\right)\times\\
P\left(Y_{n+1}^{n+{\cal L}-1}=\mathbf{y}\middle|X_{n+{\cal L}-1}=0,X_{n}^{n+{\cal L}-2}=\mathbf{x}\right)
\end{array}\right\} \\
 & =\sum_{\mathbf{x}}\left\{ \begin{array}{c}
P\left(X_{n+{\cal L}-1}^{n+{\cal L}}=11,X_{n}^{n+{\cal L}-2}=\mathbf{x}\right)\times\\
P\left(Y_{n+1}^{n+{\cal L}-1}=\mathbf{y}\middle|X_{n+{\cal L}-1}=1,X_{n}^{n+{\cal L}-2}=\mathbf{x}\right)
\end{array}\right\} 
\end{align*}
$\forall\mathbf{y}\in\mbox{GF}(2)^{\mathcal{L}-1}$, which accounts
for $2^{\mathcal{L}-1}$ equations. We further have $2^{\mathcal{L}}$
equations for the condition $P\left(X_{n+{\cal L}}\neq X_{n+{\cal L}-1}\right)=0$,
where half of the input strings are assigned probability $0$, and
another equation for $\sum_{\mathbf{x}}P_{X_{n}^{n+{\cal L}}}^{*}\left(\mathbf{x}\right)=1$.
Then we are left with a system of $2^{\mathcal{L}+1}$ variables and
$2^{\mathcal{L}}+2^{\mathcal{L}-1}+1$ equations totally. It is easy
to see that $2^{\mathcal{L}+1}\geq2^{\mathcal{L}}+2^{\mathcal{L}-1}+1$
for any $\mathcal{L}\geq1$, which means we may able to find at least
one input distribution $P_{X_{n}^{n+{\cal L}}}^{*}$ that results
in $C_{\mathcal{L}}^{ub}\left(P_{X_{n}^{n+{\cal L}}}^{*}\right)=1$,
implying again that $U_{{\cal L}}=1$. While this is not guaranteed
due to the non-negativity condition ${\displaystyle P_{X_{n}^{n+{\cal L}}}^{*}\left(\mathbf{x}\right)\geq0}$
$\forall\mathbf{x}\in\mbox{GF}(2)^{\mathcal{L}+1}$, simulations show
that it could be indeed the case for $p_{i}\geq p_{d}$, and when
$p_{i}<p_{d}$, the upper bound is also worse (see Fig. \ref{Fig_Capacity_3_Upper_Bounds}).

\section{DID Channel Capacity at Low Noise\label{sec:DID-Capacity-at-Low-Noise}}

Observe that our bounds visually match up with one another and with
the lower bound on the i.u.d. achievable rate $C_{iud}^{lb}$ (from
Fig. \ref{Fig_Capacity_Markov_Lower_Bound}) nicely for $p_{id}<0.1$.
It is thus clear that for this noise range, the computed values are
the true capacity. In this section, we will be interested in a finite-lettered
characterization of the capacity at such low noise. Again, we will
restrict the analysis to the case $p_{i}=p_{d}=p_{id}$, in particular
$p_{id}<0.5$. In principle, other cases can be resolved in a similar
fashion.

Along this line of work, Kanoria and Montanari \cite{Kanoria-2010}
achieved the same goal for the deletion channel. Their approach and
ours share certain similarities: both find a lower bound and an upper
bound, and the lower bound is simply the achievable rate with a specific
input distribution, which is the i.u.d. input%
\footnote{The journal version \cite{Kanoria-2013} of their work extends further
the low-noise expansion by considering a broader class of input distributions
that encompasses the i.u.d. input.%
}. The key difference lies in the upper-bounding technique. In their
analysis, first, an intermediate input distribution class is shown
to yield an ``upper bound'' on the rate achieved with any input
distribution%
\footnote{In fact, in \cite{Kanoria-2010}, the rate of any input is upper-bounded
by the rate of this class plus a quantity. This quantity is almost
input-independent and small at low-noise levels.%
}; second, the rate achieved by this class is then expressed in terms
of the channel parameters. The choice of the intermediate input class
is not arbitrary. On one hand, properties of this class should offer
sufficient analytical advantages. On the other hand, the class should
be sufficiently broad so that it is possible to find an upper-bounding
relation with any input distributions. Indeed the choice in \cite{Kanoria-2010}
appears to be quite specific to the deletion channel. 

\begin{figure}
\begin{centering}
\includegraphics[width=0.6\columnwidth]{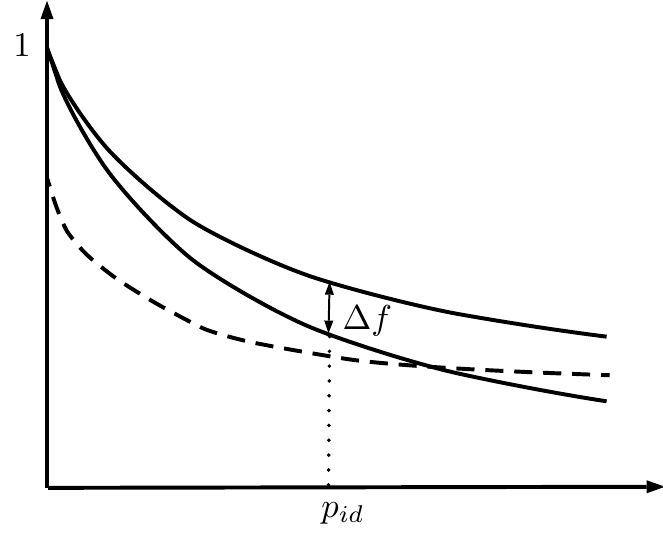}
\par\end{centering}

\protect\caption{Illustration of the low-noise calculation strategy. The upper and
lower solid curves are $f$ corresponding to a capacity-achieving
input and the i.u.d. input respectively. A different input deviation
from the i.u.d. input would yield a different curve, e.g. the dashed
curve, which may not come close to the lower solid curve as $p_{id}\to0$.}

\label{Fig_Low_Noise_Strategy_Illustration}
\end{figure}

In our case, it is not obvious how to pinpoint such an input class.
Our problem hence calls for a different approach. Suppose that $f\left(\mu\right)$
is a good upper bound on the rate achieved by an input $\mu$. At
each fixed $p_{id}$, we find the deviation $\Delta f=f\left(\mu^{*}\right)-f\left(\mu_{iud}\right)$,
which results from the deviation of $\mu^{*}$, one of the capacity-achieving
inputs, from $\mu_{iud}$ the i.u.d. input%
\footnote{There can be many inputs that achieve the capacity.%
}, via the Taylor expansion theorem. Then the order of magnitude of
$\Delta f$ is evaluated w.r.t. $p_{id}$. To make this feasible,
an important observation is that since we know the i.u.d. input is
the only capacity-achieving input when $p_{id}=0$, $\mu^{*}\to\mu_{iud}$
as $p_{id}\to0$. The use of $\Delta f$ hence circumvents the need
for a specific intermediate input class. Note that $f$ is, strictly
speaking, a function of $\mu$ and $p_{id}$. However an expansion
about both $\mu_{iud}$ and $p_{id}=0$ is expected to yield the trivial
rate of $1$, which is not useful. This explains why we need to treat
$p_{id}$ as a fixed parameter in the expansion of $f$. Fig. \ref{Fig_Low_Noise_Strategy_Illustration}
illustrates pictorially our strategy.

As observed from Fig. \ref{Fig_Low_Noise_Strategy_Illustration},
the two curves close their gap and hence $\Delta f\to0$ as $p_{id}\to0$,
which makes sense since the i.u.d. input achieves capacity at $p_{id}=0$.
This is the basis for us to evaluate the order of magnitude of $\Delta f$
in terms of $p_{id}$.

We first define the ${\cal O}$ and $\Theta$ notations. Since $p_{id}$
is small, we say
\begin{itemize}
\item a quantity $Q$ scales as ${\cal O}\left(g\left(p_{id}\right)\right)$,
or $Q={\cal O}\left(g\left(p_{id}\right)\right)$, if $\lim_{p_{id}\to0}\left(\left.Q\left/g\left(p_{id}\right)\right.\right.\right)$
is finite,
\item a quantity $Q$ scales as $\Theta\left(g\left(p_{id}\right)\right)$,
or $Q=\Theta\left(g\left(p_{id}\right)\right)$, if $\lim_{p_{id}\to0}\left(\left.Q\left/g\left(p_{id}\right)\right.\right.\right)$
is finite and not equal to $0$,
\item a quantity $Q$ is more significant than ${\cal O}\left(g\left(p_{id}\right)\right)$
if $\lim_{p_{id}\to0}\left(\left.Q\left/g\left(p_{id}\right)\right.\right.\right)$
is infinite.
\end{itemize}
should the limits exist.

\subsection{Low-Noise Lower Bound}

$C_{iud}^{lb}$ is sufficient for this purpose. The first term in
$C_{iud}^{lb}$, given in Eq. \eqref{eq:Lower_bound_term_1}, can
be expressed as follows:
\begin{align*}
 & H\left(Y_{n}\middle|Y_{n-1},X_{n-2},Z_{n-2}\right)\\
 & =\frac{1}{4}h_{2}\left(\frac{1}{2}-\frac{1}{2}p_{id}\right)\\
 & \quad+\left(\frac{1}{2}-\frac{1}{4}p_{id}\right)h_{2}\left(\frac{1}{2}-\frac{1}{2}p_{id}+{\cal O}\left(p_{id}^{2}\right)\right)\\
 & \quad+\left(\frac{1}{4}+\frac{1}{4}p_{id}\right)h_{2}\left(\frac{1}{2}-\frac{1}{2}p_{id}+{\cal O}\left(p_{id}^{2}\right)\right)\\
 & =1+{\cal O}\left(p_{id}^{2}\right)
\end{align*}
since $h_{2}\left(\frac{1}{2}+x\right)=1+\mathcal{O}\left(x^{2}\right)$.
The second term, given in Eq. \eqref{eq:Lower_bound_term_2}, can
be shown to be more significant than ${\cal O}\left(p_{id}^{2}\right)$.
Therefore, we leave the whole term as it is, and obtain the following:
\begin{equation}
C\geq C_{iud}^{lb}=1-\left[\sum_{k=1}^{\infty}\frac{1}{2^{k+1}}R\left(p_{id};k\right)\right]+{\cal O}\left(p_{id}^{2}\right)\label{eq:Capacity_Lower_Bound_Expansion}
\end{equation}
where
\[
R\left(p_{id};k\right)=h_{2}\left(\frac{1}{2}+\frac{1}{2}\left(1-2p_{id}\right)^{k}\right)
\]

\subsection{Low-Noise Upper Bound\label{sub:SubSection_Low-Noise-Upper-Bound}}

\begin{figure*}
\normalsize
\setcounter{MyTempEquationCounter}{\value{equation}}
\setcounter{equation}{21}
\begin{align}
f\left(\delta_{1},\delta_{2},p_{id}\right) & =f\left(0,0,p_{id}\right)+A_{1}\left(p_{id}\right)\delta_{1}+A_{2}\left(p_{id}\right)\delta_{2}\nonumber \\
 & \quad-B_{2,0}\left(c\delta_{1},c\delta_{2},p_{id}\right)\delta_{1}^{2}-2B_{1,1}\left(c\delta_{1},c\delta_{2},p_{id}\right)\delta_{1}\delta_{2}-B_{0,2}\left(c\delta_{1},c\delta_{2},p_{id}\right)\delta_{2}^{2}\label{eq:Taylor-expansion}
\end{align}
\begin{align*}
f\left(0,0,p_{id}\right) & =1-\left[\sum_{k=1}^{\infty}\frac{1}{2^{k+1}}R\left(p_{id};k\right)\right]+{\cal O}\left(p_{id}^{2}\right)\\
A_{1}\left(p_{id}\right) & =\left(1-p_{id}\right)\log\frac{2-p_{id}}{2+p_{id}}-\sum_{k=1}^{\infty}\frac{k-3}{2^{k}}R\left(p_{id};k\right)\\
A_{2}\left(p_{id}\right) & =\left(1-2p_{id}\right)\log\frac{2-p_{id}}{2+p_{id}}-\sum_{k=1}^{\infty}\frac{k-3}{2^{k}}R\left(p_{id};k\right)\\
B_{2,0}\left(c\delta_{1},c\delta_{2},p_{id}\right) & =\frac{2\left(1-p_{id}\right)^{2}}{{\displaystyle \left[1-\left(\frac{1}{2}p_{id}+r\right)^{2}\right]}\ln2}+\frac{1}{2}\sum_{k=1}^{\infty}q^{k-3}\left[\left(q-1\right)^{2}k^{2}+\left(q^{2}+2q-3\right)k+2\right]R\left(p_{id};k\right)\\
B_{0,2}\left(c\delta_{1},c\delta_{2},p_{id}\right) & =\frac{2\left(1-2p_{id}\right)^{2}}{{\displaystyle \left[1-\left(\frac{1}{2}p_{id}+r\right)^{2}\right]}\ln2}+\frac{1}{2}\sum_{k=1}^{\infty}q^{k-3}\left[\left(q-1\right)^{2}k^{2}+\left(q^{2}+2q-3\right)k+2\right]R\left(p_{id};k\right)\\
B_{1,1}\left(c\delta_{1},c\delta_{2},p_{id}\right) & =\frac{2\left(1-p_{id}\right)\left(1-2p_{id}\right)}{{\displaystyle \left[1-\left(\frac{1}{2}p_{id}+r\right)^{2}\right]}\ln2}+\frac{1}{2}\sum_{k=1}^{\infty}q^{k-3}\left[\left(q-1\right)^{2}k^{2}+\left(q^{2}+2q-3\right)k+2\right]R\left(p_{id};k\right)\\
r & =2\left(1-p_{id}\right)c\delta_{1}+2\left(1-2p_{id}\right)c\delta_{2}\\
q & =\frac{1}{2}+c\delta_{1}+c\delta_{2}
\end{align*}
\setcounter{equation}{\value{MyTempEquationCounter}}\vspace*{4pt}

\rule[0.5ex]{1\textwidth}{1pt}
\end{figure*}

A suitable upper bound not only is sufficiently simple to analyze,
but also well retains its second term $\lim_{n\to\infty}H\left(Y_{n}\middle|Y^{n-1},X^{n}\right)$
so that it can match up with the summand in Eq. \eqref{eq:Capacity_Lower_Bound_Expansion}.
We now develop an upper bound with $\lim_{n\to\infty}H\left(Y_{n}\middle|Y^{n-1}\right)\leq H\left(Y_{n}\middle|Y_{n-1}\right)$,
which is Eq. \eqref{eq:Eq_Upper Bound, Bound on Firsrt Term} with
${\cal L}=2$ (which should suffice since the first term on the left-hand
side of this inequality decreases slowly for low $p_{id}$, as suggested
by Fig. \ref{Fig_Capacity_Entropy_Term}), while leaving the second
term unchanged as before.

Let 
\begin{align*}
p_{1} & =P\left(X_{n}=X_{n-1}=X_{n-2}\right)\\
p_{2} & =P\left(X_{n}=X_{n-1}\neq X_{n-2}\right)\\
p_{3} & =P\left(X_{n}\neq X_{n-1}=X_{n-2}\right)\\
p_{4} & =P\left(X_{n}=X_{n-2}\neq X_{n-1}\right)
\end{align*}
The bit-symmetry condition gives us
\begin{eqnarray*}
P\left(X_{n-2}^{n}=000\right)= & P\left(X_{n-2}^{n}=111\right)= & p_{1}/2\\
P\left(X_{n-2}^{n}=100\right)= & P\left(X_{n-2}^{n}=011\right)= & p_{2}/2\\
P\left(X_{n-2}^{n}=110\right)= & P\left(X_{n-2}^{n}=001\right)= & p_{3}/2\\
P\left(X_{n-2}^{n}=010\right)= & P\left(X_{n-2}^{n}=101\right)= & p_{4}/2
\end{eqnarray*}
The stationarity condition, as in Lemma \ref{lem:Lemma_Stationarity Condition},
requires that
\[
P\left(X_{n-1}^{n}=x_{1}x_{2}\right)=P\left(X_{n-2}^{n-1}=x_{1}x_{2}\right)
\]
for any $x_{1},x_{2}\in\left\{ 0,1\right\} $, leading to $p_{2}=p_{3}$.
With the fact that $p_{1}+p_{2}+p_{3}+p_{4}=1$, we are left with
2 free variables $p_{1}$ and $p_{2}$ such that $p_{4}=1-p_{1}-2p_{2}$.
Then:
\begin{align*}
 & H\left(Y_{n}\middle|Y_{n-1}\right)=\\
 & h_{2}\left(\left(1-\frac{2p_{i}p_{d}}{p_{i}+p_{d}}\right)p_{1}+\left(1-\frac{4p_{i}p_{d}}{p_{i}+p_{d}}\right)p_{2}+\frac{2p_{i}p_{d}}{p_{i}+p_{d}}\right)
\end{align*}
For the term $\lim_{n\to\infty}H\left(Y_{n}\middle|Y^{n-1},X_{0}^{n}\right)$,
in Eq. \eqref{eq:Lower_bound_term_2}, we have $\alpha=P\left(X_{n}\neq X_{n-1}\right)=1-p_{1}-p_{2}$.
Let $\delta_{1}=p_{1}-\nicefrac{1}{4}$, $\delta_{2}=p_{2}-\nicefrac{1}{4}$,
and 
\[
f\left(\delta_{1},\delta_{2},p_{id}\right)=H\left(Y_{n}\middle|Y_{n-1}\right)-{\displaystyle \lim_{n\to\infty}H\left(Y_{n}\middle|Y^{n-1},X_{0}^{n}\right)}
\]
The non-negativity condition of the input distribution (i.e. $p_{1}\geq0$,
$p_{2}\geq0$, $p_{4}=1-p_{1}-2p_{2}\geq0$) translates to
\begin{equation}
{\displaystyle \frac{1}{4}\geq\delta_{1}}+2\delta_{2},\quad\delta_{1}\geq-\frac{1}{4},\quad\delta_{2}\geq-\frac{1}{4}\label{eq:Eq_Validity_Condition_on_Delta_Low_noise}
\end{equation}
The Taylor expansion theorem with Lagrange remainder applied to $f$
about the i.u.d. input (i.e. $\delta_{1}=\delta_{2}=0$) with $p_{id}$
fixed as a parameter gives Eq. \eqref{eq:Taylor-expansion}\addtocounter{equation}{1}
for some $c\in\left[0,1\right]$ (see next page). Let $\delta_{1}^{*}$
and $\delta_{2}^{*}$ be values that correspond to a capacity-achieving
input. As an upper bound on the capacity $C=C\left(p_{id}\right)$
at $p_{id}$,
\[
C\left(p_{id}\right)\leq f\left(\delta_{1}^{*},\delta_{2}^{*},p_{id}\right)
\]
Observe that $f\left(0,0,p_{id}\right)$ matches up with $C_{iud}^{lb}$
up to ${\cal O}\left(p_{id}^{2}\right)$, which is expected for the
zero-order term of a good upper bound expanded about the i.u.d. input.
Therefore we only need to estimate how the rest of the terms, which
represent $\Delta f$ in the opening discussion of this section, scale
with $p_{id}$, without knowing their exact expression in terms of
$p_{id}$.

Allowing approximations to reduce unnecessarily complex algebra and
noticing that $\delta_{1}^{*}$ and $\delta_{2}^{*}$ are functions
of $p_{id}$, we have: 
\begin{align*}
A_{1}\left(p_{id}\right) & \approx A_{2}\left(p_{id}\right)=A\left(p_{id}\right)\\
B_{2,0}\left(c\delta_{1}^{*},c\delta_{2}^{*},p_{id}\right) & \approx{\displaystyle B_{0,2}\left(c\delta_{1}^{*},c\delta_{2}^{*},p_{id}\right)}\\
 & \approx{\displaystyle B_{1,1}\left(c\delta_{1}^{*},c\delta_{2}^{*},p_{id}\right)}={\displaystyle B\left(p_{id}\right)}
\end{align*}
One can show that $A\left(p_{id}\right)=\Theta\left(p_{id}\right)$
with elementary calculations. The analysis of $B\left(p_{id}\right)$
requires more care, since it involves $\delta_{1}^{*}$ and $\delta_{2}^{*}$.
Let $b_{1}\left(p_{id}\right)$ denote the first term in $B\left(p_{id}\right)$,
i.e.
\[
b_{1}\left(p_{id}\right)=\frac{2\left(1-p_{id}\right)\left(1-2p_{id}\right)}{{\displaystyle \left[1-\left(\frac{1}{2}p_{id}+r^{*}\right)^{2}\right]}\ln2}
\]
where $r^{*}=2\left(1-p_{id}\right)c\delta_{1}^{*}+2\left(1-2p_{id}\right)c\delta_{2}^{*}$,
and $b_{2}\left(p_{id}\right)=B\left(p_{id}\right)-b_{1}\left(p_{id}\right)$.
For $\delta_{1}^{*}$ and $\delta_{2}^{*}$ under the constraint \eqref{eq:Eq_Validity_Condition_on_Delta_Low_noise}
and $p_{id}<0.5$, noticing that $\delta_{1}^{*}$ and $\delta_{2}^{*}$
approach $0$ as $p_{id}\to0$, one can see easily that $b_{1}\left(p_{id}\right)>0$
and $b_{1}\left(p_{id}\right)=\Theta\left(1\right)$. Also, $b_{1}\left(p_{id}\right)$
is undefined if and only if $c=1$, $\delta_{1}^{*}=0.75$ and $\delta_{2}^{*}=-0.25$,
in which case the capacity-achieving input only allows the all-zeros
and all-ones input strings and hence the capacity is trivially $0$.
However $C_{iud}^{lb}$ is always away from $0$ for any $p_{id}$.
We therefore exclude such large deviations from $0$ of $\delta_{1}^{*}$
and $\delta_{2}^{*}$.

To evaluate $b_{2}\left(p_{id}\right)$, we first note that $q^{*}=\frac{1}{2}+c\delta_{1}^{*}+c\delta_{2}^{*}\in\left[0,1\right]$
under the constraint \eqref{eq:Eq_Validity_Condition_on_Delta_Low_noise}.
$q^{*}=1$ if and only if $c=1$, $\delta_{1}^{*}=0.75$ and $\delta_{2}^{*}=-0.25$.
So for the same reason above, we can bound $q^{*}\leq\epsilon<1$.
In addition, since $q^{*}$ is finite, there exists a polynomial $P_{2}\left(k\right)$
of degree $2$ over the real numbers such that $P_{2}\left(k\right)\geq\left|\left(q^{*}-1\right)^{2}k^{2}+\left(q^{*2}+2q^{*}-3\right)k+2\right|$
for any $k\geq1$. Letting $M_{k}=\epsilon^{k-3}P_{2}\left(k\right)$,
we then have $M_{k}\geq\left|q^{*k-3}\left[\left(q^{*}-1\right)^{2}k^{2}+\left(q^{*2}+2q^{*}-3\right)k+2\right]R\left(p_{id};k\right)\right|$
for any $p_{id}$, since $\left|R\left(p_{id};k\right)\right|\leq1$.
We also have that $\sum_{k=1}^{\infty}M_{k}$ converges as a straightforward
application of the ratio test for series convergence. Then by the
Weierstrass M-test, we have that $b_{2}\left(p_{id}\right)$ converges
uniformly. Consequently,
\begin{align*}
 & \lim_{p_{id}\to0}b_{2}\left(p_{id}\right)\\
 & =\frac{1}{2}\sum_{k=1}^{\infty}\lim_{p_{id}\to0}\left\{ \begin{array}{c}
q^{*k-3}R\left(p_{id};k\right)\times\\
\left[\left(q^{*}-1\right)^{2}k^{2}+\left(q^{*2}+2q^{*}-3\right)k+2\right]
\end{array}\right\} \\
 & =0
\end{align*}
That is, $b_{2}\left(p_{id}\right)={\cal O}\left(1\right)$ (and in
fact, of order of magnitude much lesser than $\Theta\left(1\right)$).

\begin{figure}
\begin{centering}
\includegraphics[width=1\columnwidth]{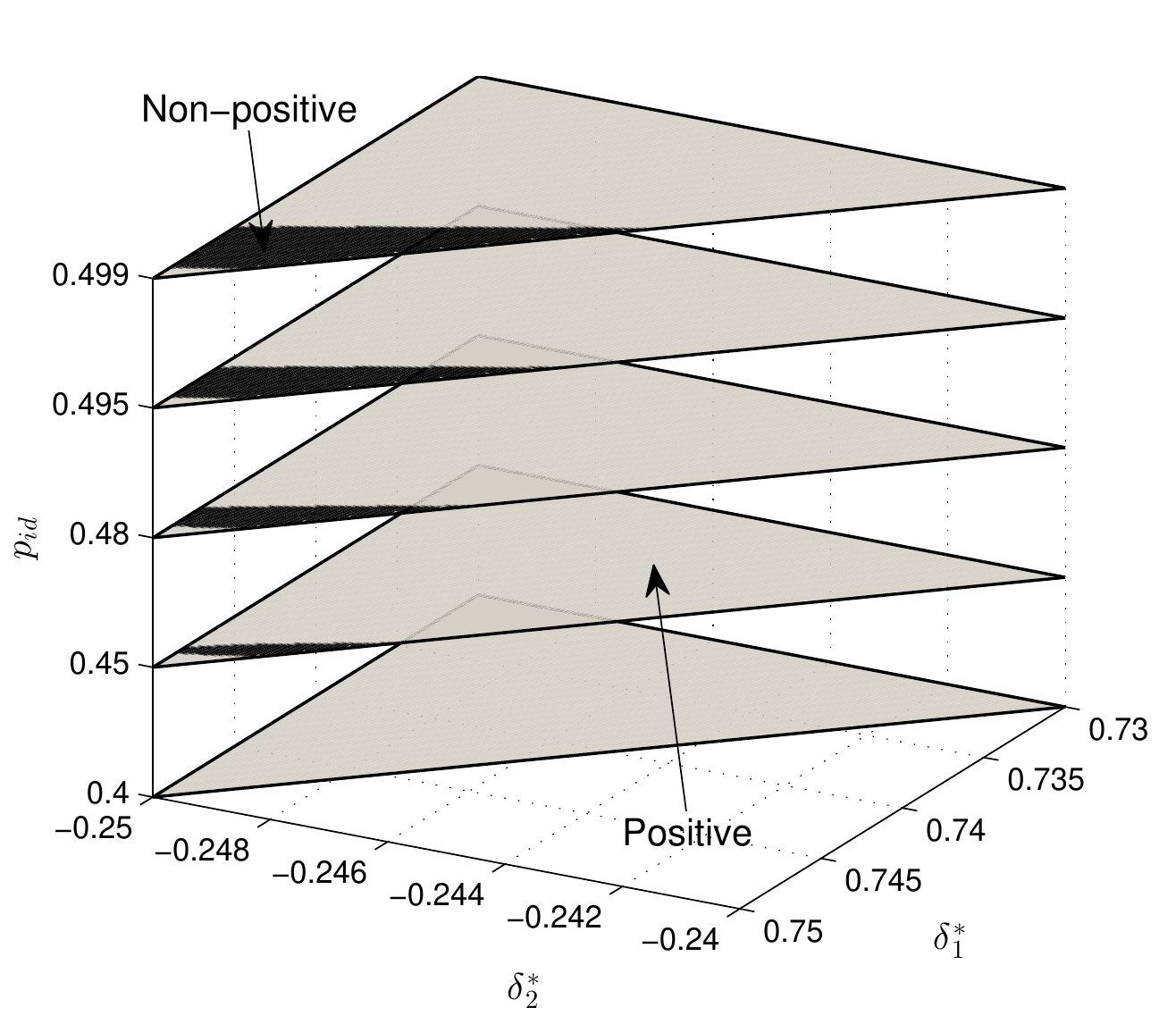}
\par\end{centering}

\protect\caption{Sign of $B\left(p_{id}\right)$ for different values of $p_{id}$,
$\delta_{1}^{*}$ and $\delta_{2}^{*}$, which satisfy the constraint
\eqref{eq:Eq_Validity_Condition_on_Delta_Low_noise}. The summands
of $b_{2}\left(p_{id}\right)$ are limited to a maximum $k$ of $2000$
for computational feasibility. In general, the ``non-positive''
region shrinks as the maximum $k$ is increased. Here $c$ is set
to $1$. However this particular choice of $c$ is irrelevant, since
values of $B\left(p_{id}\right)$ with $\left(c,\delta_{1}^{*},\delta_{2}^{*}\right)=\left(c',d_{1},d_{2}\right)$,
where $c'<1$, are equal to those with $\left(c,\delta_{1}^{*},\delta_{2}^{*}\right)=\left(1,c'd_{1},c'd_{2}\right)$.}

\label{Fig_Low_Noise_B(p)_Negative_Positive}
\end{figure}

As a result, $B\left(p_{id}\right)=\Theta\left(1\right)$ for $p_{id}<0.5$.
Since $b_{2}\left(p_{id}\right)\ll b_{1}\left(p_{id}\right)$ and
$b_{1}\left(p_{id}\right)>0$, we can conclude that $B\left(p_{id}\right)>0$
at sufficiently low $p_{id}$. In fact, Fig. \ref{Fig_Low_Noise_B(p)_Negative_Positive}
suggests that $B\left(p_{id}\right)$ is potentially non-positive
only in extreme cases where $p_{id}$ is close to $0.5$, $\delta_{1}^{*}$
is close to $0.75$ and $\delta_{2}^{*}$ is close to $-0.25$. For
the purpose of obtaining low-noise approximations, we can therefore
say that $B\left(p_{id}\right)>0$. We then have:
\begin{align}
 & C\left(p_{id}\right)-f\left(0,0,p_{id}\right)\nonumber \\
 & \leq f\left(\delta_{1}^{*},\delta_{2}^{*},p_{id}\right)-f\left(0,0,p_{id}\right)\nonumber \\
 & \approx\left(\delta_{1}^{*}+\delta_{2}^{*}\right)A\left(p_{id}\right)-\left(\delta_{1}^{*}+\delta_{2}^{*}\right)^{2}B\left(p_{id}\right)\nonumber \\
 & \leq\sup_{t\in\mathbb{R}}\left[tA\left(p_{id}\right)-t^{2}B\left(p_{id}\right)\right]\nonumber \\
 & =\frac{A^{2}\left(p_{id}\right)}{4B\left(p_{id}\right)}=\Theta\left(p_{id}^{2}\right)\label{eq:Capacity_Upper_Bound_Expansion}
\end{align}
using the identity $\sup_{t\in\mathbb{R}}\left(at^{2}+bt+c\right)=\left(4ac-b^{2}\right)/\left(4a\right)$
for $a<0$. This completes the derivation of the low-noise upper bound.

The same conclusion, $C\left(p_{id}\right)\leq f\left(0,0,p_{id}\right)+\Theta\left(p_{id}^{2}\right)$,
can be reached without the approximations by doing some tedious algebra.
Approximations do not affect our result anyway, since in this analysis
only the order of magnitude of $A\left(p_{id}\right)$ and $B\left(p_{id}\right)$
matters. We also see that $\Delta f$ in the opening discussion corresponds
to $\Theta\left(p_{id}^{2}\right)$, which tends to $0$ as $p_{id}\to0$.
This concurs with the discussed observation from Fig. \ref{Fig_Low_Noise_Strategy_Illustration}.

As a note, it is clear why the expansion of $f$ is up to the second
order. The first-order expansion would make the supremum in our final
step unbounded, while any higher-order expansions would complicate
the analysis.

\subsection{Low-Noise Characterization of the Capacity}

\begin{figure}
\begin{centering}
\includegraphics[width=1\columnwidth]{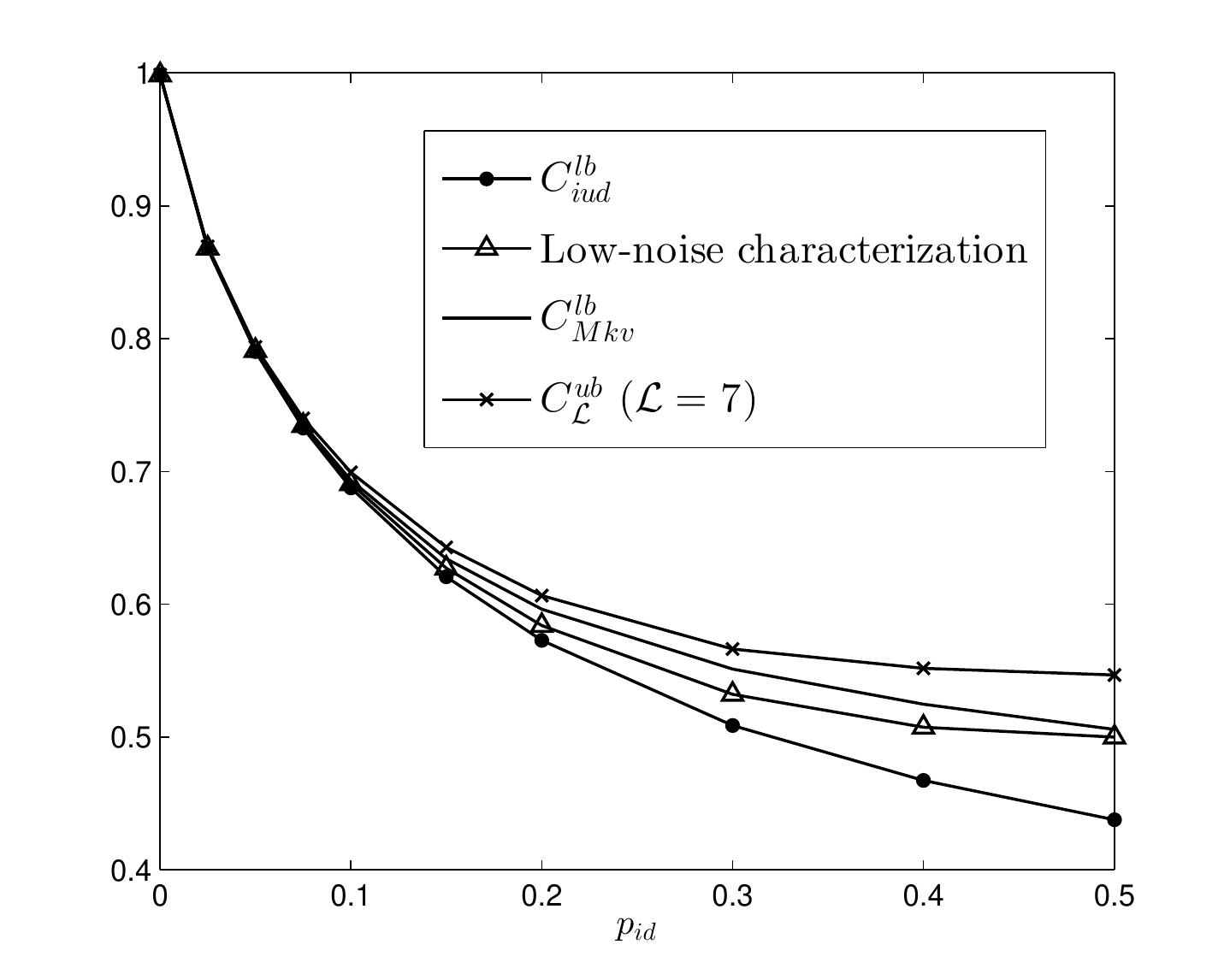}
\par\end{centering}

\protect\caption{The DID channel capacity at low noise.}

\label{Fig_Capacity_Low_Noise}
\end{figure}

Putting Eq. \eqref{eq:Capacity_Lower_Bound_Expansion} and \eqref{eq:Capacity_Upper_Bound_Expansion}
together,
\[
C=1-\left[\sum_{k=1}^{\infty}\frac{1}{2^{k+1}}R\left(p_{id};k\right)\right]+{\cal O}\left(p_{id}^{2}\right)
\]
This characterization, ignoring ${\cal O}\left(p_{id}^{2}\right)$,
is plotted in Fig. \ref{Fig_Capacity_Low_Noise}.

It can be easily shown that $R\left(p_{id};k\right)$ is concave in
$p_{id}$ for every $k\geq1$ and $p_{id}\in\left(0,0.5\right)$.
Therefore this analysis suggests that the capacity is a convex function
of the channel parameter $p_{id}$ in the low-noise regime.

\section{Concluding Remarks\label{sec:Conclusions}}

This work has illustrated the use of the stationarity condition of
the supremizing input set in the capacity formula in a case study
of the DID channel model. Input stationarity has been identified in
both ergodic-theoretic and Shannon-theoretic capacity formulations.
Evidently this condition is pivotal to many of our results. Without
it, the new upper bounds could become trivial as discussed in Section
\ref{sub:Importance-of-Stationary (Upper bounds)}. Moreover this
condition helps describe finite-dimensional marginal distributions
$P_{X_{n+a}^{n+b}}$, with $a\leq b$, for all $n$'s in only a finite
number of variables, which is key to establishing Lemma \ref{lem:Lemma_Second_Term_Formula_for_Stationary_Input}
and subsequently the low-noise characterization.

Our simultaneous treatment of the two separate, seemingly unrelated
frameworks is necessary. Input stationarity arises naturally in the
ergodic-theoretic framework. The reason is that this framework allows
only infinitely long input sequences, an example of whose source is
stationary inputs. On the contrary, in the Shannon-theoretic framework,
allowable input sequences have finite (block) lengths, and only when
reliable communications is of concern (i.e. the error probability
is driven to zero) are the block lengths increased to infinity. This
setting makes it harder to see whether stationary inputs and the likes
should play any role in capacity formulations. As witnessed in Section
\ref{sub:Shannon-Applicability-DID}, via a connection to results
in the ergodic-theoretic framework, developments to a formulation
with stationary inputs could be realized. With further scrutiny, one
sees that the difference between the two frameworks lies mainly in
their operational structures, whereas the said connection is purely
on information-theoretic quantities, which are dictated by the joint
input-output distribution, shared by both frameworks. Curiously the
Shannon-theoretic framework is not the only beneficiary. An argument
in \nameref{sec:Appendix A} shows that so is the ergodic-theoretic
framework.

Next we discuss a few relevant directions concerning the capacity
evaluation techniques for further investigations.

\paragraph{Channels with High Local Dependency}

The DID channel output has relatively low \textit{local dependency}:
$Y_{i}$ is determined only through the present and the immediate
past, $\left\{ Z_{i},X_{i-1}^{i}\right\} $. This may give an intuitive
explanation of why the upper bounds in Fig. \ref{Fig_Capacity_Upper_Bound}
converge very quickly at low $\mathcal{L}$.

We however cannot expect this to be the case for other practical channels,
e.g. the model in \cite{Wu-2013}. As mentioned in Section \ref{sub:DID-Channel-Model},
the DID channel only captures partially key features of the BPMR write
process. For example, it can be observed that in the DID channel model,
an output sub-sequence either remains synchronized with its corresponding
input sub-sequence or leads it by $1$ bit. The scenario in which
the output sub-sequence lags behind the input sub-sequence by $1$
bit is therefore missing. This was considered by the model in \cite{Wu-2013},
which sets the alphabet of $Z_{i}$ to ${\cal Z}=\left\{ -1,0,1\right\} $.
While such ${\cal {\cal Z}}$ makes the channel non-causal, one can
redefine the input to be $\hat{X}_{i}=X_{i-1}$ and consequently $Y_{i}=\hat{X}_{i+1-Z_{i}}$,
which is a mathematically causal channel and to which Eq. \eqref{eq:Equation_C_2}
is again applicable. The scenario where misalignment by more than
$1$ bit is allowed was discussed in \cite[Section VI]{Iyengar-2011},
in which case ${\cal Z}$ also has larger sizes.

All those considerations lead to higher local dependency, which gives
rise to numerous difficulties. At first glance, the convergence would
be slower. To resolve this, one may increase $\mathcal{L}$ (or some
similar parameters that control the bounds) to obtain the desired
accuracy to the true capacity. However the larger ${\cal L}$ is,
higher the computational complexity. The intrinsic tradeoff between
accuracy and complexity is therefore more stringent in this case,
which may call for a modification or a different formulation of the
upper bound.

\paragraph{Channels with Substitution Errors}

Substitution errors are usually inevitable in practical systems. For
example, it was noted in \cite{Wu-2013} that insertion and deletion
errors, underlied by the Markov channel state $\left\{ Z_{i}\right\} _{i=1}^{\infty}$,
are not sufficient to faithfully describe all imperfections that arise
in the context of BPMR. Their model considered burst substitution
errors that accompany insertion and deletion events, in addition to
random substitution errors caused by a localized random phase drift
between the desired and actual window to write each bit.

In general, we may model substitution errors by $Y_{i}=X_{i-Z_{i}}\oplus B_{i}$,
where $B_{i}$ is a binary random variable representing the additive
noise, independent of the input. A straightforward approach is to
take into account $B_{i}$ and do the exact calculation. Another way
is to view the model as two subsystems, in which one is $A_{i}=X_{i-Z_{i}}$,
concatenated with the other $Y_{i}=A_{i}\oplus B_{i}$. Then an upper
bound on the capacity is given by the data processing inequality,
$I\left(\mathbf{X};\mathbf{Y}\right)\leq\min\left\{ I\left(\mathbf{X};\mathbf{A}\right),I\left(\mathbf{A};\mathbf{Y}\right)\right\} $,
suggesting that $C\leq\min\left\{ C_{1},C_{2}\right\} $, where $C$,
$C_{1}$ and $C_{2}$ are the true capacity, the first subsystem's
capacity and that of the second subsystem respectively. This gives
a simple benchmark upper bound.

\paragraph{Finite-Block-Length Regime}

One natural question is how the fundamental limit behaves when the
block length is finite (see e.g. \cite{Polyanskiy-2010,Tan-FiniteBlocklength}).
In this analysis, the capacity is known as the first-order coding
rate, and the quest is to find higher-order coding rates given a finite
block length $n$ and some non-zero tolerable error probability $\epsilon$.
Note that the finite-block-length analysis is usually developed under
the Shannon-theoretic framework. It is unclear how the ergodic-theoretic
channel definition could extend itself to encompass finite block lengths.

While finite-lettered characterizations of the second-order coding
rate have been determined precisely for certain channels, they are
not known in general for channels with memory or complex structures.
Since the DID channel's first-order coding rate has been determined
approximately under the Shannon-theoretic framework as discussed above,
it would be interesting to see how one can approximate its second-order
coding rate as a series expansion in the channel parameters.

\section*{Appendix A\label{sec:Appendix A}}

We consider the ergodic-theoretic ``capacity'' formulation with
general initializations of the state $Z_{1}$. The DID channel is
an FSC in the Shannon-theoretic framework. It was pointed out  by
Kieffer and Rahe \cite{Kieffer-1981} (and also by Gray et al. \cite{Gray-1987})
that FSCs can be naturally described as a special case of (one-sided)
Markov channels in the ergodic-theoretic setting. Furthermore they
showed that Markov channels are asymptotically mean stationary (AMS).
Hence so is the DID channel.

For any $p_{i}$ and $p_{d}$ in $\left(0,1\right)$, the process
$\left\{ Z_{i}\right\} _{i=1}^{\infty}$ is mixing. As such, the DID
channel is output strongly mixing, in light of Section \ref{sub:Ergodic-theoretic Capacity}.
Since the channel is (one-sided) Markov, by \cite[Theorem 2]{Gray-1987}
and \cite[Lemma 3]{Gray-1987}, it is ergodic.

By \cite[Theorem 12.6.1]{Gray-Entropy}, the following rate is achievable:
\begin{align}
C^{*}\left(P_{\mathbf{Z}}\right) & =\sup_{\text{AMS }P_{X_{0}^{\infty}}}\lim_{n\to\infty}\frac{1}{n}I_{AMS}\left(X_{0}^{n};Y_{1}^{n}\right)\label{eq:Eq_Capacity_4}
\end{align}
Here the subscript $AMS$ means that it is calculated w.r.t. the stationary
mean of the joint input-output distribution, and the notation $C^{*}\left(P_{\mathbf{Z}}\right)$
emphasizes the dependence of the rate on the distribution $P_{\mathbf{Z}}$
of the process $\left\{ Z_{i}\right\} _{i=1}^{\infty}$.

All that is left is to establish an equality between Eq. \eqref{eq:Equation_C_0}
(which assumes $P\left(Z_{1}=0\right)=p_{d}/\left(p_{i}+p_{d}\right)$
and $P\left(Z_{1}=1\right)=p_{i}/\left(p_{i}+p_{d}\right)$) and Eq.
\eqref{eq:Eq_Capacity_4}. We digress from the task temporarily. The
notion of AMS processes does not closely describe the processes associated
with the DID channel. We hence appeal to the following definition.
\begin{defn}
A random process $\{V_{n}\}_{n=1}^{\infty}$, defined over a probability
space $(\Omega,\mathcal{F},P)$, is asymptotically stationary (AS)
if $\forall E\in\mathcal{F}$, the limit $P^{\infty}(E)=\lim_{n\rightarrow\infty}P(T^{-n}E)$
exists. $P^{\infty}$ is then called the asymptotic stationary probability
measure.
\end{defn}
One can easily prove that a stationary measure is AS and AMS. In addition,
we have the following lemma.
\begin{lem}
\label{lem:Lemma_AS_process_is_AMS}An AS process is AMS. Furthermore,
the stationary mean is its asymptotic stationary probability measure.\end{lem}
\begin{IEEEproof}
The claim follows easily from the definitions and the Cesàro mean
theorem.
\end{IEEEproof}
Consider two independent processes $\{A_{n}\}_{n=1}^{\infty}$ and
$\{B_{n}\}_{n=1}^{\infty}$, whose underlying probability measures
are respectively $P_{A}$ and $P_{B}$. Let $P_{AB}$ denote the probability
measure of the joint process $\left\{ \left(A_{n},B_{n}\right)\right\} _{n=1}^{\infty}$.
For any event $E$ on this joint process, we have:
\begin{align*}
P_{AB}\left(T^{-n}E\right) & =\intop_{\left(\mathbf{a},\mathbf{b}\right)\in E}\sum_{\substack{\mathbf{a}'\in{\cal A}^{n}\\
\mathbf{b}'\in{\cal B}^{n}
}
}dP_{AB}\left(\left[\mathbf{a}',\mathbf{a}\right],\left[\mathbf{b}',\mathbf{b}\right]\right)\\
 & =\intop_{\left(\mathbf{a},\mathbf{b}\right)\in E}\sum_{\substack{\mathbf{a}'\in{\cal A}^{n}\\
\mathbf{b}'\in{\cal B}^{n}
}
}d\left(P_{A}\times P_{B}\right)\left(\left[\mathbf{a}',\mathbf{a}\right],\left[\mathbf{b}',\mathbf{b}\right]\right)\\
 & =\intop_{\left(\mathbf{a},\mathbf{b}\right)\in E}d\left(P_{A}\times P_{B}\right)\left(T^{-n}\left\{ \mathbf{a}\right\} ,T^{-n}\left\{ \mathbf{b}\right\} \right)
\end{align*}
This shows that if $\{A_{n}\}_{n=1}^{\infty}$ and $\{B_{n}\}_{n=1}^{\infty}$
are AS, $\left\{ \left(A_{n},B_{n}\right)\right\} _{n=1}^{\infty}$
is also AS. Moreover $P_{AB}^{\infty}$ is determined completely by
$P_{A}^{\infty}$ and $P_{B}^{\infty}$, which denote the associated
asymptotic stationary probability measures. 

Next, consider a process $\{V_{n}\}_{n=1}^{\infty}$, defined under
$P_{V}$. Define another process $\{W_{n}:\; W_{n}=\phi(V_{n},V_{n-1},...,V_{n-k})\}$,
where $\phi$ is a deterministic and time-invariant function and $k$
is finite. Let $P_{W}$ be the underlying probability measure of $\{W_{n}\}_{n=k+1}^{\infty}$,
and consider an event $E_{W}$ on the process. Let $\mathcal{E}_{i}=\left\{ u_{i}:\;\mathbf{u}\in E_{W}\right\} $.
Let $E_{W}^{i}=\left\{ \mathbf{w}:\; w_{i}\in\mathcal{E}_{i}\right\} $
and $E_{V}^{i}=\left\{ \mathbf{v}:\;\phi\left(v_{i-k}^{i}\right)\in\mathcal{E}_{i}\right\} $.
Then for any $j\geq0$:
\begin{align*}
P_{W}\left(T^{-j}E_{W}\right) & =P_{W}\left(T^{-j}\bigcap_{i=k+1}^{\infty}E_{W}^{i}\right)\\
 & =P_{W}\left(\bigcap_{i=k+1}^{\infty}\left\{ \mathbf{w}:\; w_{i+j}\in\mathcal{E}_{i}\right\} \right)\\
 & =P_{V}\left(\bigcap_{i=k+1}^{\infty}\left\{ \mathbf{v}:\;\phi\left(v_{i+j-k}^{i+j}\right)\in\mathcal{E}_{i}\right\} \right)\\
 & =P_{V}\left(T^{-j}\bigcap_{i=k+1}^{\infty}E_{V}^{i}\right)
\end{align*}
where the third equality is because firstly, any $\mathbf{v}\in\bigcap_{i=k+1}^{\infty}\left\{ \mathbf{v}:\;\phi\left(v_{i+j-k}^{i+j}\right)\in\mathcal{E}_{i}\right\} $,
under $\phi$, is transformed into some $\mathbf{w}\in\bigcap_{i=k+1}^{\infty}\left\{ \mathbf{w}:\; w_{i+j}\in\mathcal{E}_{i}\right\} $,
and secondly, for any $\mathbf{v}\notin\bigcap_{i=k+1}^{\infty}\left\{ \mathbf{v}:\;\phi\left(v_{i+j-k}^{i+j}\right)\in\mathcal{E}_{i}\right\} $,
there exists at least one index $r\geq k+1$ such that $\phi\left(v_{r+j-k}^{r+j}\right)\notin\mathcal{E}_{r}$
and so $\mathbf{v}$ is not transformed into any member of $\bigcap_{i=k+1}^{\infty}\left\{ \mathbf{w}:\; w_{i+j}\in\mathcal{E}_{i}\right\} $.
This shows that if $\{V_{n}\}_{n=1}^{\infty}$ is AS, so is $\{W_{n}\}_{n=k+1}^{\infty}$.
Moreover $P_{W}^{\infty}$ is completely determined by $P_{V}^{\infty}$.

We return to our original problem. Consider AMS $P_{X_{0}^{\infty}}$,
since Eq. \eqref{eq:Eq_Capacity_4} involves only AMS inputs. It is
known that $\{Z_{i}\}_{i=1}^{\infty}$ is AS for any $p_{i},p_{d}\in\left(0,1\right)$;
moreover it has a unique $P_{\mathbf{Z}}^{\infty}$ which coincides
with its distribution $P_{\mathbf{Z}}^{*}$ when $P\left(Z_{1}=0\right)=p_{d}/\left(p_{i}+p_{d}\right)$
and $P\left(Z_{1}=1\right)=p_{i}/\left(p_{i}+p_{d}\right)$, i.e.
when the initialization allows it to be stationary, for the same $p_{i}$
and $p_{d}$. Notice that for the DID channel, we can express $\left(X_{i},Y_{i}\right)$
as deterministic and time-invariant functions of $\left(\left(X_{i},X_{i-1}\right),Z_{i}\right)$.
Let $\left(X_{i},X_{i-1}\right)$, $Z_{i}$, $\left(\left(X_{i},X_{i-1}\right),Z_{i}\right)$
and $\left(X_{i},Y_{i}\right)$ play the role of $A_{i}$, $B_{i}$,
$V_{i}$ and $W_{i}$, respectively, in the above discussion. We see
that the joint input-output distribution is AS. By Lemma \ref{lem:Lemma_AS_process_is_AMS},
this implies that $C^{*}\left(P_{\mathbf{Z}}\right)$ can be computed
w.r.t. $P_{X^{\infty}\mathbf{Y}}^{\infty}$. Furthermore $P_{X^{\infty}\mathbf{Y}}^{\infty}$
is completely determined by the input distribution, which is AMS,
and $P_{\mathbf{Z}}^{\infty}$, which is the same as $P_{\mathbf{Z}}^{*}$.
Therefore, instead of using the channel law in Eq. \eqref{eq:Eq_channelLaw_ergodic}
which asserts
\[
P_{\left.\mathbf{Y}\middle|X_{0}^{\infty}\right.}\left(E_{Y}\middle|\mathbf{x}\right)=P_{\mathbf{Z}}\left({\cal E}\left(\mathbf{x},E_{Y}\right)\right)
\]
we can compute $C^{*}\left(P_{\mathbf{Z}}\right)$ as if the channel
law is
\[
P_{\left.\mathbf{Y}\middle|X_{0}^{\infty}\right.}\left(E_{Y}\middle|\mathbf{x}\right)=P_{\mathbf{Z}}^{*}\left({\cal E}\left(\mathbf{x},E_{Y}\right)\right)
\]
i.e. $C^{*}\left(P_{\mathbf{Z}}\right)=C^{*}\left(P_{\mathbf{Z}}^{*}\right)$.
When the channel assumes $P_{\mathbf{Z}}^{*}$, it is stationary as
shown in Section \ref{sub:Ergodic-theoretic Capacity}. By \cite[Lemma 12.4.2]{Gray-Entropy},
\begin{align*}
C^{*}\left(P_{\mathbf{Z}}^{*}\right) & =\sup_{\text{Stationary }P_{X_{0}^{\infty}}}\lim_{n\to\infty}\frac{1}{n}I\left(X_{0}^{n};Y_{1}^{n}\right)
\end{align*}
whose right-hand side coincides with that of Eq. \eqref{eq:Equation_C_0}.
This completes our argument.

\section*{Appendix B\label{sec:Appendix-Proof_of_SE_Capacity_Shannon-theoretic_Framework}}

We prove Theorem \ref{thm:Theorem_consistent_SE_Channel_Capacity (Shannon-theoretic)}.
As a reminder, this proof is under the Shannon-theoretic framework.
\begin{lem}
\label{lem:Lemma_SE_Capacity}The SE capacity of a consistent SE channel
with finite alphabets is given by
\[
C_{SE}=\sup_{\text{SE }\left\{ X_{1-\lambda_{-}}^{n+\lambda_{+}}\right\} _{n=1}^{\infty}}\lim_{n\to\infty}\frac{1}{n}I\left(X_{1-\lambda_{-}}^{n+\lambda_{+}};Y^{n}\right)
\]
\end{lem}
\begin{IEEEproof}
Given any SE input, a single joint distribution $P_{X_{1-\lambda_{-}}^{\infty}\mathbf{Y}}$
exists and is SE. With an abuse of notation, we therefore drop all
subscripts and use $p$ to denote the respective probability measure.
Let $W_{n}=\left(Y_{n},X_{n-\lambda_{-}}^{n+\lambda_{+}}\right)$;
then $\left\{ W_{n}\right\} _{n=1}^{\infty}$ is SE.
\begin{align*}
 & \frac{1}{n}i^{\left(n\right)}\left(X_{1-\lambda_{-}}^{n+\lambda_{+}};Y^{n}\right)\\
 & =\frac{1}{n}\log p\left(W^{n}\right)-\frac{1}{n}\log p\left(X_{1-\lambda_{-}}^{n+\lambda_{+}}\right)-\frac{1}{n}\log p\left(Y^{n}\right)\\
 & \xrightarrow{n\to\infty}-\lim_{n\to\infty}\frac{1}{n}H\left(W^{n}\right)+\lim_{n\to\infty}\frac{1}{n}H\left(X_{1-\lambda_{-}}^{n+\lambda_{+}}\right)\\
 & \qquad\quad+\lim_{n\to\infty}\frac{1}{n}H\left(Y^{n}\right)\\
 & =\lim_{n\to\infty}\frac{1}{n}I\left(X_{1-\lambda_{-}}^{n+\lambda_{+}};Y^{n}\right)
\end{align*}
where the convergence is almost-sure in $P_{X_{1-\lambda_{-}}^{\infty}\mathbf{Y}}$,
by invoking the Shannon-McMillan-Breiman theorem. Since $\underline{\mathbf{I}}\left(\mathbf{X};\mathbf{Y}\right)$
is the left end point of the support of the distribution of $\frac{1}{n}i^{\left(n\right)}\left(X_{1-\lambda_{-}}^{n+\lambda_{+}};Y^{n}\right)$
at the limit $n\to\infty$, this convergence implies the claim.
\end{IEEEproof}
The main proof is a modification of Feinstein's work \cite{Feinstein-1959}\textcolor{blue}{,
}which is under the ergodic-theoretic framework. We exploit his construction
of an SE probability measure from a finite-dimensional probability
measure. For some fixed $r\in\mathbb{N}^{+}$ and $s=r+\lambda_{+}+\lambda_{-}$,
let us consider an arbitrary probability measure $P^{\left(r\right)}$
of $X_{1-\lambda_{-}}^{r+\lambda_{+}}$. Define a probability measure
$\mu$ such that for any two integers $k_{1}$ and $k_{2}$ where
$0\leq k_{1}<k_{2}$:
\[
\mu\left(X_{k_{1}s+1-\lambda_{-}}^{k_{2}s-\lambda_{-}}=x_{k_{1}s+1}^{k_{2}s}\right)=\prod_{k=k_{1}}^{k_{2}-1}P^{\left(r\right)}\left(x_{ks+1}^{\left(k+1\right)s}\right)
\]
It is easy to see that $\mu$ is $s$-stationary. Let us define another
probability measure $\hat{\mu}$ such that for any event $E$ on the
input:
\[
\hat{\mu}\left(E\right)=\frac{1}{s}\left[\mu_{0}\left(E\right)+\mu_{1}\left(E\right)+...+\mu_{s-1}\left(E\right)\right]
\]
where $\mu_{k}$ is defined by $\mu_{k}\left(E\right)=\mu\left(T^{-k}E\right)$
$\forall E$, for $k=0,\ldots,s-1$. It can be easily established
in a similar manner to \cite{Feinstein-1959} that $\hat{\mu}$ is
SE. Then we immediately see that $\mu$ is ergodic, since for any
invariant event $E$, $\mu\left(E\right)=\hat{\mu}\left(E\right)$,
which is equal to either $0$ or $1$.

We shall need the following lemma.
\begin{lem}
\label{lem:Lemma_Mutual_Info_Inequality}For any four random variables
$X_{1}$, $X_{2}$, $Y_{1}$ and $Y_{2}$, in which $X_{1}$ is independent
of $X_{2}$, we have:
\[
I\left(X_{1},X_{2};Y_{1},Y_{2}\right)\geq I\left(X_{1};Y_{1}\right)+I\left(X_{2};Y_{2}\right)
\]
\end{lem}
\begin{IEEEproof}
This lemma can be found in \cite[Problem 2.4(e)]{ElGamal-Kim-2012}.
We provide a proof here for completeness.
\begin{align*}
 & I\left(X_{1},X_{2};Y_{1},Y_{2}\right)\\
 & =I\left(X_{1};Y_{1},Y_{2}\right)+I\left(X_{2};Y_{1},Y_{2}\middle|X_{1}\right)\\
 & =I\left(X_{1};Y_{1},Y_{2}\right)+I\left(X_{2};Y_{1},Y_{2},X_{1}\right)-I\left(X_{1};X_{2}\right)\\
 & =I\left(X_{1};Y_{1},Y_{2}\right)+I\left(X_{2};Y_{1},Y_{2},X_{1}\right)\\
 & \geq I\left(X_{1};Y_{1}\right)+I\left(X_{2};Y_{2}\right)
\end{align*}
where $I\left(X_{1};X_{2}\right)=0$ since $X_{1}$ and $X_{2}$ are
independent.
\end{IEEEproof}

\begin{IEEEproof}[Proof of Theorem \ref{thm:Theorem_consistent_SE_Channel_Capacity (Shannon-theoretic)}]
Let us consider an arbitrary finite-dimensional distribution sequence
$\left\{ P^{\left(n\right)}\right\} _{n=1}^{\infty}$, which should
be understood as an arbitrary input sequence and is not necessarily
consistent, and some $r\in\mathbb{N}^{+}$. We construct the aforementioned
probability measures $\mu$ and $\hat{\mu}$ from $P^{\left(r\right)}$.
By passing the input generated by $\mu$ (resp. $\hat{\mu}$) through
the channel, we obtain the joint input-output distribution $\omega$
(resp. $\hat{\omega}$) and the output distribution $\eta$ (resp.
$\hat{\eta}$). A single measure $\omega$ (resp. $\hat{\omega}$)
exists, and consequently $\eta$ (resp. $\hat{\eta}$) exists, since
the channel is consistent. Since the relations are linear,
\begin{align*}
\hat{\omega} & =\frac{1}{s}\left(\omega_{0}+\omega_{1}+...+\omega_{s-1}\right)\\
\hat{\eta} & =\frac{1}{s}\left(\eta_{0}+\eta_{1}+...+\eta_{s-1}\right)
\end{align*}
where $\omega_{k}$ and $\eta_{k}$ are the distributions corresponding
to the input $\mu_{k}$, for $k=0,\ldots,s-1$. It is easy to see
that $\mu_{k}$ is $s$-stationary since $\mu$ is $s$-stationary.
Therefore $\omega_{k}$ and $\eta_{k}$ are $s$-stationary, since
the channel is weakly stationary. Notice the following:
\begin{align*}
 & \omega_{k}\left(X_{1-\lambda_{-}}^{n+\lambda_{+}}=\mathbf{x},Y^{n}=\mathbf{y}\right)\\
 & =P_{\left.Y^{n}\middle|X_{1-\lambda_{-}}^{n+\lambda_{+}}\right.}\left(\mathbf{y}\middle|\mathbf{x}\right)\mu_{k}\left(X_{1-\lambda_{-}}^{n+\lambda_{+}}=\mathbf{x}\right)\\
 & =\sum_{\mathbf{\tilde{x}\in{\cal X}^{k}}}P_{\left.Y^{n}\middle|X_{1-\lambda_{-}}^{n+\lambda_{+}}\right.}\left(\mathbf{y}\middle|\mathbf{x}\right)\mu\left(\left\{ X_{1-\lambda_{-}}^{n+k+\lambda_{+}}=\left[\mathbf{\tilde{x}},\mathbf{x}\right]\right\} \right)\\
 & \stackrel{\left(a\right)}{=}\sum_{\mathbf{\tilde{x}\in{\cal X}^{k}}}P_{\left.Y^{n+k}\middle|X_{1-\lambda_{-}}^{n+k+\lambda_{+}}\right.}\left(\left\{ Y_{k+1}^{n+k}=\mathbf{y}\right\} \middle|\left[\mathbf{\tilde{x}},\mathbf{x}\right]\right)\\
 & \qquad\qquad\times\mu\left(\left\{ X_{1-\lambda_{-}}^{n+k+\lambda_{+}}=\left[\mathbf{\tilde{x}},\mathbf{x}\right]\right\} \right)\\
 & =\omega\left(T^{-k}\left\{ X_{1-\lambda_{-}}^{n+\lambda_{+}}=\mathbf{x},Y^{n}=\mathbf{y}\right\} \right)
\end{align*}
where $\left(a\right)$ is because the channel is stationary. This
implies that $\omega_{k}=\omega T^{-k}$ and consequently $\eta_{k}=\eta T^{-k}$.

With these facts, similar to \cite{Feinstein-1959}, one can prove
the following limits exist:
\begin{align*}
{\displaystyle \lim_{n\rightarrow\infty}\frac{1}{n}H_{\hat{\mu}}\left(X_{1-\lambda_{-}}^{n+\lambda_{+}}\right)} & ={\displaystyle \lim_{n\rightarrow\infty}\frac{1}{n}H_{\mu}\left(X_{1-\lambda_{-}}^{n+\lambda_{+}}\right)}\\
{\displaystyle \lim_{n\rightarrow\infty}\frac{1}{n}H_{\hat{\omega}}\left(X_{1-\lambda_{-}}^{n+\lambda_{+}},Y^{n}\right)} & ={\displaystyle \lim_{n\rightarrow\infty}\frac{1}{n}H_{\omega}\left(X_{1-\lambda_{-}}^{n+\lambda_{+}},Y^{n}\right)}\\
{\displaystyle \lim_{n\rightarrow\infty}\frac{1}{n}H_{\hat{\eta}}\left(Y^{n}\right)} & =\lim_{n\rightarrow\infty}\frac{1}{n}H_{\eta}\left(Y^{n}\right)
\end{align*}
and consequently,
\[
\lim_{n\rightarrow\infty}\frac{1}{n}I_{\hat{\mu}}\left(X_{1-\lambda_{-}}^{n+\lambda_{+}};Y^{n}\right)=\lim_{n\rightarrow\infty}\frac{1}{n}I_{\mu}\left(X_{1-\lambda_{-}}^{n+\lambda_{+}};Y^{n}\right)
\]
where the probability measure subscript in the entropy quantity implies
that the quantity is calculated w.r.t. that measure, and that in the
mutual information quantity implies the ``rate'' achieved by using
the respective measure as the input distribution. The superscript
$\left(n\right)$ is dropped without ambiguity.

Without loss of generality, let $n=ts-\lambda_{+}-\lambda_{-}$, in
which $t\rightarrow\infty$. Notice that, due to the structure of
$\mu$, the blocks $\left\{ X_{(i-1)s+1-\lambda_{-}}^{is-\lambda_{-}}\right\} _{i=1}^{t}$
are independent of each other under $\mu$.\textbf{ }Then: 
\begin{align*}
 & \frac{1}{n}I_{\mu}\left(X_{1-\lambda_{-}}^{n+\lambda_{+}};Y^{n}\right)\\
 & \geq\frac{1}{n}I_{\mu}\left(X_{1-\lambda_{-}}^{ts-\lambda_{-}};Y_{1}^{r},Y_{s+1}^{s+r},\ldots,Y_{\left(t-1\right)s+1}^{\left(t-1\right)s+r}\right)\\
 & \stackrel{\left(a\right)}{\geq}\frac{1}{n}\sum_{i=1}^{t}I_{\mu}\left(X_{\left(i-1\right)s+1-\lambda_{-}}^{is-\lambda_{-}};Y_{\left(i-1\right)s+1}^{\left(i-1\right)s+r}\right)\\
 & \stackrel{\left(b\right)}{=}\frac{t}{n}I_{\mu}\left(X_{1-\lambda_{-}}^{r+\lambda_{+}};Y_{1}^{r}\right)\\
 & \stackrel{\left(c\right)}{=}\frac{t}{n}I^{\left(r\right)}\left(X_{1-\lambda_{-}}^{r+\lambda_{+}};Y_{1}^{r}\right)\\
 & \geq\frac{1}{r+\lambda_{+}+\lambda_{-}}I^{\left(r\right)}\left(X_{1-\lambda_{-}}^{r+\lambda_{+}};Y_{1}^{r}\right)
\end{align*}
Here $\left(a\right)$ is by applying repeatedly Lemma \ref{lem:Lemma_Mutual_Info_Inequality};
$\left(b\right)$ is because the channel is weakly stationary, $\mu$
is $s$-stationary and consequently $\omega$ is $s$-stationary;
$\left(c\right)$ is due to the fact that the distribution of $X_{1-\lambda_{-}}^{r+\lambda_{+}}$
under $\mu$ is simply $P^{\left(r\right)}$ and the channel is consistent.

Next we indicate the choice of $r$. We have $\forall\epsilon>0$,
there exists $r=r\left(\epsilon\right)$ such that 
\begin{align*}
 & \frac{1}{r+\lambda_{+}+\lambda_{-}}I^{\left(r\right)}\left(X_{1-\lambda_{-}}^{r+\lambda_{+}};Y_{1}^{r}\right)\\
 & \geq\liminf_{N\rightarrow\infty}\frac{1}{N+\lambda_{+}+\lambda_{-}}I^{\left(N\right)}\left(X_{1-\lambda_{-}}^{N+\lambda_{+}};Y^{N}\right)-\epsilon\\
 & =\liminf_{N\rightarrow\infty}\frac{1}{N}I^{\left(N\right)}\left(X_{1-\lambda_{-}}^{N+\lambda_{+}};Y^{N}\right)-\epsilon
\end{align*}
Then: 
\begin{align*}
 & \lim_{n\rightarrow\infty}\frac{1}{n}I_{\hat{\mu}}\left(X_{1-\lambda_{-}}^{n+\lambda_{+}};Y^{n}\right)\\
 & \geq\liminf_{N\rightarrow\infty}\frac{1}{N}I^{\left(N\right)}\left(X_{1-\lambda_{-}}^{N+\lambda_{+}};Y^{N}\right)-\epsilon\\
 & \geq\underline{\mathbf{I}}\left(\mathbf{X};\mathbf{Y}\right)-\epsilon
\end{align*}
where the last inequality is from \cite[Theorem 8.h]{Verdu-Han-1994}.
Maximizing the respective input on each side (i.e. $\hat{\mu}$ on
the left-hand side and $\left\{ X_{1-\lambda_{-}}^{n+\lambda_{+}}\right\} _{n=1}^{\infty}$
on the right-hand side), we obtain:
\begin{align*}
C-\epsilon & \leq\sup_{\hat{\mu}}\lim_{n\rightarrow\infty}\frac{1}{n}I_{\hat{\mu}}\left(X_{1-\lambda_{-}}^{n+\lambda_{+}};Y^{n}\right)\\
 & \leq\sup_{\text{SE }\left\{ X_{1-\lambda_{-}}^{n+\lambda_{+}}\right\} _{n=1}^{\infty}}\lim_{n\to\infty}\frac{1}{n}I\left(X_{1-\lambda_{-}}^{n+\lambda_{+}};Y^{n}\right)\\
 & =C_{SE}
\end{align*}
by the fact that $\hat{\mu}$ is SE and Lemma \ref{lem:Lemma_SE_Capacity}.
Since $\epsilon$ is arbitrary and we know that $C\geq C_{SE}$, we
conclude $C=C_{SE}$.
\end{IEEEproof}

\section*{Appendix C\label{sec:Appendix-Proof_Bit-symmetry}}

\subsection{Proof of Proposition \ref{prop:Proposition_Bit-symmetry (ergodic)}}

Consider a stationary input $\mu$. Let $\bar{\mu}$ be such that
$\bar{\mu}\left(E\right)=\mu\left(\neg E\right)$ for any $E\subseteq\mathrm{GF}\left(2\right)^{\infty}$.
Let $\mu_{0}=\left(\mu+\bar{\mu}\right)/2$. It is easy to see that
$T^{-1}\left(\neg E\right)=\neg\left(T^{-1}E\right)$. Then:
\begin{align*}
\mu_{0}\left(T^{-1}E\right) & =\frac{1}{2}\left(\mu\left(T^{-1}E\right)+\bar{\mu}\left(T^{-1}E\right)\right)\\
 & =\frac{1}{2}\left(\mu\left(T^{-1}E\right)+\mu\left(\neg\left(T^{-1}E\right)\right)\right)\\
 & =\frac{1}{2}\left(\mu\left(T^{-1}E\right)+\mu\left(T^{-1}\left(\neg E\right)\right)\right)\\
 & =\frac{1}{2}\left(\mu\left(E\right)+\mu\left(\neg E\right)\right)=\mu_{0}\left(E\right)
\end{align*}
That is, $\mu_{0}$ is stationary. Furthermore, 
\[
\mu_{0}\left(E\right)=\frac{1}{2}\left(\mu\left(E\right)+\mu\left(\neg E\right)\right)=\mu_{0}\left(\neg E\right)
\]
and therefore $\mu_{0}$ is bit-symmetric.

Since the channel is fixed, let $I_{n}\left(\mu\right)$ denote $I\left(X_{0}^{n};Y_{1}^{n}\right)$
when the input admits $\mu$. Let $\omega$, $\eta$, $\bar{\omega}$
and $\bar{\eta}$ be the joint input-output distributions and the
output distributions of $\mu$ and $\bar{\mu}$ respectively. We have:
\begin{align*}
\bar{\omega}\left(E_{X},E_{Y}\right) & =\int_{E_{X}}\nu_{\mathbf{x}}\left(E_{Y}\right)d\bar{\mu}\left(\mathbf{x}\right)\\
 & =\int_{E_{X}}\nu_{\neg\mathbf{x}}\left(\neg E_{Y}\right)d\mu\left(\neg\mathbf{x}\right)\\
 & =\int_{\neg E_{X}}\nu_{\mathbf{x}}\left(\neg E_{Y}\right)d\mu\left(\mathbf{x}\right)=\omega\left(\neg E_{X},\neg E_{Y}\right)
\end{align*}
Consequently,
\begin{align*}
\bar{\eta}\left(E_{Y}\right) & =\bar{\omega}\left(\mathrm{GF}\left(2\right)^{\infty},E_{Y}\right)\\
 & =\omega\left(\neg\mathrm{GF}\left(2\right)^{\infty},\neg E_{Y}\right)=\eta\left(\neg E_{Y}\right)
\end{align*}
We then have the following:
\begin{align*}
H_{n}\left(\mu\right) & =-\sum_{\mathbf{x}\in\mathrm{GF}\left(2\right)^{n+1}}\mu\left(\left\{ X_{0}^{n}=\mathbf{x}\right\} \right)\log\mu\left(\left\{ X_{0}^{n}=\mathbf{x}\right\} \right)\\
 & =-\sum_{\mathbf{x}\in\mathrm{GF}\left(2\right)^{n+1}}\bar{\mu}\left(\left\{ X_{0}^{n}=\neg\mathbf{x}\right\} \right)\log\bar{\mu}\left(\left\{ X_{0}^{n}=\neg\mathbf{x}\right\} \right)\\
 & =-\sum_{\mathbf{x}\in\mathrm{GF}\left(2\right)^{n+1}}\bar{\mu}\left(\left\{ X_{0}^{n}=\mathbf{x}\right\} \right)\log\bar{\mu}\left(\left\{ X_{0}^{n}=\mathbf{x}\right\} \right)\\
 & =H_{n}\left(\bar{\mu}\right)
\end{align*}
and similarly, $H_{n}\left(\omega\right)=H_{n}\left(\bar{\omega}\right)$,
$H_{n}\left(\eta\right)=H_{n}\left(\bar{\eta}\right)$. Consequently,
$I_{n}\left(\mu\right)=H_{n}\left(\mu\right)+H_{n}\left(\eta\right)-H_{n}\left(\omega\right)=I_{n}\left(\bar{\mu}\right)$.
Given a fixed channel, it has been shown that $I_{n}\left(\mu\right)$
is a concave function of $\mu$ \cite[Corollary 5.5.5]{Gray-Entropy}.
Therefore, 
\[
I_{n}\left(\mu_{0}\right)=I_{n}\left(\frac{1}{2}\left(\mu+\bar{\mu}\right)\right)\geq\frac{1}{2}I_{n}\left(\mu\right)+\frac{1}{2}I_{n}\left(\bar{\mu}\right)=I_{n}\left(\mu\right)
\]
This is true for every $n\geq1$, which implies
\begin{align*}
\sup_{\text{Stationary }\mu}\lim_{n\to\infty}\frac{1}{n}I\left(X_{0}^{n};Y_{1}^{n}\right) & \leq\sup_{\substack{\text{Bit-symmetric,}\\
\text{stationary }\mu
}
}\lim_{n\to\infty}\frac{1}{n}I\left(X_{0}^{n};Y_{1}^{n}\right)
\end{align*}
The proof is complete.

\subsection{Proof of Proposition \ref{prop:Proposition_Bit-symmetry (Shannon-theoretic)}}

Consider a fixed $n\geq1$. Since the channel is fixed, let $I\left(Q^{\left(n\right)}\right)$
denote $I\left(X_{1-\lambda_{-}}^{n+\lambda_{+}};Y^{n}\right)$ when
$X_{1-\lambda_{-}}^{n+\lambda_{+}}\sim Q^{\left(n\right)}$. Define
$\bar{Q}^{\left(n\right)}$ where $\bar{Q}^{\left(n\right)}\left(\mathbf{x}\right)=Q^{\left(n\right)}\left(\neg\mathbf{x}\right)$
$\forall\mathbf{x}\in\mathrm{GF}\left(2\right)^{n+\lambda_{+}+\lambda_{-}}$.
Similar to the proof of Proposition \ref{prop:Proposition_Bit-symmetry (ergodic)},
it can be proven that $I\left(Q^{\left(n\right)}\right)=I\left(\bar{Q}^{\left(n\right)}\right)$.
Next, construct an input distribution $P^{\left(n\right)}$ on $X_{1-\lambda_{-}}^{n+\lambda_{+}}$
such that $P^{\left(n\right)}\left(\mathbf{x}\right)=\left(Q^{\left(n\right)}\left(\mathbf{x}\right)+\bar{Q}^{\left(n\right)}\left(\mathbf{x}\right)\right)/2$
$\forall\mathbf{x}$. It is easy to see that $P^{\left(n\right)}$
satisfies $P^{\left(n\right)}\left(\mathbf{x}\right)=P^{\left(n\right)}\left(\neg\mathbf{x}\right)$
$\forall\mathbf{x}$. Given a fixed channel, $I\left(P^{\left(n\right)}\right)$
is a concave function of $P^{\left(n\right)}$ \cite[Theorem 2.7.4]{Cover-2006}.
Therefore, 
\begin{alignat*}{1}
I\left(P^{\left(n\right)}\right) & =I\left(\frac{1}{2}\left(Q^{\left(n\right)}+\bar{Q}^{\left(n\right)}\right)\right)\\
 & \geq\frac{1}{2}I\left(Q^{\left(n\right)}\right)+\frac{1}{2}I\left(\bar{Q}^{\left(n\right)}\right)=I\left(Q^{\left(n\right)}\right)
\end{alignat*}
This is true for every $n\geq1$.

Now for every stationary input process $\text{IP}_{Q}\equiv\left\{ X_{1-\lambda_{-}}^{n+\lambda_{+}}:\; X_{1-\lambda_{-}}^{n+\lambda_{+}}\sim Q^{\left(n\right)}\right\} _{n=1}^{\infty}$
(with a single underlying probability measure $Q$), if we can find
a stationary input process $\text{IP}_{P}\equiv\left\{ X_{1-\lambda_{-}}^{n+\lambda_{+}}:\; X_{1-\lambda_{-}}^{n+\lambda_{+}}\sim P^{\left(n\right)}\right\} _{n=1}^{\infty}$
with $P^{\left(n\right)}$ defined as above, then the proposition
is proven since $P^{\left(n\right)}$ exhibits bit-symmetry. That
is, we need to justify that $\text{IP}_{P}$ is consistent and stationary.
To show Kolmogorov consistency:
\begin{align*}
 & \sum_{x_{n+1+\lambda_{+}}}P^{\left(n+1\right)}\left(x_{1-\lambda_{-}}^{n+1+\lambda_{+}}\right)\\
 & =\sum_{x_{n+1+\lambda_{+}}}\frac{1}{2}Q^{\left(n+1\right)}\left(x_{1-\lambda_{-}}^{n+1+\lambda_{+}}\right)+\frac{1}{2}Q^{\left(n+1\right)}\left(\neg x_{1-\lambda_{-}}^{n+1+\lambda_{+}}\right)\\
 & =\frac{1}{2}Q^{\left(n\right)}\left(x_{1-\lambda_{-}}^{n+\lambda_{+}}\right)+\frac{1}{2}Q^{\left(n\right)}\left(\neg x_{1-\lambda_{-}}^{n+\lambda_{+}}\right)=P^{\left(n\right)}\left(x_{1-\lambda_{-}}^{n+\lambda_{+}}\right)
\end{align*}
since $\text{IP}_{Q}$ is consistent. We can then assign a single
underlying probability measure $P$ to $\text{IP}_{P}$. To show stationarity,
note that $P\left(E\right)=\left(Q\left(E\right)+Q\left(\neg E\right)\right)/2$
$\forall E$ from the relation between $P^{\left(n\right)}$ and $Q^{\left(n\right)}$.
Then similar to the proof of Proposition \ref{prop:Proposition_Bit-symmetry (ergodic)},
we have $P$ is stationary since $\text{IP}_{Q}$ is stationary.

\section*{Acknowledgement}

The authors would like to thank the editor and anonymous reviewers
for their helpful comments from which this paper greatly benefits. 

\bibliographystyle{IEEEtran}
\bibliography{IEEEabrv,paperBibTex}

\begin{IEEEbiographynophoto}{Phan-Minh Nguyen}
 received his B.Eng degree in electrical engineering from the National
University of Singapore in 2014. He is currently working towards a
Ph.D. degree at Stanford University. His research interest includes
information and coding theory and related fields.
\end{IEEEbiographynophoto}

\begin{IEEEbiographynophoto}{Marc A. Armand}
 received his B.Eng and Ph.D. degree in electrical engineering from
the University of Bristol, U.K., in 1995 and 1999, respectively. He
is currently an Associate Professor at the Department of Electrical
and Computer Engineering, National University of Singapore, which
he joined in Dec 2000 as an Assistant Professor. He is a Senior Member
of the IEEE and has published over 50 journal and conference papers
in the area of coding theory and techniques, as well as coding, detection
and channel modelling for magnetic recording and underwater communications.\end{IEEEbiographynophoto}

\end{document}